\documentclass[12pt,twoside]{article}

\usepackage{color}
\usepackage[
  left=3cm,right=2.5cm,top=3.1cm,bottom=2cm,headheight=2cm,headsep=1.2cm,footskip=1.4cm,twoside
]{geometry}
\usepackage[margin=1cm,font=small]{caption}
\usepackage{amsmath,amssymb}
\usepackage{afterpage}
\usepackage{lineno}
\usepackage{url}
\usepackage{pstricks}
\newcmykcolor{darkgreen}{1 0 0.6 0.5}  %% cyan magenta yellow black
\newcommand{\rd}{}
\usepackage{epsfig}
\usepackage{cite}   % collapse adjacent citations
\usepackage{fancyhdr}
\advance \headheight by 3.0truept       % for 12pt mandatory...

\newlength{\nseparation}
\newenvironment{nfigure}[1]
        {\begin{figure}[#1]\begin{center}\hrule\vspace{\nseparation}\par}
        {\vspace{\nseparation}\par \hrule\end{center} \end{figure}}

\input{babarsymMerge-chka}

\newcommand{\GF}{G_{\rm F}}

\newcommand{\lt}{\left}
\newcommand{\rt}{\right}
\newcommand{\no}{\nonumber}
\newcommand{\nn}{\nonumber \\}
\newcommand{\ov}[1]{\overline{#1}}
\newcommand{\eq}[1]{Eq.~(\ref{#1})}
\newcommand{\eqsand}[2]{Eqs.~(\ref{#1}) and (\ref{#2})}
\newcommand{\eqsto}[2]{Eqs.~(\ref{#1}--\ref{#2})}
\newcommand{\imag}{\mathrm{Im}\,}
\newcommand{\real}{\mathrm{Re}\,}

\newcommand{\unit}[1]{\;[\mathrm{#1}]}
\newcommand{\BRB}[1]{\mathcal{B}(B \to #1)}

\newcommand{\epmuli}[2]{
 \raisebox{-0.5ex}{\shortstack[l]{$\scriptstyle+#1$\\$\scriptstyle-#2$}}}

\newcommand{\bbduli}{\ensuremath{B_d\!-\!\Bbar{}_d\,}}
\newcommand{\bb}{\ensuremath{B\!-\!\Bbar{}\,}}
\newcommand{\bbsuli}{\ensuremath{B_s\!-\!\Bbar{}_s\,}}
\newcommand{\bbq}{\ensuremath{B_q\!-\!\Bbar{}_q\,}}
\newcommand{\kk}{\ensuremath{K\!-\!\Kbar \,}}
\newcommand{\dduli}{\ensuremath{D\!-\!\Dbar \,}}
\newcommand{\ddm}{\dduli\ mixing}
\newcommand{\bbm}{\bb\ mixing}
\newcommand{\bbms}{\bbsuli\ mixing}
\newcommand{\bbmd}{\bbduli\ mixing}
\newcommand{\bbmq}{\bbq\ mixing}
\newcommand{\kkm}{\kk\ mixing}

\newcommand{\bra}[1]{\ensuremath{\langle #1 |}}
\newcommand{\ket}[1]{\ensuremath{| #1 \rangle }}
\newcommand{\fig}[1]{Fig.~\ref{#1}}
\newcommand{\tab}[1]{Tab.~\ref{#1}}

\newcommand{\dm}{\ensuremath{\Delta M}}
\newcommand{\dg}{\ensuremath{\Delta \Gamma}}

\newcommand\Amptpbar{\kern 0.18em\overline{\kern -0.18em {\cal A}}_{3\pi}}

\newcommand\Amptpbarkappa{\kern 0.18em\overline{\kern -0.18em A}^{\kappa}{}}
\newcommand\Amptpbarsigma{\kern 0.18em\overline{\kern -0.18em A}^{\sigma}{}}

\renewcommand\Im{{\rm Im}}
\newcommand\Nbpm{{\kern 0.18em\overline{\kern -0.18em N}}^{+-}}
\newcommand\Nbmp{{\kern 0.18em\overline{\kern -0.18em N}}^{-+}}

\newcommand\BRpmb{{\cal \kern 0.18em\overline{\kern -0.18em  B}}{}_{\rho\pi}^{+-}}
\newcommand\BRmpb{{\cal \kern 0.18em\overline{\kern -0.18em  B}}{}_{\rho\pi}^{-+}}

\newcommand\BRipmb{{\cal \kern 0.18em\overline{\kern -0.18em  B}}{}_{\rho^+\pi^-}}
\newcommand\BRimpb{{\cal \kern 0.18em\overline{\kern -0.18em  B}}{}_{\rho^-\pi^+}}

\newcommand\Abar{\kern 0.18em\overline{\kern -0.18em A}{}}

\def\journalL#1#2#3#4#5{\journal{#1 #2}{#3}{#4}{#5}}
\def\journal#1#2#3#4{#1~{\bf #2}, #3 (#4)}

\def\PLB#1#2#3{\journal{Phys.\ Lett. B}{#1}{#2}{#3}}

\def\NPB#1#2#3{\journal{Nucl.\ Phys. B}{#1}{#2}{#3}}

\def\PRD#1#2#3{\journal{Phys.\ Rev. D}{#1}{#2}{#3}}
\def\PRL#1#2#3{\journal{Phys.\ Rev. Lett.}{#1}{#2}{#3}}

\newcommand{\arxiv}[1]{{arxiv:{#1}}}

\newcommand{\etal}{\emph{et al.}\xspace}

\newcommand{\Bag}{\mathcal{B}}
\newcommand{\HatBag}{\hat{\mathcal{B}}}

\newcommand{\Dzero}{D\O\xspace}
\newcommand{\TeVatron}{Tevatron\xspace}

\newcommand{\BabarColl}{[\babar\ collaboration]\xspace}
\newcommand{\BelleColl}{[Belle collaboration]\xspace}
\newcommand{\DzeroColl}{[\Dzero collaboration]\xspace}
\newcommand{\CDFColl}{[CDF collaboration]\xspace}

\newcommand{\Order}{\mathrm{O}}
\newcommand{\percent}{~\%}

\newcommand{\braOket}[3]{\langle#1|#2|#3\rangle}

\newcommand{\expe}{\mathrm{e}}

\newcommand{\ed}{\mathrm{d}}
\newcommand{\ii}{\mathrm{i}}

\begin{document}
%\linenumbers
\thispagestyle{plain}
\begin{titlepage}
   {\noindent HU-EP-10/43 \hfill \raggedleft TTP10-33\\
   DO-TH 10/05 \hfill \raggedleft SFB/CPP-10-68\\
   LPT-ORSAY/10-59   \hfill \raggedleft CPT-P040-2010\\
  }
  \vskip 2em
  {\centering \Huge \bf Anatomy of New Physics \\ in
    $\boldsymbol{B}$--$\boldsymbol{\overline{B}}$ mixing \\}
  \vskip 2.5em
  {\centering
    \large 
    A.~Lenz$^{a,b}$, U.~Nierste$^c$\\[0.6em]
    and \\[0.6em]
    J.~Charles$^d$, S.~Descotes-Genon$^e$, A.~Jantsch$^f$, C.~Kaufhold$^g$, H.~Lacker$^h$, S.~Monteil$^i$, V.~Niess$^i$, S.~T'Jampens$^g$\\[0.2cm]
    [CKMfitter Group]\\
  }
    \vskip 2em \noindent
{\small \em $^{a}$Institut f\"ur Physik, Technische Universit\"at Dortmund, 
                   D-44221 Dortmund, Germany}\\[0.2cm]
{\small \em $^{b}$Institut f\"ur Theoretische Physik, Universit\"at Regensburg, 
                   D-93949 Regensburg, Germany,
                {e-mail: Alexander.Lenz@physik.uni-regensburg.de}} \\[0.2cm]
{\small \em $^{c}$Institut f\"ur Theoretische Teilchenphysik, Universit\"at Karlsruhe, Karlsruhe Institute of Technology, 
                   D-76128 Karlsruhe, Germany,
                {e-mail: nierste@particle.uni-karlsruhe.de}} \\[0.2cm]
{\small \em $^{d}$Centre de Physique Th\'eorique, 
                   Campus de Luminy, 
                   Case 907, F-13288 Marseille Cedex 9, France
                   (UMR 6207 du CNRS associ\'ee aux
                   Universit\'es d'Aix-Marseille I et II et
                   Universit\'e du Sud Toulon-Var; laboratoire
                   affili\'e \`a la FRUMAM-FR2291),
                {e-mail: charles@cpt.univ-mrs.fr}} \\[0.2cm]
{\small \em $^{e}$Laboratoire de Physique Th\'eorique d'Orsay,
                  UMR8627, CNRS/Univ. Paris-Sud 11, 91405 Orsay Cedex, France,
                {email: sebastien.descotes-genon@th.u-psud.fr}} \\[0.2cm]
{\small \em $^{f}$Max-Planck-Institut f\"ur Physik 
                   (Werner-Heisenberg-Institut),
		   F\"ohringer Ring 6, 80805 M\"un\-chen, Germany,
                {e-mail: jantsch@mppmu.mpg.de}} \\[0.2cm]
{\small \em $^{g}$Laboratoire d'Annecy-Le-Vieux de Physique des Particules, 
                   9 Chemin de Bellevue, BP 110, F-74941
                   Annecy-le-Vieux Cedex, France
                   (UMR 5814 du CNRS-IN2P3 associ\'ee \`a
                   l'Universit\'e de Savoie),
                {e-mail: kaufhold@lapp.in2p3.fr, tjampens@lapp.in2p3.fr}} \\[0.2cm]
{\small \em $^{h}$Humboldt-Universit\"at zu Berlin,
                   Institut f\"ur Physik,
                   Newtonstr. 15,
                   D-12489 Berlin, Germany,
                {e-mail: lacker@physik.hu-berlin.de}}\\[0.2cm]
{\small \em $^{i}$Laboratoire de Physique Corpusculaire de Clermont-Ferrand,
                  Universit\'e Blaise Pascal,
                  24 Avenue des Landais F-63177 Aubi\`ere Cedex,
		  (UMR 6533 du CNRS-IN2P3 associ\'ee \`a
                   l'Universit\'e Blaise Pascal),
		  {e-mail: monteil@clermont.in2p3.fr, niess@clermont.in2p3.fr}}
		  
 \newpage\thispagestyle{plain} 
  {\centerline{\textbf{\Large Abstract}}
  \vskip 2em
    \noindent
  We analyse 
  three different New Physics scenarios for 
  $\Delta F=2$ flavour-changing neutral currents in the quark sector in the light
  of recent data on neutral-meson mixing. 
  We parametrise generic New Physics contributions to  
  \bbmq, $q\!=\!d,s$, in terms of one complex 
  quantity $\Delta_q$, while three parameters $\Delta_K^{tt}$,
  $\Delta_K^{ct}$ and
  $\Delta_K^{cc}$ are needed to describe \kkm.  
  In Scenario I, we consider uncorrelated 
  New Physics contributions in the $B_{d}$, $B_{s}$, and $K$ sectors. 
  In this scenario, it is only possible to constrain 
  the parameters $\Delta_d$ and $\Delta_s$ whereas there are no non-trivial
  constraints on the kaon parameters.
  In Scenario II, we study the case of 
  Minimal Flavour Violation (MFV)
  and  
  small bottom Yukawa coupling, where
  $\Delta\equiv\Delta_d=\Delta_s=\Delta_K^{tt}$. We show that
  $\Delta$ must then be real, so that no new CP phases can be accomodated,
  and express the remaining parameters $\Delta_K^{cc}$ and $\Delta_K^{ct}$
  in terms of $\Delta$ in this scenario.
%  In addition, 
%  $\Delta_K^{cc}$ is very close to the Standard Model value 
%  $\Delta_K^{cc}=1$ and $\Delta_K^{ct}$ 
%  can be approximately expressed in terms of $\Delta_K^{tt}$.  
  Scenario III is the generic MFV case with
  large bottom Yukawa couplings. 
  In this case, the Kaon sector is uncorrelated to the $B_d$ and 
  $B_s$ sectors.  As in the second scenario 
  one has $\Delta_d=\Delta_s\equiv\Delta$, however, 
  now with a complex parameter $\Delta$. 
  Our quantitative analyses consist of global CKM fits within the Rfit frequentist statistical approach, 
  determining the Standard Model parameters and the
  new physics parameters of the studied scenarios simultaneously. 
We find that the recent
measurements indicating discrepancies with the Standard Model are well
accomodated in Scenarios I and III with new mixing phases, 
with a slight preference for Scenario I
%However, Scenario III
%is slightly disfavoured with respect to the first one 
that permits 
different new CP phases in the $B_d$ and $B_s$ systems. 
Within our statistical framework, we find evidence of New Physics in both $B_d$ and $B_s$ systems.
The Standard-Model hypothesis $\Delta_d=\Delta_s=1$ is disfavoured 
with p-values of $3.6\sigma$ and $3.3\sigma$ in Scenarios I and III, respectively.
We also present an exhaustive list of numerical predictions in each
     scenario. In particular, we predict the
     CP phase in $B_s \to J/\psi \phi$ and the difference between the
     $B_s$ and $B_d$ semileptonic asymmetries, which
     will be both measured by the LHCb experiment.
}

\end{titlepage}

%%% Local Variables: 
%%% mode: latex
%%% TeX-master: "main"
%%% End: 

\clearpage
\tableofcontents 
\clearpage

\section{Introduction}\label{sec:intro}

Considerations of the stability of the electroweak scale lead to the
general belief that there is New Physics with particle masses below 1
TeV. While the high-$p_T$ experiments at the LHC should
produce these new particles directly, one can study their dynamics also
indirectly, through their impact on precision measurements at lower
energies. To this end flavour-changing neutral current (FCNC)
processes are extremely useful. On one hand they are highly
suppressed in the Standard Model and are therefore very sensitive to new
physics. On the other hand FCNC processes of $K$, $B_d$ and $B_s$ mesons
are still large enough to be studied with high statistics in dedicated
experiments. Here meson--antimeson mixing plays an outstanding
role. First, meson--antimeson oscillations occur at time scales which
are sufficiently close to the meson lifetimes to permit their
experimental investigation. Second, the Standard Model contribution to
meson--antimeson mixing is loop-suppressed and comes with two or more
small elements of the Cabibbo-Kobayashi-Maskawa (CKM) matrix \cite{ckm}.
Third, the decays of oscillating mesons give access to many
mixing-induced \CP asymmetries through the time-dependent study of
decays into CP-eigenstates, which in some cases one can relate to the
parameters of the underlying theory with negligible hadronic uncertainties.

The $B$-factories have revealed that the dominant $b\to d$ and $b\to u$ 
transitions fit into the pattern of the CKM mechanism and are in agreement 
with the information on $s\to d$ transitions gained in more than fourty 
years of kaon physics. The success of the CKM picture is evident from 
the many different measurements combining into a consistent and precise
determination of the apex $(\rhobar,\etabar)$ of the $B$-meson
unitarity triangle (in terms of the Wolfenstein parameterization of the
CKM matrix \cite{Wolfenstein,BurasLautenbacherOstermaier}).  As a
consequence, any contribution from the expected new TeV-scale physics to
the measured flavour-changing processes must be suppressed
compared to the established CKM mechanism.

Models with only CKM-like flavour violation are said to respect the
principle of \emph{minimal flavour violation (MFV)}
\cite{dgis,kpvz}. This principle is often invoked in an ad-hoc way
  to suppress excessive FCNC amplitudes for model-building purposes.  
   In the Minimal Supersymmetric
  Standard Model (MSSM), new sources of flavour violation solely stem
  from the supersymmetry-breaking sector. A sufficient condition for MFV
  are supersymmetry-breaking terms which are flavour-blind at a given
  energy scale. This situation occurs in supergravity with a flat K\"ahler
  metric \cite{Bertolini:1990if} or if supersymmetry breaking is
  mediated by gauge interactions \cite{gmsb}. The overall picture of
experimental data does not require sizeable corrections to MFV. Still it
is difficult to probe the CKM picture with a better accuracy than, say,
30\%, because most quantities entering the global fit of the unitarity
triangle suffer from sizeable hadronic uncertainties. It should also be
stressed that the accuracy of the determination of the CKM parameters
decreases notably when one assumes that one or several crucial input(s)
could be affected by New Physics contributions. Interestingly, several
authors have detected possible hints of New Physics in the data. For
example it has been argued in the literature that one starts to see a
discrepancy between the measurement of $\sin2\beta$ and the region
preferred by $|V_{ub}|$ from semileptonic decays on one hand, and
$|\varepsilon_K|$ on the other hand~\cite{smtensions,bg}.  Also the
recently improved measurement of the $B\to\tau\nu$ branching ratio
deviates from its indirect CKM fit prediction~\cite{CKMfitter08}.  In
addition there are anomalies in the data on $b\to s$ transitions.  The
latter processes do not involve $\rhobar$ and $ \etabar$ (to a good
accuracy) and therefore directly probe the CKM mechanism. An ongoing
debate addresses an extra contribution to $b\to s \ov{q} q$, $q=u,d,s$,
decay amplitudes with a \CP phase different from $\arg(V_{ts}^*
V_{tb}^{\phantom{*}})$ that can alleviate the pattern of shifts between
the measured \CP asymmetries in these $b\to s$ penguin modes and the
Standard Model predictions (see e.g. \cite{cppeng}).

However, the first place to look for New Physics in $b\to s$ transitions
is \bbms, where New Physics can be parameterised in terms of just
one complex parameter (or two real parameters) in a
model-independent way, as we will discuss in great detail below. At the
end of 2006 a combined analysis of several observables has pointed to
the possibility of a new-physics contribution with a \CP phase different
from that of the Standard-Model box diagram \cite{ln}. Models of
supersymmetric grand unification can naturally accommodate new
contributions to $b \to s$ transitions \cite{cmm}: right-handed quarks
reside in the same quintuplets of SU(5) as left-handed neutrinos, so
that the large atmospheric neutrino mixing angle could well affect
squark-gluino mediated $b\to s$ transitions.  At the same time the GUT
models of Refs.~\cite{cmm} do not induce too dangerous contributions to
the well-measured rare decay $B\to X_s \gamma$.  \bbms\ has been further
investigated in other supersymmetric scenarios with \cite{othergut} and
without \cite{SUSY,bsltb} GUT boundary conditions, in unparticle physics
scenarios in Ref.~\cite{unparticle}, in multi-Higgs-doublet models
\cite{MoreHiggsDoubletts}, in models with extra gauge bosons $Z'$
\cite{Zprime}, warped extra dimensions \cite{WarpedExtraDimensions},
left-right symmetry \cite{LRmodels}, anomalous tWb-couplings
\cite{AnomaloustWbcouplings}, additional quark families
\cite{fourthfamily} or an additional singlet quark
\cite{AdditionalsingletQuark}, and in a little-Higgs model
\cite{dmphenburas}.  

On the experimental side, the understanding of $b\to s$ transitions has
made a tremendous progress in the past years. The \TeVatron experiments
have discovered and precisely quantified \bbms\ oscillations
\cite{dmsexp,dmsexpvalue} whose frequency is in good agreement with the
Standard Model prediction, and presented first determinations of the associated
\CP-violating phase from tagged analyses of $B_s\to \jpsi \phi$
decays~\cite{taggedphaseCDF,taggedphaseD0,taggedphaseCDF_2,taggedphaseCDF_3,taggedphaseD0_2}.
Recently, possible New Physics in the \CP\ phase of the \bbms\ amplitude
has received new attention: The D\O\ collaboration has reported a
measurement of the dimuon charge asymmetry which disagrees with the Standard Model
prediction by 3.2 standard deviations \cite{dimuon_evidence_d0}. The CP
asymmetry in semileptonic or, more generally, any flavour-specific
decays, involves the \bbms\ phase just as $B_s\to \jpsi \phi$, so that
both pieces of experimental information can be combined to constrain
this phase.  The new measurement of the dimuon charge asymmetry has
already triggered considerable theoretical interest.  Besides predictions
for the \CP\ phase of \bbms\ in specific models, as quoted above, also
model-independent analyses of New Physics effects have appeared
\cite{paperafterD0}. Due to the large size of the dimuon asymmetry it
was also investigated whether sizeable New Physics contributions to the
decay of $B_s$ mesons are possible. This alternative is however strongly
constrained by the lifetime ratios of $B$-mesons (see
e.g. \cite{lifetimes}) as well as the semileptonic branching ratios and
the the average number of charm quarks per b-decays (see
e.g. \cite{nc}).  

In the present article, we analyse generic scenarios of New Physics which are
compatible (at different levels) with the experimental picture sketched
above. We set up our notation and define our theoretical framework in
Sect.~\ref{sec:scene}, the relevant updated experimental and theoretical
inputs to our global analysis are presented in Sect.~\ref{ssec:inputs}.
In Sect.~\ref{ssec:SMfit} we first present the current status of the
Standard Model fit. In Sect.~\ref{ssec:bdbs} we perform a fit in which
we allow for New Physics in the \bbmd\ and the \bbms\ systems and we
project the results onto the New Physics parameters that describe the
\bbmd\ and the \bbms\ systems. Sects.~\ref{ssec:mfv1} and \ref{ssec:mfv2}
are dedicated to two MFV scenarios with correlated effects in all
meson--antimeson mixing amplitudes. Finally, we conclude and list a few
perspectives {\rd for} the close future.
 
\section{Setting the scene}\label{sec:scene}
\boldmath
\subsection{\bbm\ basics}\label{ssec:bbmbasics}
\unboldmath
\bbq\ oscillations (with $q=d$ or $q=s$) are described by a
Schr\"odinger equation:
\begin{equation}
\ii \frac{\ed}{\ed t}
\left(
\begin{array}{c}
\ket{B_q(t)} \\ \ket{\bar{B}_q (t)}
\end{array}
\right)
=
\left( M^q - \frac{\ii}{2} \Gamma^q \right)
\left(
\begin{array}{c}
\ket{B_q(t)} \\ \ket{\bar{B}_q (t)} 
\end{array}
\right)\label{sch}
\end{equation} 
with the mass matrix $M^q=M^{q\dagger}$ and the decay matrix 
$\Gamma^q=\Gamma^{q\dagger}$. 
The physical eigenstates $\ket{B^q_H}$ and $\ket{B^q_L}$ with masses
$M^q_H,\,M^q_L$ and decay rates $\Gamma^q_H,\,\Gamma^q_L$ are obtained
by diagonalizing $M^q-\ii \, \Gamma^q/2$.  The \bbq\ oscillations in
\eq{sch} involve the three physical quantities $|M_{12}^q|$,
$|\Gamma_{12}^q|$ and the \CP phase
\begin{eqnarray}
\phi_q &=& \arg(-M_{12}^q/\Gamma_{12}^q) . \label{defphi}
\end{eqnarray}
We denote the average $B_q$ mass and
width by $M_{B_q}$ and $\Gamma_{B_q}$, respectively.
The mass and width differences
between $B^q_L$ and $B^q_H$ are related to them as
\begin{eqnarray}
\dm_q &=& M^q_H -M^q_L \; = \; 2\, |M_{12}^q|,
\qquad \dg_q \; =\; \Gamma^q_L-\Gamma^q_H \; =\;
        2\, |\Gamma_{12}^q| \cos \phi_q, \label{dmdg}
\end{eqnarray}
up to numerically irrelevant corrections of order $m_b^2/M_W^2$.
$\dm_q$ simply equals the frequency of the \bbq\ oscillations (for
details see e.g.\ \cite{run2}). 
A third quantity probing mixing is 
\begin{eqnarray}
a^q_\text{fs} \; =\; 
2\, \lt( 1- \lt| \frac{q}{p}\rt| \rt)
     &=&
    \imag \frac{\Gamma_{12}^q}{M_{12}^q}
    \; = \; \frac{|\Gamma_{12}^q|}{|M_{12}^q|} \sin \phi_q
    \; = \; \frac{\dg_q}{\dm_q} \tan \phi_q
 . \label{defafs}
\end{eqnarray}
$a_\text{fs}^q$ is the \CP asymmetry in \emph{flavour-specific} $B_q\to
f$ decays, i.e., the decays $\bar B_q \to f$ and $B_q \to \bar f$ are
forbidden.  The standard way to measure $a_\text{fs}^q$ uses $B_q \to X
\ell^+ \nu_\ell$ decays, which explains the common name
\emph{semileptonic \CP asymmetry} for $a_\text{fs}^q$, with the
corresponding notation $a_\text{SL}^q$ (for more details see e.g.\
\cite{n}).  In theoretical contexts, we use the notation
  $a_\text{fs}^q$ in this paper, while we write $a_{\rm SL}^q$ when
  referring to the specific experimental observable inferred from
  semileptonic decays. Further
\begin{equation}\label{eq:qp}
\frac qp = {\rd -}
\sqrt{ \frac{2 \, M_{12}^{*}\,-\,i \Gamma_{12}^{*}}{2 \, M_{12}\,-\, i\Gamma_{12}}}.
\end{equation}
Let us now discuss our theoretical understanding of the off-diagonal terms 
of the evolution matrix, which are responsible for \bbq mixing. The 
dispersive term $M_{12}^q$ is completely dominated by box diagrams 
involving virtual top quarks, and it is related to the effective 
$|\Delta B|=2$ Hamiltonian $H_q^{|\Delta B|=2}$ as
\begin{eqnarray}\label{eq:m12q}
M_{12}^q &=& \frac{\braOket{B_q}{H_q^{|\Delta B|=2}}{\ov B_q}}{2 M_{B_q}} \, .
\end{eqnarray}
The Standard Model expression for $H_q^{|\Delta B|=2}$ is \cite{hw}
\begin{eqnarray}
H_q^{|\Delta B|=2} &=& (V_{tq}^* V_{tb}^{\phantom{*}})^2 \, C \, Q \; +\;  
                       \mbox{h.c.}\label{defheff} 
\end{eqnarray}
with the four-quark operator 
\begin{eqnarray}
Q &=& \ov q_L \gamma_\mu b_L \, \ov q_L \gamma^\mu b_L \label {defq} \, ,
\\
  && q_L = \frac{1}{2}  \left( 1 - \gamma_5 \right) q\, ,
\end{eqnarray}
and the Wilson coefficient $C$, which depends on the heavy 
mass scales of the theory. In a wider class of models $H_q^{|\Delta B|=2}$ 
maintains the form of \eq{defheff} (meaning that there is no other operator 
than $Q$ involved), but with a value of $C$ different from the value in 
the Standard Model:
\begin{eqnarray}
C^{\rm SM} &=& \frac{G_\text{F}^2}{4\pi^2} M_W^2 \, \widehat \eta_B\, 
      S\lt( \frac{\ov m_t^2}{M_W^2} \rt) . \label{defc}
\end{eqnarray}
Here $\ov m_t$ is the top quark mass defined in the $\ov{\text{MS}}$ scheme, 
related to the pole mass $m_t^\text{pole}$ determined at the \TeVatron as 
$\ov m_t(\ov m_t) =  0.957\ m_t^\text{pole}$  (at next-to-leading order of QCD). 
The Inami-Lim function $S$~\cite{il} is calculated from the box 
diagram with two internal top quarks and evaluates to 
$S(\ov{m}_t^2/M_W^2)=2.35$ for the central value of $\ov{m}_t$ listed in 
Table~\ref{tab:TheoreticalInputs}.
QCD corrections are comprised in \cite{bjw,Buras:1990fn}
\begin{eqnarray}
\widehat \eta_B &=& 0.8393 \pm 0.0034 . \label{defeta2}
\end{eqnarray}
The hadronic matrix element involved is usually parameterised as 
\begin{eqnarray}
\braOket{B_q}{Q(\mu_B)}{\ov B_q}  
&=& \frac{2}{3} M^2_{B_q}\, f^2_{B_q} \Bag_{B_{q}} (\mu_B) \, , 
      \label{defb} 
\end{eqnarray}
with the decay constant $f_{B_q}$ and the `bag' factor $\Bag_{B_{q}}$. 
The product $\widehat \eta_B \Bag_{B_{q}}$ is scale and scheme invariant.
Our convention in \eq{defeta2} corresponds to a scale dependent bag parameter
with $\Bag_{B_{q}} = 1$ in vacuum insertion approximation. Typical values for 
the bag parameter obtained on the lattice are e.g. $\Bag_{B_{s}} \approx 0.84$, see 
Sect.~\ref{ssec:MethodofAveraging}.
Sometimes a different normalisation with a scale independent bag parameter 
$\widehat \Bag_{B_{q}} = b_B(\mu_B) \Bag_{B_{q}}(\mu_B)$ is used. The 
corresponding quantities 
$\eta_B= \widehat \eta_B(\mu_B) /  b_B(\mu_B) = 0.551 $ and 
$\widehat\Bag_{B_{s}} \approx 1.28$ obviously satisfy 
$\eta_B \widehat \Bag_{B_{s}}= \widehat \eta_B \Bag_{B_{s}}$. 
The analytic formula for  $b_B(\mu_B)$ can be found e.g. 
in Eq.~(XIII.5) of \cite{BBL95}.

The absorptive term $\Gamma_{12}^q$ is dominated by on-shell
charmed intermediate states, and it can be expressed as a two-point
correlator of the $|\Delta B|=1$ Hamiltonian $H_q^{|\Delta B|=1}$. 
By performing a $1/m_b$-expansion of this two-point
correlator, one can express $\Gamma_{12}^q$ in terms of $Q$ and
another four-quark operator
\begin{eqnarray}
\widetilde{Q}_S & = &  \ov{q}_L^\alpha b_R^\beta \, 
                       \ov{q}_L^\beta  b_R^\alpha,
\label{defqst}
\end{eqnarray}
where $S$ stands for ``scalar'' and $\alpha,\beta=1,2,3$ are colour indices, 
see \cite{ln}. The matrix element is expressed as 
\begin{eqnarray}
\braOket{B_q}{\widetilde{Q}_S}{\ov B_q} &=& \frac{1}{12}  M^2_{B_q}\,
                  f^2_{B_q} \widetilde \Bag_{S,B_q} 
                  \left( \frac{M_{B_q}}{\bar m_b + \bar m_q}\right)^2
                  =: \frac{1}{12}  M^2_{B_q}\,
                  f^2_{B_q} \widetilde \Bag_{S,B_q}^\prime . 
    \label{defbstp} 
\end{eqnarray}
The prediction of $\Gamma_{12}^q$ involves also operators which are
subleading in the heavy quark expansion, the matrix elements of which
are parameterised by the `bag' factors $\Bag_{R_{0,1,2,3}}$ and
$\Bag_{\tilde{R}_{1,2,3}}$~\cite{ln}, which satisfy two relations
in the heavy quark limit \cite{dega,ln}:
\begin{equation} \label{eq:HQETBagR}
\Bag_{R_2}=\Bag_{\tilde{R}_{2}}, \qquad
\Bag_{R_3}=\frac{5}{7}\Bag_{\tilde{R}_{3}} +\frac{2}{7}\Bag_{\tilde{R}_{2}}.
\end{equation}
Even though we have not included the flavour of the light-quark in our notation,
we consider these $1/m_b$-suppressed operators to have different values for
 $B_d$ and $B_s$ mesons.

Finally, we discuss the relative phase $\phi_q$ between the two off-diagonal terms. 
In contrast to $M_{12}^{q}$, $\Gamma_{12}^{q}$ receives 
non-negligible contributions from subleading $u$ and $c$ CKM
couplings, which implies that $\phi_q$ is not a pure CKM phase in the 
Standard Model. The Standard Model contribution to $\phi_q$ reads \cite{bbln,ln},
with our updated inputs (see Table~\ref{tab:fitResults_SM2})
\begin{eqnarray}
\phi_d^{\text{SM}} & = & (-10.1^{+3.7}_{-6.3})\cdot 10^{-2}\,,\no\\
\phi_s^{\text{SM}} & = & (+7.4^{+0.8}_{-3.2})\cdot 10^{-3} \,, \label{finphi}
\end{eqnarray}
and thus in $\phi_s$ the Standard Model contribution is clearly subleading in the presence of generic 
New Physics effects. 

The previous quantities are expected to be affected by New Physics in different ways. 
While $M_{12}^q$ coming from box diagrams is very sensitive to New Physics 
both for $B_d$ and $B_s$, $\Gamma_{12}^s$ stems from Cabibbo-favoured
tree-level decays and possible New Physics effects are expected to be smaller 
than the hadronic uncertainties.
In the case of $\Gamma_{12}^d$ though, the contributing decays are Cabibbo-suppressed. 
In this paper, we only consider scenarios where New Physics does not enter tree-level 
decays. More specifically, we assume that  $B$ decays 
proceeding through a  four-flavour change ($i.e.$, 
$b \to q_1\bar q_2q_3$, $q_1 \ne q_2 \ne q_3$) obtain only Standard Model contributions 
(\emph{SM4FC})~\cite{Goto,CKMfitter2}. This assumption is better defined than
just the neglect of New Physics contributions to tree-mediated decays since on the 
non-perturbative level tree and penguin amplitudes can not be well separated.
Our class of  four-flavour-change decays  includes  $b\to d$ decays in which the 
strong isospin changes by 3/2 units, i.e.\ we use strong isospin as the
flavour quantum number of the first quark generation.
Then the following inputs used in the fit are considered to be free from New Physics 
contributions in their extraction from data: $|V_{ud}|$, $|V_{us}|$,
$|V_{ub}|$, $|V_{cb}|$ and $\gamma$. Also
the leptonic decays $B\to \tau \nu$ (or $D_{s} \to \tau \nu$ and $D_{s} \to \mu \nu$),
which could be significantly affected by charged Higgs exchange 
contributions (see~\cite{Deschamps:2009rh} and references within), are assumed to be Standard Model-like. Using these inputs a reference 
unitarity triangle can be constructed (see the first two articles in Ref.~\cite{Goto}), as will be discussed further in 
Sect.~\ref{ssec:InputsWithoutNP} (in Ref.~\cite{Blanke:2006ig}, this triangle is compared with the universal unitarity triangle for models of constrained MFV introduced in Ref.~\cite{Buras}).

In addition, in order to take advantage of the measurement of the {\rd
  width differences $\dg_q$}, and of the time-dependent CP-asymmetries
in dominant $b\to c$ decays, we neglect possible non standard
contributions to the $b\to c\bar cq$ ($q=d,s$) transitions, although
they do not strictly enter the SM4FC family. Finally, we assume that the
unitarity of the $3 \times 3$-CKM matrix still holds in the presence of
New Physics, which ensures that the Standard Model contribution to the neutral meson mixing
keeps its usual expression as a function of $(\bar \rho,\bar \eta)$ and
other parameters.  Hence, our discussion would not hold in the case of
an additional sequential fourth fermion family, which however is
not excluded yet by experimental 
constraints~(see Refs.~\cite{fourthfamily} or \cite{menzellacker} and references therein).

Thus, New Physics can find its way 
into the quantities studied in this paper only by changing magnitude and/or phase 
of $M_{12}^q$. It is convenient to define the New Physics complex parameters $\Delta_q$ and 
$\phi^\Delta_q$ ($q=d,s$) through
\begin{eqnarray}
M_{12}^q & \equiv & M_{12}^{\text{SM},q} \cdot  \Delta_q \, ,
\qquad\qquad  \Delta_q \; \equiv \;  |\Delta_q| \expe^{\ii \phi^\Delta_q} , 
 \label{defdel}
\end{eqnarray}
see e.g.~\cite{ln}. With the definition in \eq{defdel} the \CP phase of \eq{defphi} reads
\begin{eqnarray}
\phi_q &=& \phi_q^{\text{SM}} + \phi^\Delta_q . \label{phism}
\end{eqnarray}
As discussed in Sect.~\ref{ssec:inputs}, New Physics in $M_{12}^q$ will
not only affect the neutral-meson mixing parameters, but also the
time-dependent analyses of decays corresponding to an interference
between mixing and decay.  

The relation of $\Delta_q$ to the parameters used e.g. in \cite{rtheta,nir2006, Laplace:2002ik} is
\(
|\Delta_q| = r_q^2\,, 
\phi^\Delta_q  = 2\theta_q 
\), and the Standard Model is of course located at $\Delta_q=1$.
It is more transparent to look at the cartesian $\imag \Delta_q$ vs.\ $\real \Delta_q$ plot
than the polar $2\theta_q$ vs.\ $r_q^2$ one, because it
visualizes the New Physics contribution more clearly and it allows a simple 
geometrical interpretation of the shape of each individual constraint.
For completeness, we note that some authors (e.g. \cite{Agashe,Ligeti:2006pm,Ball:2006xx}, 
see also~\cite{Goto}) 
prefer to split the Standard Model 
contribution from the pure New Physics one in a polar parametrization.
 The two New Physics parameters $h_q$ and 
$2\sigma_{q}$ introduced in this way are defined by
\begin{eqnarray}
\frac{M_{12}^{q}}{M_{12}^{\text{SM},q}} =  1 + \frac{M_{12}^{\text{NP},q}}{M_{12}^{\text{SM},q}} 
= \Delta_q 
         \:=\: 1+h_q\,\expe^{\ii 2\sigma_{q}}~.
\end{eqnarray}
We will study the case of the neutral kaon system in the following section, defining
ana\-logous parameters $\Delta_K^{tt}$, $\Delta_K^{ct}$ and $\Delta_K^{cc}$. But 
in this paper, we will not consider the neutral $D$ meson system. Indeed,
in the scenarios we consider here \ddm\ is severely GIM-suppressed and
gives no useful constraint, as is already the case within the Standard
Model, see e.g.~\cite{DmixHQE}.

\boldmath
\subsection{\kkm\ basics}\label{ssec:kkmbasics}
\unboldmath
The effective $|\Delta S|=2$ Hamiltonian describing \kkm\ resembles 
the  $|\Delta B|=2$ Hamiltonian of \eq{defheff}, with the important 
distinction that now also contributions from internal charm quarks are 
important: 
\begin{eqnarray}
H^{|\Delta S|=2} & = & \lt[ 
      ( V_{ts} V_{td}^* )^2 \, C_{tt}   \; + \;   
      2 V_{ts} V_{td}^* V_{cs} V_{cd}^* \, C_{ct} 
      \; + \; ( V_{cs} V_{cd}^* )^2 \, C_{cc}  \rt] \,
      Q 
  \; + \; h.c. \label{deshs2}
\end{eqnarray}
with the operator $Q=\ov d_L \gamma_\mu s_L \, \ov d_L \gamma^\mu s_L$~\footnote{We 
use the same notation for operators in the $B$ and $K$ systems (cf.~\eq{defq}), 
implying the corresponding flavours ($b$, $s$ or $d$) of the quark fields.}.
As for the case of the $B_d$ and $B_s$ mesons, the contribution
from $H^{|\Delta S|=2}$ to $M_{12}^K$ is found from \eq{eq:m12q}. A
new feature is an additive poorly calculable long-distance contribution 
involving $H^{|\Delta S|=1}$ (see e.g.\ \cite{run2,Nierste:2009wg}). 
The Wilson coefficients $ C_{ij}$, $i,j=c,t$, and the operator $Q$ depend on 
the renormalisation scale $\mu_K$ at which we evaluate these coefficients 
and the hadronic matrix element $\bra{K}Q\ket{\Kbar}$.  
We parametrise the hadronic matrix element as
\begin{eqnarray}
 \bra{K} Q(\mu_K) \ket{\,\ov{\!K}} & =&
  \frac{2}{3} m_{K}^2 \, f_K^2 \,
  \frac{\widehat \Bag_K}{ b_K(\mu_K)} . \label{melk}%
\end{eqnarray}
Here $f_K\simeq 156\,\mev$ and $m_K$ are the decay constant and  mass 
of the kaon, respectively, and $\widehat\Bag_K$ is the bag parameter, 
from  which a factor $ b_K(\mu_K)$ is stripped off. 
Analogous to the case of $B$ mixing $ b_K(\mu_K)$ contains the 
dependence of $ \bra{K} Q(\mu_K) \ket{\,\ov{\!K}}$ on the renormalisation
scheme and the renormalisation scale $\mu_K$. 
The Standard Model values of the Wilson coefficients are 
\begin{eqnarray}
 C_{tt}^{\rm SM} & = & \frac{G_F^2}{4 \pi^2}\, M_W^2 \,
            S\lt( \frac{\ov{m}_t^2}{M_W^2} \rt)
            \eta_{tt} \,  b_K(\mu_K)  \, , \nn 
 C_{ct}^{\rm SM} & = & \frac{G_F^2}{4 \pi^2}\, M_W^2 \,
              S\lt( \frac{\ov{m}_c^2}{M_W^2},  \frac{\ov{m}_t^2}{M_W^2} \rt) 
              \eta_{ct} \, b_K(\mu_K)  \, ,\nn 
 C_{cc}^{\rm SM} & = & \frac{G_F^2}{4 \pi^2}\, M_W^2 \, 
       S\lt( \frac{\ov{m}_c^2}{M_W^2} \rt) \eta_{cc} \,  b_K(\mu_K) 
        \label{wcs2} \, ,
\end{eqnarray}
with the Inami-Lim functions $S$ calculated from the usual box diagrams.
By comparing \eqsand{defc}{wcs2} we verify the MFV feature of the
Standard Model $C^{\rm SM}= C_{tt}^{\rm SM}$. In $ C_{cc}^{\rm SM}$
the Inami-Lim function can be expanded in terms of the tiny
quantity $\ov{m}_c^2/M_W^2$ to find $S(x_c)= x_c + O(x_c^2)$. Likewise
$S(x_c,x_t)\simeq -x_c \log x_c +x_c F(x_t)$ with
$F(\ov{m}_t^2/M_W^2)=0.56$.  From
\eqsand{melk}{wcs2}, %one realises that
$b_K(\mu_K)$ drops out if $ \bra{K} H^{|\Delta S|=2} \ket{\,\ov{\!K}}$
is expressed in terms of $\widehat\Bag_K$.  The QCD correction factors
$\eta_{cc}$ \cite{Nierste1}, $\eta_{ct}$ \cite{Nierste2} and $\eta_{tt}$
\cite{bjw} are listed in Table~\ref{tab:TheoreticalInputs}. The dominant
sources of uncertainty in these quantities are higher-order QCD
corrections\footnote{Very recently a NNLO calculation of $\eta_{ct}$ 
was performed \cite{Brod:2010mj}, leading to a value, which is $5\%$ larger than the 
value used here. This result is not yet included in our analysis.} 
($\eta_{cc}$ also depends on $\alpha_s$ and $\ov m_c$ in a
sizeable way).  The latter dependence is made explicit in
Table~\ref{tab:TheoreticalInputs}.

In analogy to \eq{defdel} we introduce complex parameters for 
New Physics in the three different contributions and write
\begin{eqnarray}
M_{12}^K & \equiv & 
\frac{\braOket{K}{H^{|\Delta S|=2}}{\Kbar}}{2 M_K} 
\; =\; ( V_{ts} V_{td}^* )^2 M_{12}^{tt} +  
2 V_{ts} V_{td}^* V_{cs} V_{cd}^* M_{12}^{ct} + 
( V_{cs} V_{cd}^* )^2 M_{12}^{cc} \,, \nn 
 M_{12}^{ij} &=&  M_{12}^{\text{SM},ij}  \Delta_K^{ij} 
    \; \equiv \;   M_{12}^{\text{SM},ij} 
 | \Delta_K^{ij} | \expe^{\ii \phi^{\Delta_K^{ij}} }
 \label{defdelk} .
\end{eqnarray}
The physical quantities associated with \kkm\ are the $K_L$--$K_S$ mass
difference $\dm_K=M_{K_L}-M_{K_S}$ and the \CP-violating quantity
  $\epsilon_K$. \CP\ violation in $H^{|\Delta S|=1}$ is characterised by
  $\epsilon'_K$. These quantities are defined as
\begin{equation}
\epsilon_K=\frac{\eta_{00}+2\eta_{+-}}{3}, \qquad \quad 
  \epsilon'_K=\frac{-\eta_{00}+\eta_{+-}}{3} \qquad\quad 
  \mbox{with }~~\eta_{ab} \equiv \frac{ {\cal A} (K_L\to \pi^a\pi^b) }{
  {\cal A} (K_S \to \pi^a\pi^b)}. \quad 
\label{ek} 
\end{equation}
Since these two quantities are defined in terms of $K_L$ and $K_S$, they
can be expressed in terms of \kkm\ parameters and the isospin
decay amplitudes $A(K_0\to (\pi\pi)_I)=A_Ie^{i\delta_I}=a_Ie^{i\theta_i}e^{i\delta_I}$, where
$a_I$, $\delta_I$ and $\theta_I$ denote the modulus, the ``strong''
(CP-even) phase and the ``weak'' (CP-odd) phase of the decay
amplitude~\cite{Chau,run2,BurasLectures}.  
$\epsilon_K$ is essentially proportional to the CP phase 
$\phi_K\equiv \arg(-M_{12}^K/\Gamma_{12}^K)$. In view of the 
phenomenological ``$\Delta I=1/2$ rule'' $a_0/a_2\approx 22$ (and 
the fact that all other decay modes come with even smaller amplitudes 
than $a_2$) one can saturate the inclusive quantity $\Gamma_{12}^K$
completely by the contribution proportional to $a_0^2$. Expanding in
various small parameters (see Ref.~\cite{run2} for an elaborate
discussion of the approximations involved) one finds:
\begin{equation}
  \epsilon_K=\sin\phi_\epsilon e^{i\phi_\epsilon}
 \left[\frac{{\rm Im}M_{12}^K}{\dm_K}+\xi \right]\,,
  \qquad \quad 
  \mbox{with }~~\tan\phi_\epsilon=\frac{2\dm_K}{\dg_K} \quad
  \mbox{and }\quad\xi=\frac{{\rm Im}\ A_0}{{\rm Re}\ A_0}\,. \label{epskm}
\end{equation}
The troublesome long-distance contribution to $M_{12}^K$ mentioned 
after \eq{deshs2} is eliminated from \eq{epskm} by trading  
$2{\rm Re} M_{12}^K$ for the experimental value of $\dm_K$. 
Long-distance contributions to ${\rm Im} M_{12}^K$ are small~\cite{bgi}.
In \eq{epskm} $\xi$ comprises the contribution from
$\arg(-\Gamma_{12}^K)$ in the limit of $A_0$ dominance discussed above. 
The corrections are of order $(a_2/a_0)^2$ and therefore negligible.
The usual expression for $\epsilon_K$ is obtained from this expression
by taking the following further approximations: 
i) use $\phi_\epsilon=45^\circ$ instead of the measured value 
   $\phi_\epsilon=43.5(7)^\circ$, 
ii) neglect $\xi$ and  
iii) compute ${\rm Im}M_{12}$ using
only the lowest-dimension $d=6$ operator in the effective
Hamiltonian of~\eq{deshs2}, which is dominated by top and charm box
diagrams. The effect of the three simplifications can be 
parameterised in terms of the parameter $\kappa_\epsilon$~\cite{bg} entering  
\begin{eqnarray}\label{eq:epsilon0}
\epsilon_K&=&\frac{\kappa_\epsilon}{\sqrt{2}} 
 e^{i\phi_\epsilon}\left[\frac{{\rm Im}M^{(6)}_{12}}{\Delta M}\right]\\
&=&C_{\epsilon}\kappa_\epsilon  e^{i\phi_\epsilon}  \hat{\Bag}_K 
\left[ \imag \lt[ \lt( V_{cs}V_{cd}^*\rt)^2\, 
                       \Delta_K^{cc}\rt] \, \eta_{cc} \, 
              S\lt( \frac{\ov{m}_c^2}{M_W^2} \rt) 
       \, +\,  \imag  \lt[\lt( V_{ts}V_{td}^*\rt)^2  \, \Delta_K^{tt}\rt]
              \, \eta_{tt} \, 
         S \lt( \frac{\ov{m}_t^2}{M_W^2} \rt) 
      \rt. \nn
&&\lt. \qquad  + \, 
        2 \, \imag  \lt( V_{ts}V_{td}^* V_{cs}V_{cd}^*\, 
                \Delta_K^{ct}            \rt)
           \, \eta_{ct} \,
          S\lt( \frac{\ov{m}_c^2}{M_W^2},  \frac{\ov{m}_t^2}{M_W^2} \rt)  
    \right] .
\end{eqnarray}
The value $\kappa_\epsilon=1$ corresponds to the approximations i)---iii) 
outlined above. The normalisation reads
\begin{equation}
  C_\epsilon=\frac{G_F^2 F_K^2 m_K M^2_W}{12\sqrt{2}\pi^2\Delta M_K}.
\end{equation}
When expressed in terms of Wolfenstein parameters to lowest order in 
$\lambda$, \eq{eq:epsilon0} defines the familiar hyperbola in the 
$\ov \rho$--$\ov \eta$ plane. 

A series of papers~\cite{run2,Andriyash:2003ym,bg,bgi} has studied how
much the factor $\kappa_\epsilon$ should deviate from 1 in order to
account for the terms neglected by the previous approximations.  We
recall the different elements in App.~\ref{app:kappaepsilon}, separating
the uncertainties coming from statistical and systematic sources, and we
obtain the estimate
\begin{equation}
\kappa_\epsilon = 0.940 \pm 0.013 \pm 0.023,
\label{kappares}
\end{equation}
in good agreement with 
$\kappa_\epsilon=0.94\pm 0.02$ in Ref.~\cite{bgi}.
We emphasize that the estimate of $\kappa_\epsilon$ in \eq{kappares} relies on
the assumption that $\epsilon'_K$ is unaffected by New Physics
(which goes beyond the SM4FC assumption which protects only $I=2$ final
states). From $\epsilon_{K,exp}=(2.229\pm 0.010)\cdot
10^{-3}$ we get the following value for the combination:
\begin{equation}
\epsilon_{K,exp}^{(0)}=\frac{\epsilon_{K,exp}}{\kappa_\epsilon}
    =(2.367 \pm 0.033 \pm 0.049)\cdot 10^{-3}.
\end{equation}
In the presence of New Physics, the relationship between the measured
$\epsilon_K$ and the $ \Delta_K^{ij}$'s is discussed after
\eq{cons}.

One can also study the semileptonic CP asymmetry 
\begin{equation}
 A_L \equiv 
  {\Gamma(K_{\rm long} \to \ell^+\nu\,\pi^-) -
          \Gamma(K_{\rm long}\to \ell^-\bar\nu\,\pi^+) \over
   \Gamma(K_{\rm long} \to \ell^+\nu\,\pi^-) +
          \Gamma(K_{\rm long}\to \ell^-\bar\nu\,\pi^+)} \nn
= \frac{1 - |q/p|^2}{1 + |q/p|^2}, \no
\end{equation}
which, however, contains the same information on fundamental parameters  
as $\real \epsilon_K$.

\subsection{Master formulae}\label{sec:masterformulae}
In this section,
% following 
we provide the master formulae of the theoretical predictions for the
observables relevant to the analysis of new-physics contributions in
mixing. These formulae reflect the dependences on the most important
parameters entering the fit and are obtained from the input values as
described in Sect.~\ref{ssec:inputs}. It should be stressed that these
numerical equations are shown for illustrative purpose only: the
complete formulae are used in the fitting code, which allow to take into
account all the contributions computed so far together with the correct
treatment of the correlations.

Combining \eqsto{dmdg}{defb} with \eq{defdel} one finds: 
\begin{eqnarray}
\dm_d & = & 
    0.502 \psinv 
        \left(  {\frac{|V_{tb}^{\phantom{*}} V_{td}|}{0.0086} } \right)^2 \, 
   \frac{S ( \ov m_t^2/M_W^2 )}{2.35} \,  
   \frac{f_{B_d}^2 \Bag_{B_d}}{(0.17 \gev)^2}  
    \cdot | \Delta_d| \, , \no\\
\dm_s & = & 17.24  \psinv \cdot 
   \left( \frac{|V_{tb}^{\phantom{*}} V_{ts}^*|}{0.04} \right)^2\,
   \frac{S ( \ov m_t^2/M_W^2 )}{2.35} \,  
   \frac{f_{B_s}^2 \Bag_{B_s}}{(0.21\gev)^2} 
    \cdot | \Delta_s| \, . \label{mastmass}
\end{eqnarray}
The remaining uncertainties in the prefactors of the above formulae are due to
the choice of the renormalization scale and the values of $\alpha_s$ and 
the top quark mass. They are at most 3 \% and therefore 
negligible compared to the theoretical error due to the 
non-perturbative and CKM parameters.

The derivation of the formulae involving $\Gamma_{12}^q$ is more 
complicated \cite{bbln,ln,dega}. For the $B_s$ system the dependence 
{\rd on the apex} % with respect to the shape
$(\bar\rho,\bar\eta)$ of the unitarity triangle is strongly suppressed, 
in contrast to the $B_d$ system.
Furthermore the Standard Model contribution to  $a_\text{fs}^s$ is tiny and remains 
below the present experimental sensitivity, while $a_\text{fs}^d$ is one order of magnitude
larger and therefore not completely negligible, 
see 
%numerical results in 
Table \ref{tab:fitResults_SM2}.
Summing up logarithms of the form $m_c^2/m_b^2 \ln m_c^2/m_b^2$ \cite{zlnz} 
and using the $\overline{\rm MS}$-scheme for the $b$ quark mass one finds from Ref.~\cite{ln}
for the decay rate differences:
\begin{eqnarray}
\! \! \! \Delta \Gamma_d  \! \! \! & = & \! \! \! 
\left( \frac{f_{B_d} \sqrt{\Bag_{B_d}}}{0.17 \gev} \right)^2 
\left[ 0.00241 
     + 0.00056 \frac{\widetilde \Bag_{S,B_d}^\prime}{\Bag_{B_d}}  
     - 0.00047 \frac{\Bag_{R,B_d}                  }{\Bag_{B_d}} \right]
      \, \cos \left( \phi_d^\text{SM} + \phi^\Delta_d \right)  ,
\\
\! \! \! \Delta \Gamma_s \! \! \!  & = & \! \! \!
\left( \frac{f_{B_s} \sqrt{\Bag_{B_s}}}{0.21 \gev} \right)^2 
\left[ 0.0797
     + 0.0278 \frac{\widetilde \Bag_{S,B_s}^\prime}{\Bag_{B_s}}  
     - 0.0181 \frac{\Bag_{R,B_s}                  }{\Bag_{B_s}} \right]
      \, \cos \left( \phi_s^\text{SM} + \phi^\Delta_s \right) \, .
\end{eqnarray}
Now the uncertainties in the coefficients are considerably larger than
in the case of the mass differences, but they are still less than
about 15 \%. The dominant theoretical error of the coefficients comes from the
renormalization scale $\mu_1$ followed by the CKM factors. 
One encounters matrix elements of higher-dimensional operators in these
expressions, denoted by $\Bag_R$, which have a power suppression
parametrised by $m_b^{pow}$.
The general (assuming unitarity of the $3\times 3$ quark mixing matrix) 
expression for the semileptonic CP asymmetries reads
\begin{eqnarray}
10^4 a_\text{fs}^q 
&= &
\left[ a_q \Im \left( \frac{\lambda_u^q}{\lambda_t^q} \right)
+
b_q \Im \left( \frac{\lambda_u^q}{\lambda_t^q} \right)^2 \right]
  \, \frac{\sin \left( \phi_q^\text{SM} + \phi^\Delta_q \right)}
          {|\Delta_q|} \, ,
\label{boundafs}
\end{eqnarray}
with $ \lambda_x^q = V_{xb} V_{xq}^*$.
The coefficients $a$, $b$, $c$ read \cite{bbln,ln}: 
\begin{eqnarray}
a_d & = &   9.2905
          + 0.2973 \frac{\widetilde{\Bag}_{S,B_d}^\prime}{\Bag_{B_d}} 
          + 0.2830 \frac{\Bag_{R,B_d}}{\Bag_{B_d}}\,, \nn
a_s & = &   9.4432
          + 0.2904 \frac{\widetilde{\Bag}_{S,B_s}^\prime}{\Bag_{B_s}} 
          + 0.2650 \frac{\Bag_{R,B_s}}{\Bag_{B_s}}\,, \nn
b_d & = &   0.0720 
          + 0.0184 \frac{\widetilde{\Bag}_{S,B_d}^\prime}{\Bag_{B_d}}
          + 0.0408 \frac{\Bag_{R,B_d}}{\Bag_{B_d}}\,, \nn
b_s & = &   0.0732 
          + 0.0180 \frac{\widetilde{\Bag}_{S,B_s}^\prime}{\Bag_{B_s}}
          + 0.0395 \frac{\Bag_{R,B_s}}{\Bag_{B_s}}\,, \nn
c_d & = & -46.8169 
          -17.0083 \frac{\widetilde{\Bag}_{S,B_s}^\prime}{\Bag_{B_s}}
          + 9.2818 \frac{\Bag_{R,B_s}}{\Bag_{B_s}}\,, 
\end{eqnarray}
again with uncertainties in the coefficients of less than 15 \%. The dominant one
comes from the renormalization scale $\mu_1$ followed by the CKM factors.
For the semileptonic CP asymmetries in the $B_d$ system we can also 
write \cite{bbln}:
\begin{eqnarray}
-10^4 \, a_\text{fs}^d &=& 
\lt[ 
c_d \; +\;  a_d \, \lt( \frac{\cos\beta}{R_t} -1 \rt)
 \; +\;  b_d \, \lt( \frac{\cos2\beta}{R_t^2} - 2\frac{\cos\beta}{R_t} +1 \rt)
\rt] 
     \frac{\sin \phi_d^\Delta }{ |\Delta_d|} \; 
\label{afsd}  \nonumber \\
&& {\rd +} \lt[ 
a_d\, \frac{\sin \beta}{R_t}
  \; {\rd +} \; 
  b_d \, \lt( \frac{\sin2\beta}{R_t^2} - 2\frac{\sin\beta}{R_t} \rt)
\rt] \frac{\cos\phi_d^\Delta }{|\Delta_d|} \, ,
\end{eqnarray}
where we have written the $(\bar\rho,\bar\eta)$ dependence in terms of
the angle $\beta$ of the
unitarity triangle and the side 
$R_t =\sqrt{ (1-\ov{\rho})^2+\ov \eta^2}$.

The mixing-induced \CP asymmetries in $B_d\to \jpsi K_S$ and 
$B_s\to \jpsi \phi$ are very important to constrain $\phi_d$ and
$\phi_s$, respectively. For the latter mode an angular analysis is needed to
separate the different \CP components \cite{dfn}. The  mixing-induced \CP asymmetries 
in the two modes determine
\begin{eqnarray}
\sin(\phi_d^\Delta + 2\beta)\qquad\qquad \mbox{and} \qquad\qquad 
\phi_s^\Delta-2\beta_s \, .
\end{eqnarray}
Here the angle $\beta_s$ is defined as positive:
\begin{eqnarray}
\beta_s & = & - \arg \lt( - \frac{V_{ts}^* V_{tb}^{\phantom{*}}}{V_{cs}^* V_{cb}^{\phantom{*}}}\rt) 
   \; =\; 0.01818\pm 0.00085. \label{defbetas}
\end{eqnarray}
(This should be compared with $\beta = \arg ( - V_{td}^*
V_{tb}^{\phantom{*}}/V_{cd}^* V_{cb}^{\phantom{*}}) \approx 0.38$.)
\footnote{It should be emphasized that the \TeVatron experiments, which
  have presented first determinations of
  $\phi_s^\Delta-2\beta_s$ from tagged analyses
  \cite{taggedphaseCDF,taggedphaseD0}, also use the notation $\phi_s$
  and $\beta_s$, but with a slightly different meaning. Comparing the
  notation of Ref.~\cite{taggedphaseCDF,taggedphaseD0} with our notation
  one gets
\begin{eqnarray}
    \phi_s^\text{\Dzero}  &=& \phi_s^\Delta-2\beta_s , \no
\\
-2 \beta_s^\text{CDF} &=& \phi_s^\Delta-2\beta_s . \no
\end{eqnarray}
Ref.~\cite{taggedphaseCDF,taggedphaseD0} has neglected $2\beta_s$ in the
relation between $\dg_s$ and $\phi_s$ in \eq{dmdg}. This is justifed in view of the large
experimental errors and the smallness of $2\beta_s$.}

The measured value $|\epsilon_K^{\rm exp}|$ implies the following 
relation among the CKM elements: 
\begin{eqnarray}
1.25\cdot 10^{-7}
&=& \hat{\Bag}_K
\left[ \imag \lt[ \lt( V_{cs}V_{cd}^*\rt)^2\, 
                       \Delta_K^{cc}\rt] \, \eta_{cc} \, 
              S\lt( \frac{\ov{m}_c^2}{M_W^2} \rt) 
       \, +\,  \imag  \lt[\lt( V_{ts}V_{td}^*\rt)^2  \, \Delta_K^{tt}\rt]
              \, \eta_{tt} \, 
         S \lt( \frac{\ov{m}_t^2}{M_W^2} \rt) 
      \rt. \nn
&&\lt. \qquad  + \, 
        2 \, \imag  \lt( V_{ts}V_{td}^* V_{cs}V_{cd}^*\, 
                \Delta_K^{ct}            \rt)
           \, \eta_{ct} \,
          S\lt( \frac{\ov{m}_c^2}{M_W^2},  \frac{\ov{m}_t^2}{M_W^2} \rt)  
    \right] . \label{cons}
\end{eqnarray}
Here the number on the LHS originates from
\begin{eqnarray}
1.25 \cdot 10^{-7}
&=&
\frac{12 \sqrt{2}\,  \pi^2\, \dm_K}{G_\text{F}^2 \, f_K^2\, m_K \, M_W^2}
  \frac{|\epsilon_K^{\rm exp}|}{\kappa_\epsilon}, 
     \label{epsnum}  
\end{eqnarray}
The peculiar hierarchy of the CKM 
elements in \eq{cons} enhances the sensitivity to the imaginary part 
of $ \Delta_K^{cc}$. Expanding to lowest non-vanishing order in the 
Wolfenstein parameter $\lambda$ shows
\begin{eqnarray} 
 \imag \lt[ \lt( V_{cs}V_{cd}^*\rt)^2\, 
                       \Delta_K^{cc}\rt] &=& 
   - 2 A^2 \lambda^6 \ov\eta \,\real \Delta_K^{cc} 
   \, +\, \lambda^2 \,\imag \Delta_K^{cc} \, , \nn 
 \imag \lt[ \lt( V_{ts}V_{td}^*\rt)^2\, 
                       \Delta_K^{tt}\rt] &=& 
 2  A^4 \lambda^{10} (1- \ov\rho) \ov\eta \,\real \Delta_K^{tt} 
  \, +\, A^4 \lambda^{10} \lt[ (1-\ov\rho)^2 - \ov\eta^2   \rt]
    \,\imag \Delta_K^{tt} \, ,\nn 
   2 \, \imag  \lt( V_{ts}V_{td}^* V_{cs}V_{cd}^*\, 
                \Delta_K^{ct}            \rt) 
   &=& 
   2  A^2 \lambda^6 \ov\eta \,\real \Delta_K^{ct} \, +\, 
   2  A^2 \lambda^6  (1-\ov\rho)  \,\imag \Delta_K^{ct} .
 \label{epswolf}
\end{eqnarray}

%\begin{eqnarray} 
% \imag \lt[ \lt( V_{cs}V_{cd}^*\rt)^2\, 
%                       \Delta_K^{cc}\rt] &=& 
%   - 2 |V_{cb}|^2 \lambda^2 \ov\eta \,\real \Delta_K^{cc} 
%   \, +\, \lambda^2 \,\imag \Delta_K^{cc} \, , \nn 
% \imag \lt[ \lt( V_{ts}V_{td}^*\rt)^2\, 
%                       \Delta_K^{tt}\rt] &=& 
% 2  |V_{cb}|^4 \lambda^2 (1- \ov\rho) \ov\eta \,\real \Delta_K^{tt} 
%  \, +\, |V_{cb}|^4 \lambda^2 \lt[ (1-\ov\rho)^2 - \ov\eta^2   \rt]
%    \,\imag \Delta_K^{tt} \, ,\nn 
%   2 \, \imag  \lt( V_{ts}V_{td}^* V_{cs}V_{cd}^*\, 
%                \Delta_K^{ct}            \rt) 
%   &=& 
%   2  |V_{cb}|^2 \lambda^2 \ov\eta \,\real \Delta_K^{ct} \, +\, 
%   2  |V_{cb}|^2 \lambda^2  (1-\ov\rho)  \,\imag \Delta_K^{ct} .
% \label{epswolf}
%\end{eqnarray}

$\dm_K$ is dominated by physics from low scales. The short-distance
contribution is dominated by the charm-charm contribution involving the 
QCD coefficient $\eta_{cc}$ \cite{Nierste1}. 
There is an additional long-distance contribution from box diagrams with 
two internal up quarks, which cannot be calculated reliably. 
For instance, one could attribute an uncertainty of order
100\% to the theory prediction of $\dm_K$ and try to extract
a constraint on  $| \Delta_K^{cc}|$ from $\dm_K$. 
While $\Delta_K^{cc}$ is very sensitive to any kind of New Physics which 
distinguishes between the first and second quark generations, we will see 
in Sect.~\ref{sec:scenarios} that in MFV scenarios 
all effects on  $\Delta_K^{cc}$ are totally negligible. 
In an unspecified non-MFV scenario both 
$\epsilon_K$ and $\dm_K$ are useless, 
because $\Delta_K^{cc}$, $\Delta_K^{ct}$ and $\Delta_K^{tt}$  are uncorrelated 
with any other observable entering the global fit of the unitarity triangle, 
while in MFV scenarios $\dm_K$ is Standard-Model-like.  
Therefore we do not
consider $\dm_K$ any further.

\subsection{Three scenarios}\label{sec:scenarios}
After having introduced our parameterisation of New Physics in terms of the
$\Delta$ parameters in \eqsand{defdel}{cons}, we can now discuss the three
different physics scenarios which we consider in this article. The common feature of
all scenarios is the assumption that all relevant effects of New Physics are
captured by the $\Delta$ parameters. As long as one only considers the
quantities entering the global fit of the unitarity triangle in conjunction
with the observables of \bbms, this property is  fulfilled in
many realistic extensions of the Standard Model~\footnote{A 
notable exception are models with large couplings of a light charged-Higgs boson
to down-type fermion. In such models $\mathcal{B}(B \to \tau \nu_\tau)$, which we
assume to be Standard-Model-like, is modified. Another exception are models with a non unitary
$3\times 3$ CKM matrix, \textit{e.g} with new fermion generations.}.  
However, once a specific model is studied, often other quantities (unrelated to the global fit of the
unitarity triangle) constrain the parameter space; prominent examples are
branching ratios of rare decays such as $\mathcal{B}(B\to X_s \gamma)$ and $\mathcal{B}(B_s\to
\mu^+ \mu^-)$. Such effects cannot be included in a model-independent
approach like ours. Still, we will see that interesting bounds on the
$\Delta$ parameters can be found within the broad classes of models defined
by our three scenarios. In any specific model covered by our scenarios, the
constraints on the $\Delta$ parameters can only be stronger, but not
weaker than those presented in this paper.
  
Two scenarios involve the MFV hypothesis. The notion of MFV means that
all flavour-violation stems from the Yukawa sector. It is usually implied that
all flavour-changing transitions in the quark sector are solely governed by
the CKM matrix, while flavour-changing transitions in the lepton sector come
with elements of the Pontecorvo--Maki--Nakagawa--Sakata (PMNS) matrix.
Strictly speaking, this conclusion is only valid if MFV is invoked at or below
the GUT scale. If MFV is built into a GUT model at a higher scale, it is well
possible that imprints of the PMNS matrix can be found in FCNC processes of
quarks. Indeed, the articles in Ref.~\cite{cmm} discuss supersymmetric GUT
models with flavour-blind soft SUSY-breaking terms near the Planck scale. The
renormalisation group evolution involving the large top Yukawa coupling then
induces FCNC transitions between right-handed bottom and strange quarks at low
energies. In our analysis this situation is a very special 
case of the scenario I discussed below.

\subsubsection{Scenario I: Non-MFV} 
In this scenario, we do not assume anything about the flavour structure 
of the New Physics interaction. Since here $\Delta_K^{cc}$, $\Delta_K^{ct}$
and $\Delta_K^{tt}$ are unrelated to other parameters, we can neither derive 
any constraints on these parameters nor use $\epsilon_K$ in the global fit.
While $\Delta_d$ and $\Delta_s$ are a-priori independent, the allowed ranges
for these parameters are nevertheless correlated through the global fit and the 
unitarity constraints on the CKM matrix.
This can be qualitatively understood  as follows. Consider a value for 
$|\Delta_s|$ which exhausts the range allowed by the hadronic uncertainties in 
$\dm_s$. The good theoretical control over the ratio $\dm_d/\dm_s$ then
fixes $|V_{td}|^2|\Delta_d |$ quite precisely. The global fit of the unitarity
triangle further constrains $|V_{td}|\propto \sqrt{(1-\ov\rho)^2+\ov \eta^2}$,
so that a-posteriori the allowed ranges for $|\Delta_d|$ and  $|\Delta_s|$ 
become correlated. Also the flavour-mixed CP-asymmetry $a_\mathrm{fs}$ measured 
at the TeVatron experiments correlates the parameters $\Delta_d$ and $\Delta_s$.
  
\subsubsection{Scenario II: MFV with small bottom Yukawa coupling} 
We adopt the symmetry-based definition of MFV of Ref.~\cite{dgis} to discuss 
our two other scenarios. Ignoring
the lepton sector here, the starting point is the $[U(3)]^3$ flavour symmetry
of the gauge sector of the Standard Model, which entails the flavour-blindness
of this sector. The gauge part of the Lagrangian is invariant under
independent unitary rotations of the left-handed quark doublets $Q_L^i$ (where
$i=1,2,3$ labels the generation), and the right-handed quark singlets $d_R^i$
and $u_R^i$ in flavour space. In the Standard Model the $[U(3)]^3$ flavour
symmetry is broken by the Yukawa interactions. This symmetry breaking permits
discriminating flavour quantum numbers, quark masses and flavour-changing
transitions. Within the Standard Model only the top Yukawa coupling $y_t$ is
large, all other Yukawa couplings are small or even tiny. These small
parameters pose a challenge to generic extensions of the Standard Model.  This
challenge is met by the MFV hypothesis which extends the Standard Model assuming that
the only sources of $[U(3)]^3$ flavour symmetry breaking remain
the Yukawa couplings. Specifying to the familiar basis of mass
eigenstates, we list the following consequences of the MFV hypothesis: 
\begin{itemize}
\item[i)] Any flavour-changing transition is governed by the same CKM elements
          as in the Standard Model.
\item[ii)] Any chirality flip $q_R \to q_L$ is proportional to the
           Yukawa coupling $y_q$ (and, by hermiticity of the Lagrangian, any $q_L \to
           q_R$ flip is proportional to $y_q^*$).
\item[iii)] Any flavour-changing transition of a right-handed quark involves a
           factor of the corresponding Yukawa coupling.
\item[iv)] FCNC transitions have the same pattern of GIM cancellations as in
           the Standard Model.
\end{itemize}
For example, property i) and iii) imply that any $b_R \to s_R$ transition
is of the form $V_{ts} V_{tb}^* y_b y_s^* f(|y_t|^2,|y_b|^2,|y_s|^2|)$, where
$f$ is some function of $|y_{t,s,b}|^2$. 
The actual power of Yukawa couplings in
the contribution from a given Feynman diagram is determined by the number of
chirality flips through property ii). Property iv) ensures that any possible
contribution proportional to  $V_{cs} V_{cb}^*$ is GIM-suppressed, i.e.\ 
proportional to $|y_c|^2$, and negligible as in the Standard Model.

However, we deviate in one important aspect from Ref.~\cite{dgis}. We
explicitly allow for \CP-violating phases which do not originate from the
Yukawa sector, i.e., we proceed as in Ref.~\cite{kpvz}.
\CP violation is an interference phenomenon and involves the
differences from otherwise unphysical phases. In order to avoid new \CP phases
one must align the phases of the Yukawa couplings with those of other
parameters which are unrelated to the Yukawa sector. For instance, in the context of the
Minimal Supersymmetric Standard Model (MSSM) this tuning of phases affects the
$\mu$ term, the gaugino mass terms, and the trilinear soft SUSY-breaking
terms. It is difficult to motivate this alignment from symmetries or through a dynamical
mechanism. We therefore explicitly permit extra \CP phases outside the Yukawa
sector, i.e.\ we consider effects from flavour-conserving \CP phases. Usually
such phases are constrained by experimental bounds on electric dipole moments, in particular for the
MSSM, where stringent bounds on flavour-conserving \CP phases can only be
avoided with quite heavy superparticles. But in our context of generic MFV
sizeable flavour-conserving \CP phases cannot be excluded a priori~\cite{Mercolli:2009ns}.

Refs.~\cite{dgis,kpvz} consider two possibilities for the dominant
flavour-symmetry breaking mechanism. While the large top Yukawa
coupling always breaks the flavour symmetries of the gauge sector, one can
consider the case that the bottom Yukawa coupling is also
large and spoils flavour blindness at an equal level (this occurs
in the popular MSSM scenarios with large $\tan\beta$ \cite{ltb}). 
Our scenario II corresponds to the case, where only the top Yukawa 
coupling is large.  That is, in scenario II we neglect all effects from 
down-type Yukawa couplings. The possible $|\Delta B|=2$ operators are
discussed in Refs.~\cite{dgis,kpvz}.  
Thanks to MFV property iii) four-quark operators with
right-handed $b$ or $s$ fields are accompanied by small (down-type)
Yukawa couplings \cite{dgis,kpvz} and no other operator than $Q$ in
\eq{defq} occurs in scenario II. Therefore, the only effect of New Physics is to
change the coefficient $C$ in \eq{defheff}. 

An important observation is that $C$ will always be real, even in the presence of
flavour-conserving \CP phases: The MFV hypothesis implies that $C$ is
independent of the flavours of the external quarks. If we interchange
$b_L$ and $q_L$, the corresponding four-quark interaction will be
governed by the same coefficient $C$ and the effective Hamiltonian will
contain the combination
\begin{eqnarray}   
   (V_{tq}^* V_{tb}^{\phantom{*}})^2 \, C \, 
   \ov q_L \gamma_\mu b_L \, \ov q_L \gamma^\mu b_L \; + \; 
   (V_{tb}^* V_{tq}^{\phantom{*}})^2 \, C \, 
   \ov b_L \gamma_\mu q_L \, \ov b_L \gamma^\mu q_L. \no
\end{eqnarray}   
Now the hermiticity of the Hamiltonian implies $C=C^*$. 
(An explicit check is provided by the MSSM, where flavour-conserving
\CP-violating parameters enter $C$ only through their moduli or real parts).
Hence our scenario II corresponds to the case 
\begin{eqnarray}
\Delta_s=\Delta_d=\Delta_K^{tt} && \qquad  \mbox{with }\qquad 
\phi_s^\Delta=\phi_d^\Delta= \phi_K^{ij\,\Delta}= 0.
\label{defsc2}
\end{eqnarray}
We next discuss an important extension of the MFV analysis of
Ref.~\cite{dgis} in the case of $\epsilon_K$, where we include effects from the
charm Yukawa coupling $y_c$. The potential relevance of these effects becomes
clear when one notices that the Inami-Lim functions $S ( \ov{m}_c^2/M_W^2 ,
\ov{m}_t^2/M_W^2 ) $ and $S ( \ov{m}_c^2/M_W^2 ) $ are proportional to $\ov
m_c^2/M_W^2$. 
Within the Standard Model a substantial contribution to
the $\epsilon_K$-hyperbola stems from terms which are quadratic in $y_c$, and 
we have to extend our analysis of MFV New Physics to order $y_c^2$. 
Splitting the Wilson coefficients as:
\begin{eqnarray}
   C&=& C^{\rm SM} + C^{\rm NP}, \qquad 
 C_{ij}\; =\;  C_{ij}^{\rm SM}  + C_{ij}^{\rm NP},  
\end{eqnarray}
MFV constrains the new contributions $ C^{\rm NP}$, $ C_{ij}^{\rm NP}$ 
to obey the following pattern: 
\begin{eqnarray}
   C_{ij}^{\rm NP} &=&  V_{is} V_{id}^* V_{js} V_{jd}^* 
                 f(|y_i|^2,|y_j|^2)  \label{dkpat}
\end{eqnarray}
where $f$ is a real-valued function with $f(0,x)=f(x,0)=0$ by the GIM mechanism 
and, of course, $ C^{\rm NP}(\mu)= C_{tt}^{\rm NP}(\mu) $.  
We must now distinguish two cases depending on whether the dominant contributions 
from New Physics affect the diagrams with the light charm or up quarks or rather
involve particles with a mass similar to or heavier than the top quark.

In the first case, we have to consider New Physics contributions $
C_{ct}^{\rm NP}$ and $ C_{cc}^{\rm NP}$ which involve the charm and
up quarks on an internal line. Such contributions occur, for
example, in box diagrams in which one or both W bosons are replaced by
charged Higgs bosons. These diagrams with only light internal quark lines 
lead to negligible effects, if the new particle exchanged between the quark lines is a
scalar (like a Higgs boson), because scalars couple left-handed to
right-handed fields and come with the penalty of small Yukawa couplings. The
extra helicity flip (if an internal quark is right-handed) or the GIM
mechanism (on an internal line with only left-handed quarks) brings in extra
Yukawa couplings and the contribution to $ C_{ct}^{\rm NP}$ and $ C_{cc}^{\rm NP}$ 
is of order $|y_c|^4$ or smaller and negligible compared to Standard Model
contributions, which are proportional to $\ov m_c^2 \propto |y_c|^2$. A
scaling like in the Standard Model, with just two powers of $ |y_c|$, could occur in principle if the
new exchanged particle is a heavy gauge boson mimicking the Standard Model couplings to
the left-handed quark doublets. We are not aware of a realistic theory with
such particles and do not consider this possibility further.  

The second case corresponds to New Physics contributions which involve
heavy particles and directly add to the coefficient $C$ in
\eq{defc}. This class of contributions includes box diagrams with 
a charged Higgs and one charm and one top quark as internal quark lines.
These diagrams indeed give a contribution to $ C_{ct}^{\rm NP}$
proportional to  $|y_c|^2$.
Another prominent example for a contribution of this type are the
chargino-squark diagrams in the MSSM and it is worthwile to discuss
this example for illustration, before returning to our generic
scenario.  The chargino-stop box diagram, which contributes to $
\Delta_K^{tt}$, is widely discussed in the literature. However, to our
knowledge, nobody has studied the corresponding effect in
$\Delta_K^{ct}$ or $\Delta_K^{cc}$. The former parameter receives
contributions from a box diagram with a scharm on one line and a stop
on the other. In the limit $y_c=0$ there is an exact GIM cancellation
between the contributions from the charm and up squarks, which has
been invoked to justify the omission of scharm effects. The first
non-vanishing contribution is proportional to $|y_c|^2$ (corresponding
to $\tilde c_L \to \tilde c_R$ and $\tilde c_R \to \tilde c_L$
flips). Since the same flip is needed on the stop line, there is also
a factor of $|y_t|^2$ involved.  Clearly, we recognise the pattern of
\eq{dkpat} with the same function $f$ as in $\Delta_K^{tt}$. Extending
to the generic MFV situation, it is easy to relate $\Delta_K^{ct}$ to
$\Delta_K^{tt}$ for theories in which $y_t$ is small enough that we
can expand $\Delta_K^{ct}$ and $\Delta_K^{tt}$ to the lowest order in
$y_t$ (like in the MSSM for moderate values of $\tan \beta$ and
not-too-large values of the trilinear breaking term
$A_t$).\footnote{The actual expansion parameter is $y_t v/M$, where
  $M$ is the mass scale of the new particles in the loop and
  $v=174\,\gev$ is the Higgs vacuum expectation value.} Then
 \begin{eqnarray}
  f(|y_i|^2,|y_j|^2) &=& f_0 \, |y_i|^2 \, |y_j|^2 + {\cal O} (|y_{i,j}|^6),
\end{eqnarray}  
and up to small corrections one has $ C_{ct}^{\rm NP}= C_{tt}^{\rm
  NP}|y_c|^2/|y_t|^2 $ and $ C_{cc}^{\rm NP}= C_{tt}^{\rm NP}|y_c|^4/|y_t|^4
$.  Since $ C_{cc}^{\rm NP}$ is real in scenario II, we can certainly neglect
it and set $\Delta_K^{cc}=1$ in \eq{cons}. To account for the situation that
$y_t$ is close to one we should vary $ C_{ct}^{\rm NP}$ around
$C_{tt}^{\rm NP}|y_c|^2/|y_t|^2 $. A realistic range for 
$ C_{ct}^{\rm NP}$ 
can be obtained from the Standard Model situation. In $S(\ov
m_{c,t}^2/M_W^2)$ the relevant quantity is $\ov m_{c,t}/M_W\sim 2 y_{c,t}$. 
However, while $S(\ov m_c^2/M_W^2)\simeq \ov m_c^2/M_W^2 $, %the corresponding quantity
$S(\ov m_t^2/M_W^2)$ differs from $\ov m_t^2/M_W^2 $ by a bit less than a
factor of 2. We take this as a conservative estimate and choose
\begin{eqnarray}
  C_{ct}^{\rm NP} &=&  \lambda_K \, 
           \frac{\ov m_c^2 (\mu_{\rm NP})}{\ov m_t^2(\mu_{\rm NP})} \, C_{tt}^{\rm NP}
\qquad\qquad  
\mbox{with}\qquad 0.5 \leq \lambda_K \leq 2 , \label{nprat}
\end{eqnarray} 
while $ C_{cc}^{\rm NP}= 0$. As indicated in \eq{nprat}, the Yukawa
  couplings $y_c$ and $y_t$, which are expressed in terms of $\ov m_c$ and
  $\ov m_t$, enter the Wilson coefficients at the scale $\mu_{\rm NP}$ at which
  the heavy particles of the new-physics scenario are integrated out.
  Since  $\ov m_c/\ov m_t$ is scale-independent, we evaluate this ratio
  at the scale $\mu=\ov m_t $ in the following.

Since no new operators occur in scenario II, our New Physics parameters are
related to the Wilson coefficient $C$ in a simple way: 
\begin{eqnarray}
  \Delta_K^{ij} &=&  \frac{C_{ij}}{C_{ij}^{\rm SM} } 
       \; =\; 1 + \frac{C_{ij}^{\rm NP}}{C_{ij}^{\rm SM} }. 
\label{delcnp}
\end{eqnarray} 
\eqsand{nprat}{delcnp} imply the relation 
\begin{eqnarray}
  \Delta_K^{ct} &=&  1 \; +\; \lambda_K \, 
           \frac{\ov m_c^2(\ov m_t)}{\ov m_t^2(\ov m_t)} \,
           \frac{C_{tt}^{\rm SM}}{C_{ct}^{\rm SM}} \, 
           ( \Delta_K^{tt} -1 )\nn
&=& 1 \;+\;  \lambda_K \,  \frac{\ov m_c^2(\ov m_t)}{\ov m_t^2(\ov m_t)} \,
           \frac{S\lt( \ov{m}_t^2/M_W^2 \rt)
            \eta_{tt}}{ S\lt( \ov{m}_c^2/M_W^2,
                             \ov{m}_t^2/M_W^2 \rt) 
              \eta_{ct} } \, 
           ( \Delta_K^{tt} -1 )
 \label{dectdett} \\
 & =&
  1 \; + \; 0.017 \, \lambda_K \, ( \Delta_K^{tt} -1 ).
\label{dectdettnum}
\end{eqnarray} 
The numerical value 0.017 is obtained for the central values of 
Table~\ref{tab:TheoreticalInputs}. The smallness of this number roots in the 
enhancement of $ S ( \ov{m}_c^2/M_W^2, \ov{m}_t^2/M_W^2 )$ by the large
leading logarithm $\log( \ov{m}_c^2/M_W^2 ) =-8.3 $, which stems from box
diagrams with internal charm and up quarks. This logarithm is absent 
in the New Physics contribution, which is formally of the order of a 
next-to-leading-order correction. In summary, \eqsand{defsc2}{dectdett}
and $\Delta_K^{cc}=1$ define our scenario II.

Potential New Physics effects governed by $y_c$ have also been studied in
  \cite{kpvz} in generic $s\to d$ transitions.  This reference
  estimates their size  as of order 1\% of the contribution
  governed by $y_t$ and finds them negligible. Our result in
  \eq{dectdettnum}, specific to \kkm, is in agreement with this
  estimate. In view of the sub-percent experimental error 
  of  $\epsilon_K$ and decreasing theoretical uncertainties,   
  the correction $\Delta_K^{ct}$ is not negligible a priori.

\subsubsection{Scenario III: MFV with a large bottom Yukawa coupling} 
In scenario III we consider a large bottom Yukawa coupling. Then 
$H_q^{|\Delta B|=2}$ in \eq{defheff} is modified to include operators 
that are not suppressed anymore
\begin{eqnarray}
H_q^{|\Delta B|=2} &=& (V_{tq}^* V_{tb}^{\phantom{*}})^2 
   \lt[ C \, Q \; +\;  C_S \, Q_S \; +\; 
        \widetilde C_S \, \widetilde Q_S \rt] 
              \; +\;   \mbox{h.c.} 
\end{eqnarray}
Here $Q_S=\ov{q}_L b_R \, \ov{q}_L b_R$ and $\widetilde Q_S$ is defined in
\eq{defqst}.  MFV does not put any constraint on $C_S$ and $\widetilde C_S$ which can be complex 
(an example of an MSSM scenario with a complex $C_S$ can be found in Ref.~\cite{bk}, 
and an up-to-date renormalisation-group analysis of $C_S$ and $\widetilde C_S$ 
can be found in the appendix of the fifth article in Ref.~\cite{bsltb}).  
In the following, we will assume that the matrix elements of $Q$, $Q_S$, $\tilde{Q}_S$
are affected in the same way by $U$-spin breaking corrections, so that
the presence of new operators in $H_q^{|\Delta B|=2}$ yields a scenario III corresponding 
to the case $\Delta_s=\Delta_d$ with generally non-zero $\phi_s^\Delta=\phi_d^\Delta$.

$C_S$ and $\widetilde C_S$ necessarily
involve at least two powers of $y_b$ because of MFV property iii). In the kaon case, 
the corresponding contribution to analogous coefficients $C_S$ and $\widetilde
C_S$ in $H^{|\Delta S|=2}$ would involve $y_s^2$ instead. Clearly, such
contributions to $\Delta_K^{tt}$ and $\Delta_K^{ct}$ will have a negligible
impact on $\epsilon_K$, but this need not be the case for $\Delta_K^{cc}$, 
in view of the big lever arm in \eq{epswolf},
meaning that the coefficient of $\imag \Delta_K^{cc}$ is larger than that of
$\real \Delta_K^{cc}$ by three orders of magnitude.
Contributions to $\Delta_K^{cc}$ from box diagrams with 
new heavy particles 
involve four powers of $|y_c|$ in addition to the two powers of $y_s$
and are negligible compared to the Standard Model contribution $\propto |y_c|^2 $ even 
when multiplied by a factor of 1000. New contributions involving internal 
charm quarks (or some new neutral scalar particle coupling a $d_L$ to an 
$s_R$) can be proportional to $|y_c|^2$ as in the Standard Model, but the 
two powers of $y_s$ are sufficient to suppress the effect below a level 
which is relevant for  $\epsilon_K$. 
In addition, in scenario III, $C$ will be a function of $|y_b|^2$, which 
complicates the relationship to the corresponding 
coefficient in $H^{|\Delta S|=2}$ and even our proof that $C$ is real 
does not hold anymore. 
In summary, our scenario III comes with 
$\Delta_s=\Delta_d$ and this parameter  is complex. The parameters 
$\Delta_K^{ij}$ are as in scenario II (that is, they are real, fulfill 
\eq{dectdett}, and $\Delta_K^{cc}=1$), but are now unrelated to 
$\Delta_{d,s}$. 

We disagree with Ref.~\cite{kpvz} on one point here, namely the possibility
  of ${\cal O}(50\%)$ effects in $\epsilon_K$ through a sizeable complex
contribution to (in our notation) $\Delta_{tt}$. The claimed effect
involves two powers of the strange Yukawa coupling $y_s$.  The MFV
property links this contribution to a similar one in \bbm, which
involves two powers of $y_b$ instead. A 50\% effect in $\epsilon_K$ from
this source would imply an enhancement of \bbm\ by almost two orders of
magnitude from this term. We do not see how extra contributions with 
more powers of $y_b$ could possibly reduce this enhancement to a factor
below 1.

One ought to mention that scenarios II and III do not exhaust the
possibilities offered by MFV. For instance, Refs.~\cite{ltb,bsltb}
consider MFV-MSSM scenarios with $M_\text{SUSY}\gg M_{A^0} \gsim v$,
where $M_\text{SUSY}$ and $M_{A^0}$ denote the masses of the
superparticles and the \CP-odd Higgs boson, respectively, and $v=174
\gev$ is the electroweak scale. In the MSSM the coefficient $C_S$ is
highly suppressed \cite{bsltb}~\footnote{The vanishing of $C_S$ in
the MFV-MSSM scenario with $M_\text{SUSY}\gg M_{A^0} \gsim v$ stems
from a softly broken Peccei-Quinn symmetry \cite{dgis}, which we
have not built into our scenarios II and III.}, while the
operator $\ov{s}_L b_R \, \ov{s}_R b_L$ occurs with a sizeable
coefficient, despite of the suppression with the small strange Yukawa
coupling. Its counterpart in the $B_d$ system comes with the even
smaller down Yukawa coupling and is negligible.
In the scenario of Refs.~\cite{ltb,bsltb} the connection between \bbms\
and \bbmd\ is lost. But large effects in \bbms\ are not allowed, due to 
the experimental bound on ${\mathcal B}(B_s\to \mu \mu) $~\cite{bsltb} 
 (for up-to-date results, see \cite{bsmumu}).

\subsubsection{Testing the Standard Model}
There are various ways to test the Standard Model. The simplest one is to
determine all the relevant parameters from a global fit, and test
the fit prediction for a given observable compared with the direct
measurement. This kind of test, which can be quantified by computing the 
relevant pull value, is independent of any underlying new
physics scenario.

The second kind of test addresses a definite new-physics scenario
extending the Standard Model, and computes the statistical significance
that the parameters take their Standard Model value. In our case it corresponds to
testing whether one or several $\Delta$ parameters are compatible
with $\Delta= 1$. 

In the relevant sections below we will perform both kind of tests, and
discuss their interpretation.

\section{Inputs}\label{ssec:inputs}

In this Section, we discuss all relevant experimental and theoretical inputs
entering the fits. The corresponding values and uncertainties are quoted in 
Tables~\ref{tab:ExperimentalInputs} and~\ref{tab:TheoreticalInputs}. In general, 
if there is only one uncertainty quoted, we understand
this error as a statistical 
one. In case of two error contributions, the first one is taken as a statistical
error while the second, theoretical, error is treated as an allowed range for 
the observable or the parameter under consideration. This kind of uncertainty is 
treated in the Rfit scheme described in Refs.~\cite{CKMfitter1,CKMfitter2}.

\subsection{Hadronic parameters and method of averaging}
\label{ssec:MethodofAveraging}

Several hadronic inputs are required for the fits presented, and we mostly rely
on lattice QCD simulations to estimate these quantities involving strong 
interactions at low energies. 
The presence of results from different lattice QCD collaborations 
with various statistics and systematics make it all 
the more necessary to combine them in a careful way. The procedure that we have 
chosen to determine these lattice averages is as follows:\\
We collect the relevant calculations of the quantity that we are interested in
and 
we take only unquenched results with 2 or 2+1 dynamical fermions, even those from 
proceedings without a companion article (flagged with a star). In these results, 
we separate the error estimates into a Gaussian part and a flat part that is 
treated \`{a} la Rfit. The Gaussian part collects the uncertainties from purely 
statistical origin, but also the systematics that can be controlled and treated 
in a similar way (e.g., interpolation or fitting in some cases). The remaining 
systematics constitute the Rfit error. If there are several sources of error in 
the Rfit category, we add them linearly \footnote{Keeping in mind that in many 
papers this combination is done in quadrature and the splitting
between different sources is not published.}.
The Rfit model is simple but also very strict. It amounts to assuming that the 
theoretical uncertainty is rigorously constrained by a mathematical bound that 
is our only piece of information. If Rfit is taken stricto sensu and the individual 
likelihoods are combined in the usual way (by multiplication), the final uncertainty 
can be underestimated, in particular in the case of marginally compatible values.
We correct this effect by adopting the following averaging recipe. 
We first combine the Gaussian uncertainties by combining the likelihoods 
restricted to their Gaussian part. Then we assign to this combination the smallest 
of the individual Rfit uncertainties. The underlying idea is twofold:
\begin{itemize}
\item the present state of the art cannot allow us to reach a better theoretical 
      accuracy than the best of all estimates.
\item this best estimate should not be penalized by less precise methods (as it 
      would happen be the case if one took the dispersion of the individual 
      central values as a guess of the combined theoretical uncertainty).
\end{itemize}
It should be stressed that the concept of a theoretical uncertainty is ill-defined, 
and the combination of them even more. Thus our approach is only one among the 
alternatives that can be found in the literature \cite{LubiczTarantino,LLVdW}. 
In contrast to some of the latter, ours is algorithmic and can be reproduced. 
Moreover, we differ from the PDG-like method advocated in Ref.~\cite{LLVdW} on 
two points. We separate systematic and statistical errors, which prevents us from 
assigning a reduced systematics to a combination of several results suffering from 
the same systematic uncertainty. We do not attempt at estimating the (partial) 
correlations between the results from different collaborations, even though we are 
aware of their existence (results from the same gauge configuration, using the same 
procedure to determine the lattice spacing\ldots). Whatever the averaging method 
chosen, one should emphasize that it relies crucially on the quality of the error 
estimation performed by each collaboration.

The following tables show the inputs used and the average obtained by applying the 
procedure described above for the following hadronic parameters: 
the decay constant $f_{B_{s}}$ for the $B_{s}$ meson (Table~\ref{tab:fBs}), 
the ratio of decay constants $f_{B_{s}}/f_{B_{d}}$ (Table~\ref{tab:fBsOverfB}), 
the scheme-invariant bag parameter $\hat{\Bag}_{B_{s}}=1.523 \, \Bag_{B_{s}}\big(m_{b}\big)$
for the $B_{s}$ meson discussed after Eq.~\eqref{defb} (Table~\ref{tab:Bshat}),
the ratio of bag parameters $\Bag_{B_{s}}/\Bag_{B_{d}}$ (Table~\ref{tab:BsOverBdhat})~\footnote{Some of the lattice collaborations~\cite{HPQCD09,RBCUKQCD10} provide only the parameter $\xi_B=(f_{B_s}/f_{B_d})\sqrt{\Bag_{B_{s}}/\Bag_{B_{d}}}$ together with the ratio of decay constants, without providing the correlation coefficient between the two values. In such a case, we have extracted the ratio of bag parameters and its uncertainties errors assuming a 100\% correlation in the systematic errors between $\xi_B$ and $f_{B_s}/f_{B_d}$.},
and the bag parameter $\Bag_{K}(2 \, \mbox{GeV})$ for the neutral kaon (Table~\ref{tab:BKMSbar2GeV}).

\begin{table}[Htp]
\renewcommand{\arraystretch}{1.3}
\centering
\begin{tabular}{|c|c|c|c|}\hline
Collaboration    & $N_f$   & $f_{B_{s}} \pm \sigma_{stat} \pm \sigma_{Rfit}$ & Reference                      \\
\hline
CP-PACS01    & 2       & $242 \pm 9 ^{+53}_{-34}$                        & \cite{CP-PACS01}             \\
MILC02       & 2       & $217 \pm 6 ^{+58}_{-31}$                        & \cite{MILC02}                \\
JLQCD03      & 2       & $215 \pm 9 ^{+19}_{-15}$                        & \cite{JLQCD03}               \\
ETMC09      & 2        &      $243 \pm 6 \pm 15$                          & \cite{ETMC09}\\
HPQCD03      & 2+1     & $260 \pm 7 \pm 39$                              & \cite{HPQCD03}               \\
FNAL-MILC09 & 2+1     & $243 \pm 6 \pm 22$                              & \cite{FNAL-MILC09}           \\
HPQCD09      & 2+1     & $231 \pm 5 \pm 30$                              & \cite{HPQCD09}               \\
\hline
Our average  &         & $231 \pm 3 \pm 15$                              &                              \\
\hline
\end{tabular}
\caption[fBS]
{Calculations and average used for the decay constant $f_{B_{s}}$.
 $N_f$ stands for the number of  dynamical flavours used in the simulation.
 The first uncertainty quotes the statistical uncertainty, the second the Rfit error.}
\label{tab:fBs}
\end{table}

\begin{table}[Htp]
\renewcommand{\arraystretch}{1.3}
\centering
\begin{tabular}{|c|c|c|c|}\hline
Collaboration    & $N_f$   & $f_{B_{s}}/f_{B_{d}} \pm \sigma_{stat} \pm \sigma_{Rfit}$ & Reference              \\
\hline
CP-PACS01    & 2       & $1.179 \pm 0.018 \pm 0.023$                               & \cite{CP-PACS01}     \\
MILC02       & 2       & $1.16  \pm 0.01 ^{+0.08}_{-0.04}$                         & \cite{MILC02}        \\
JLQCD03      & 2       & $1.13  \pm 0.03 ^{+0.17}_{-0.02}$                         & \cite{JLQCD03}       \\
ETMC09      & 2 &  $1.27 \pm 0.03 \pm 0.04$ & \cite{ETMC09}\\
FNAL-MILC09 & 2+1     & $1.245 \pm 0.028 \pm 0.049$                               & \cite{FNAL-MILC09}   \\
HPQCD09      & 2+1     & $1.226 \pm 0.020 \pm 0.033$                               & \cite{HPQCD09}       \\
RBC/UKQCD10 & 2+1 & $1.15 \pm 0.05 \pm 0.20$ & \cite{RBCUKQCD10} \\
\hline
Our average  &         & $1.209 \pm 0.007 \pm 0.023$                               &                      \\
\hline
\end{tabular}
\caption[fBsOverfB]
{Calculations and average used for the ratio of decay constants $f_{B_{s}}/f_{B_{d}}$.
 $N_f$ stands for the number of  dynamical flavours used in the simulation.
 The first uncertainty quotes the statistical uncertainty, the second the Rfit error.
}
\label{tab:fBsOverfB}
\end{table}

\begin{table}[Htp]
\renewcommand{\arraystretch}{1.3}
\centering
\begin{tabular}{|c|c|c|c|}\hline
Collaboration     & $N_f$   & $\hat{\Bag}_{B_{s}} \pm \sigma_{stat} \pm \sigma_{Rfit}$ & Reference          \\
\hline
JLQCD03       & 2       & $1.299  \pm 0.034^{+0.122}_{-0.095}$                     & \cite{JLQCD03}       \\
HPQCD06       & 2+1     & $1.168  \pm 0.105 \pm 0.140$                             & \cite{HPQCD06}       \\
RBC/UKQCD07  & 2+1     & $1.21   \pm 0.05  \pm 0.05$                              & \cite{RBC/UKQCD07*}  \\
HPQCD09       & 2+1     & $1.326  \pm 0.04  \pm 0.03$                              & \cite{HPQCD09}       \\
\hline
Our average   &         & $1.28   \pm 0.02  \pm 0.03$                              &                      \\
\hline
\end{tabular}
\caption[Bshat]
{Calculations and average used for the bag parameter $\hat{\Bag}_{B_{s}}$.
 $N_f$ stands for the number of  dynamical flavours used in the simulation.
 The first uncertainty quotes the statistical uncertainty, the second the Rfit error.
}
\label{tab:Bshat}
\end{table}

\begin{table}[Htp]
\renewcommand{\arraystretch}{1.3}
\centering
\begin{tabular}{|c|c|c|c|}\hline
Collaboration     & $N_f$   & $\Bag_{B_{s}}/\Bag_{B_{d}} \pm \sigma_{stat} \pm \sigma_{Rfit}$ & Reference          \\
\hline
JLQCD03       & 2       & $1.017 \pm 0.016 ^{+0.076}_{-0.017}$                            & \cite{JLQCD03}   \\
HPQCD09       & 2+1     & $1.053 \pm 0.020 \pm 0.030$                                     & \cite{HPQCD09}   \\
RBC/UKQCD10 &   2+1 & $0.96 \pm 0.02 \pm 0.03$ &\cite{RBCUKQCD10}\\
\hline
Our average   &         & $1.006  \pm 0.010 \pm 0.030$                                       &                  \\
\hline
\end{tabular}
\caption[BsOverBdhat]
{Calculations and average used for the bag parameter ratio $\Bag_{B_{s}}/\Bag_{B_{d}}$.
 $N_f$ stands for the number of dynamical flavours used in the simulation.
 The first uncertainty quotes the statistical uncertainty, the second the Rfit error.
}
\label{tab:BsOverBdhat}
\end{table}

\begin{table}[Htp]
\renewcommand{\arraystretch}{1.3}
\centering
\begin{tabular}{|c|c|c|c|}\hline
Collaboration     & $N_f$      & $\Bag_{K}(2 \, \mbox{GeV}) \pm \sigma_{stat} \pm \sigma_{Rfit}$ & Reference              \\
\hline
JLQCD08       & 2          & $0.537 \pm 0.004  \pm 0.072$                                    & \cite{JLQCD08}       \\
HPQCD/UKQCD06 & 2+1        & $0.618 \pm 0.018  \pm 0.179$                                    & \cite{HPQCD/UKQCD06} \\
RBC/UKQCD07   & 2+1        & $0.524 \pm 0.010  \pm 0.052$                                    & \cite{RBC/UKQCD07}   \\
ALVdW09       & 2+1        & $0.527 \pm 0.006  \pm 0.049$                                    & \cite{ALVdW09}   \\
\hline
Our average   &            & $0.527 \pm 0.0031 \pm 0.049$                                    &                  \\
\hline
\end{tabular}
\caption[BKMSbar2GeV]
{Calculations and average used for the bag parameter $\Bag_{K}(2 \, \mbox{GeV})$.
 $N_f$ stands for the number of  dynamical flavours used in the simulation.
 The first uncertainty quotes the statistical uncertainty, the second the Rfit error.
}
\label{tab:BKMSbar2GeV}
\end{table}

We are not aware of lattice estimates for the power-suppressed matrix elements corresponding to
 $\Bag_{R_i}$ and  $\Bag_{\tilde{R}_i}$. 
We will assign them a default value of $1\pm 0.5$, taking a flat uncertainty for the two 
bag parameters contributing the most to $\Delta\Gamma^q$ and $a_{SL}^q$ 
($\Bag_{\tilde{R}_2}$ and $\Bag_{\tilde{R}_3}$ respectively), whereas the remaining ones
are either assigned a Gaussian uncertainty ($\Bag_{R_0}$, $\Bag_{R_1}$, $\Bag_{\tilde{R}_1}$), or
determined from the HQET relations Eq.~(\ref{eq:HQETBagR}) ($\Bag_{R_2}$, $\Bag_{R_3}$) for which we allow a 20\% power correction
modelized as a flat error. All these bag parameters vary independently for $B_d$ and $B_s$ mesons, \textit{i.e.} we do not assume the exact SU(3) symmetry for them.

\subsection{Observables not affected by New Physics in mixing}\label{ssec:InputsWithoutNP}
In this section, we first discuss the observables allowing us to establish an 
universal preferred region in the $\bar\rho-\bar\eta$ subspace, independent of any New Physics 
contributions in mixing.
\begin{itemize}
\item The CKM matrix element $|V_{ud}|$ has been determined from three different 
      methods: superallowed nuclear $\beta$-decays, neutron $\beta$-decay and 
      pion $\beta$-decay.
      Currently, the best determination of $|V_{ud}|$ comes from superallowed 
      $\beta$-decays where the uncertainty is dominated by the theoretical 
      error, see e.g. Ref.~\cite{CKM05Vud,PDG2008}. 
      An analysis by Towner and Hardy~\cite{Hardy:2008gy} focusing on an 
      improvement of the isospin-symmetry-breaking terms finds a central value
      for $|V_{ud}|$ which is larger though still compatible when compared to 
      values quoted in the past, with a slightly reduced uncertainty:
      $|V_{ud}|=0.97425 \pm 0.00022$.

\item The matrix element $|V_{us}|$ can be determined from $K_{\ell 3}$ decays, 
      from hadronic $\tau$ decays, and from semileptonic hyperon decays. 
      We are using the $K_{\ell 3}$ average quoted by Flavianet~\cite{flavianet}.
      The experimental number for $f_{+}(q^{2}=0) \cdot |V_{us}|$ obtained by averaging
      results from ISTRA+, KLOE, KTeV, and NA48(/2), as quoted by Flavianet, is 
      $f_{+}(q^{2}=0) \cdot |V_{us}| = 0.2163 \pm 0.0005$
      leading to %~\cite{flavianet}
      %Taking the %most recent 
      %Lattice QCD result for the $K \to \pi$ 
      %form factor $f_{+}$ of $0.959 \pm 0.005$~\cite{VusUKQCDRBC}, this
      %translates into 
      $|V_{us}| = 0.2254 \pm 0.0013$~\cite{VusUKQCDRBC,flavianet}.

\item The matrix element $|V_{cb}|$ is obtained from semileptonic decays
      $B \to X_{c} \ell \nu$, where $X_c$ is either a $D^{*}$ meson (exclusive 
      method) or a sum over all hadronic final states containing charm 
      (inclusive method). For several years the most precise value has been 
      provided by the inclusive method where the theoretical uncertainties 
      have been pushed below the $2\percent$ level by determining the relevant 
      non-perturbative Heavy Quark Expansion (HQE) parameters from moment 
      measurements in $B \to X_{c}\ell\nu$ and $B \to X_{s}\gamma$ decays.
      The inclusive $|V_{cb}|$ value used in our analysis is taken from the
      Heavy Flavour Averaging Group (HFAG)~\cite{HFAG10},
      $|V_{cb,\text{incl}}|=(41.85\pm0.43\pm0.59) \times 10^{-3}$, where 
      the first error contains the experimental and HQE uncertainties and 
      the second reflects the theoretical uncertainty on the total rate 
      prediction for $B \to X_{c}\ell\nu$.\\
      The theoretical uncertainty on the exclusive $|V_{cb}|$ determination 
      in the calculation of the form factor value at zero recoil $F(1)$ has 
      not been competitive so far.
      A recent calculation provides a significantly smaller error budget: 
      $F(1)=0.921 \pm 0.013 \pm 0.020$~\cite{Bernard:2008dn}, although the exclusive
      $|V_{cb}|$ determination still gives a larger uncertainty.
      Using the average value for the product $F(1) |V_{cb}|=(36.04 \pm 0.52) \times 10^{-3}$
      from HFAG~\cite{HFAG10} and applying
      a $0.7\percent$ QED correction~\cite{Sirlin}, one finds 
      $|V_{cb,\text{excl}}|=(38.85\pm 0.56_\text{exp}\pm 0.55_\text{theostat}\pm 0.84_\text{theosys}) \times 10^{-3} 
                           =(38.85 \pm 0.77 \pm 0.84) \times 10^{-3}$,
      which has a smaller central value than the inclusive result.\\
      We average the two $|V_{cb}|$ values in such a way that the smallest 
      theoretical uncertainty is preserved similarly to our procedure to average lattice  
      inputs and we obtain $|V_{cb}| = (40.89 \pm 0.38 \pm0.59) \times 10^{-3}$, 
      keeping in mind, however, that the inclusive and exclusive numbers
      are not in perfect agreement.
      
    \item The two methods to extract $|V_{ub}|$, the inclusive and the
      exclusive ones (using the theoretically cleanest $B \to \pi \ell \nu$
      decays), both suffer from significant theoretical uncertainties.
      The exclusive measurements prefer values around $3.5 \times
      10^{-3}$~\cite{HFAG10}. The numbers quoted are from partial rates
      measured at large $q^{2}$ ($>16\gev^2$) or at small $q^{2}$
      ($<16\gev^{2}$), using form factor calculations from
      lattice QCD~\cite{FNAL,HPQCD}, or Light Cone Sum Rules
      (LCSR)~\cite{LCSR:BallZwicky}, respectively. 
      The fit input is the average of these numbers, following the same procedure as
      for the lattice QCD parameters:
      $|V_{ub,\text{excl}}|=(3.51\pm0.10\pm0.46) \times 10^{-3}$.
      \\
      The average of inclusive results quoted by HFAG using~\cite{HFAG10} 
      the Shape Function (SF) scheme~\cite{BLNP} yields
      $(4.32\pm0.16^{+0.22}_{-0.23}) \times 10^{-3}$, where the first
      uncertainty contains the statistical and experimental systematic
      uncertainty as well as the modelling errors for $b \to u \ell \nu$ 
      and $b \to c \ell \nu$ transitions.
      Compared to HFAG, we 
      modify the assignment of the uncertainties as follows. We add
      the following uncertainties in quadrature: statistical uncertainty,
      experimental systematics, $b \rightarrow c$ and $b \rightarrow u$
      modeling and the error from the HQE parameters ($b$-quark mass $m_b$
      and $\mu_{\pi}^{2}$). 
      Several theoretical uncertainties can only be guestimated: the shape
      function uncertainty, contributions from subleading shape functions,
      weak annihilations and the procedure of scale matching. We assign an 
      additional uncertainty on $m_b$, which reflects higher order corrections 
      not accounted for in the partial rate predictions for 
      $B \rightarrow X_u \ell \nu$~\cite{NeubertPrivateCommunication}.
      We choose $50~{\rm MeV}$ as the additional uncertainty.
      All these 
      uncertainties of a second type are added linearly. As a result, 
      we obtain a significantly larger theoretical uncertainty compared to the 
      uncertainty quoted by the HFAG: $(4.32^{+0.21}_{-0.24} \pm 0.45) \times 10^{-3}$.\\
      The exclusive and the inclusive inputs are then averaged using the
      same recipe as for the lattice QCD parameters and we obtain: 
      $|V_{ub}|=(3.92\pm0.09\pm0.45) \times 10^{-3}$.

\item A measurement of the branching fraction for $B^{+} \to\tau^+ \nu_\tau$
      allows one to constrain the product 
      $|V_{ub}| \cdot f_{B}$ where $f_{B}$
      is the decay constant of the charged $B$ meson. %\\
      The theoretical prediction for this branching fraction is given by
      \begin{equation}
       \BRB{\tau\nu}
       = \frac{G_F^2 m_{B^{+}} m_{\tau}^2 }{ 8 \pi }
       \left( 1 - \frac{m_{\tau}^2}{m_{B^{+}}^2} \right)^2 
           |V_{ub}|^2 f_{B}^2 \tau_{B^{+}}.
      \end{equation}
      We use the experimental value $\tau_{B^{+}}=1.639 \cdot 
          10^{-12} s$ in our analysis. 
      $\BRB{\tau\nu}$ combined with the constraint from the
      oscillation frequency $\Delta m_{d}$ (see
      Sect.~\ref{ssec:InputsBdWithNP}) removes the
      dependence on the decay constant $f_{B}$ (assuming 
      that the decay constant for the charged and neutral $B$
      meson is the same, i.e., neglecting
      isospin-breaking effects of order $1\%$).\\
      First evidence for the decay $B^{+} \rightarrow \tau^{+} \nu_{\tau}$ has been 
      made by the Belle collaboration~\cite{btaunu_belle1} by reconstructing the decay
      on the recoil of fully-reconstructed B-meson decays. Using the same technique
      the \babar collaboration found then a $2.2~\sigma$ excess~\cite{btaunu_babar1}. 
      By searching for $B^{+} \rightarrow \tau^{+} \nu_{\tau}$ on the recoil of 
      semileptonic B-meson decays Belle~\cite{btaunu_belle2} found also evidence
      for $B^{+} \rightarrow \tau^{+} \nu_{\tau}$ while \babar found a $2.3~\sigma$ 
      excess~\cite{btaunu_babar2}.
      Recently, the Belle analysis first presented in Ref.~\cite{btaunu_belle2}
      has been submitted for publication~\cite{btaunu_belle3} reporting a slightly
      shifted central value and the \babar analysis described in Ref.~\cite{btaunu_babar1} 
      has been updated to the full data set~\cite{btaunu_babar1} in which evidence has 
      been found for the decay $B^{+} \rightarrow \tau^{+} \nu_{\tau}$.

      The world average for ${\mathcal{B}}(B^{+} \rightarrow \tau^{+} \nu_{\tau})$ 
      that is used in the analysis calculated from the measurements performed by 
      \babar\ and Belle~\cite{btaunu_belle1,btaunu_belle3,btaunu_babar2,btaunu_babar3}
      is ${\mathcal{B}}(B^{+} \rightarrow \tau^{+} \nu_{\tau})=(1.68 \pm 0.31) \times 10^{-4}$.

\item The input for the CKM angle $\gamma$ 
      ($=\arg [ -V_{ud}^{\phantom{*}}V_{ub}^*/(V_{cd}^{\phantom{*}}V_{cb}^*) ]$) 
      is taken from a combined full frequentist analysis of the CKMfitter 
      group using \CP-violating asymmetries in charged $B$ decays to 
      neutral $D^{(*)}$ mesons plus charged $K^{(*)}$ mesons.
      The data are taken from HFAG using the three different methods proposed 
      by Gronau, London, Wyler (GLW)~\cite{GLW}, and Atwood, Dunietz, Soni (ADS)~\cite{ADS}, and 
      including also the Dalitz plot approach developed by Giri, Grossman, 
      Soffer and Zupan (GGSZ), and independently by the Belle collaboration~\cite{GGSZ}. 
      At the $68.3\percent$ confidence level (CL), the result of this analysis is
      $(71^{+21}_{-25})^{\circ}$ with a second solution
      at $\gamma+\pi$. The constraint is shown in Figure~\ref{fig-gamma}.

      The above determination of $\gamma$ raises interesting statistical issues. 
      The angle $\gamma$ actually appears as the complex phase of a suppressed
      ratio $r_B$ of decay amplitudes (different $r_B$'s appear for different final states);
      in other words in the limit $r_B\to 0$ there is no constraint left on the CKM
      phase, and for finite $r_B$ the error on $\gamma$ is roughly inversely
      proportional to $r_B$ itself. It turns out that the current data do not exclude
      tiny values for the $r_B$'s, one obtains for
      the $DK$ final state $r_B=0.103^{+0.015}_{-0.024}$ at 68.3\% CL,
      the $D^{*}K$ final state $r_B=0.116^{+0.025}_{-0.025}$ at 68.3\% CL,
      and for the final state $DK^{*}$ final state $r_B=0.111^{+0.061}_{-0.047}$ at 68.3\% CL.
      Because the ratio of amplitudes is related to the bare observables non-linearly, 
      its maximum-likelihood estimate is biased, and
      it can be shown that this bias overestimates the value of $r_B$ which in turn
      implies an underestimate of the uncertainty on $\gamma$. In the statistical
      language, this effect yields a significant \textit{undercoverage} of the
      na\"{\i}ve 68.3 \% CL interval for $\gamma$ computed from the log-likelihood
      variation.

      A better estimate of the statistical uncertainty on $\gamma$ can be obtained
      by inspecting the deviation  of the distribution of 
      the log-likelihood among a large number of toy experiments from its asymptotic limit.
      Problems arise because 
      this distribution is not only non asymptotic, but also depends on
      \textit{nuisance parameters}, that is other parameters than $\gamma$ that
      are necessary to compute the toy experiments. In such a situation the most
      conservative approach is called the \textit{supremum} one, since it maximizes
      the uncertainty over all possible values of the nuisance parameters. To date
      this method which guarantees the coverage properties by construction is the
      default one for the treatment of $\gamma$ in CKMfitter, but it must be kept
      in mind that it actually leads to overcoverage in general~\cite{karimCKM08}.

      Another determination of $\gamma$ which is unaffected by New Physics 
      in \bbm\ is obtained by combining measurements of $\alpha$ and $\beta$: 
      The corresponding quantities are changed into
      $\alpha-\phi_d^{\Delta}/2$ and $\beta+\phi_d^{\Delta}/2$ in the presence of a new
      physics phase $\phi_d^{\Delta}$~
%      \footnote{potential non standard electroweak
%      penguin $b\to d$ transitions are excluded by our SM4FC hypothesis}, 
      which therefore drops out of $\gamma=\pi-(\alpha-\phi_d^{\Delta}/2)- (\beta+\phi_d^{\Delta})$. 
      This procedure leads to a significantly more precise determination of $\gamma$ 
      than $B\to D^{(*)} K^{(*)}$ decays. The individual measurements of
      $\alpha-\phi_d^{\Delta}/2$ and $\beta+\phi_d^{\Delta}/2$ are described 
      in the next section. 
\end{itemize}

\begin{nfigure}{Htb}
 \includegraphics[width=10cm]{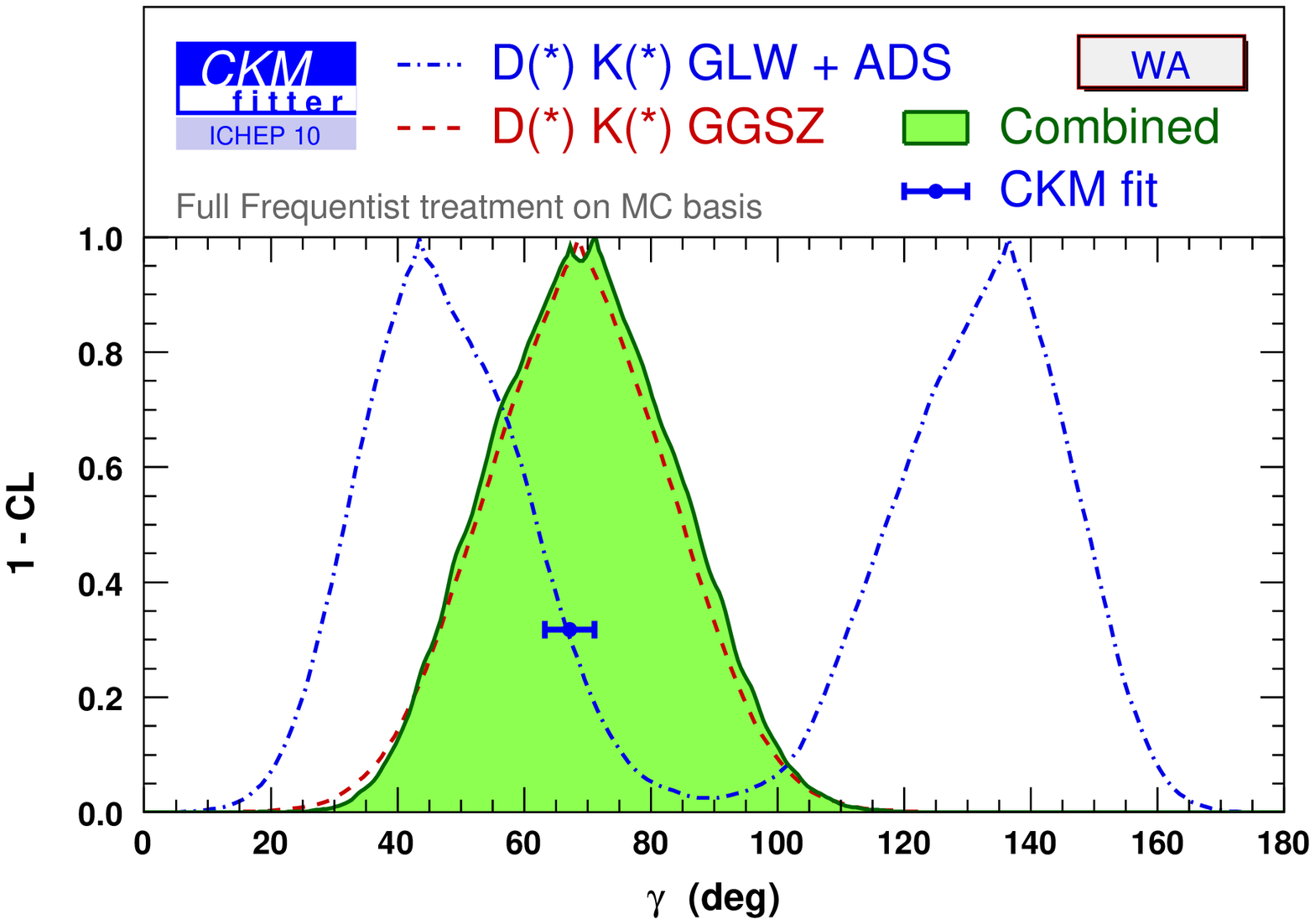}
 \caption{\small Constraint on the angle $\gamma$ from a combined analysis of
   $B\to D^{(*)}K^{(*)}$ decays.}\label{fig-gamma}\end{nfigure}

%\noindent 
%Figure~\ref{fig-universalUT} displays the profile of the preferred $(\bar\rho,\bar\eta)$ 
%plane when the fit is performed using only the inputs $|V_{ud}|$, $|V_{us}|$,
%$|V_{cb}|$, $|V_{ub}|$, $\gamma$, $\alpha-\phi_d^{\Delta}/2$ and $\beta+\phi_d^{\Delta}/2$.
%Two additional branches from $\alpha-\phi_d^{\Delta}/2$ and $\beta+\phi_d^{\Delta}/2$ exist 
%in the negative $\bar{\eta}$ plane which are further discussed in Sect.~\ref{ssec:bdbs}. 

%\begin{nfigure}{htb}
%  \includegraphics[width=10cm]{rhoeta_small_tree_bothgam.eps}
%  \caption{\small Constraint on the CKM $\bar\rho-\bar\eta$ coordinates, using only the 
%    inputs which are not affected by NP in $B_d$ mixing: $|V_{ud}|$, $|V_{us}|$, 
%    $|V_{cb}|$, $|V_{ub}|$ from semileptonic decays and from $B \rightarrow \tau \nu$, 
%    $\gamma$ and $\alpha-\phi_d^{\Delta}/2$ and $\beta+\phi_d^{\Delta}/2$.
%    Regions outside the coloured areas have ${\rm CL} > 95.45~\%$. For the combined fit 
%    the yellow area inscribed by the contour line represents points with ${\rm CL} < 95.45~\%$. 
%    The shaded area inside this region represents points with ${\rm CL} < 68.3~\%$.
%    \label{fig-universalUT}}
%\end{nfigure}

\boldmath
\subsection
{Observables in the {\bbduli} system affected by New Physics in mixing}\label{ssec:InputsBdWithNP}
\unboldmath
\noindent 
The following observables  can be affected by $\vert \Delta F\vert =2$ New Physics 
contributions in the $B_d^0 - {\bar{B}}_d^0$ system:

\begin{itemize}
\item The oscillation frequency $\Delta m_d$ in the $B_{d}$ sector has been measured 
      with $\Order(1\percent)$ precision mainly due to the $B$-factory data~\cite{HFAG08}.
      The translation of the measured value for $\Delta m_d$ into constraints 
      on the CKM parameter combination ${|V_{td}^{\phantom{*}} V_{tb}^{*}|}^{2}$ 
      suffers from significant uncertainties in the theoretical calculation of the 
      product $f_{B_{d}} \sqrt{\Bag_{B_{d}}}$ of hadronic parameters. These hadronic 
      parameters can be obtained from lattice QCD computation, computing
      $f_{B_{d}} \sqrt{\Bag_{B_d}}$ 
      from $f_{B_{s}}$ and $\Bag_{B_{s}}$, and the flavour-symmetry breaking ratios 
      $f_{B_{s}}/f_{B_{d}}$ and $\Bag_{B_{s}}/\Bag_{B_{d}}$ as quoted in 
      Section~\ref{ssec:MethodofAveraging} and summarized in Table~\ref{tab:TheoreticalInputs}. 
      The value and uncertainty for the perturbative QCD correction ${\hat{\eta}}_{B}$ 
      has been originally estimated in Ref.~\cite{BBL95}. With up-to-date values and 
      uncertainties for $\alpha_{s}$ and the top-quark mass one obtains 
      $\hat{\eta}_{B} = 0.8393 \pm 0.0034$.
      For the top-quark mass we take the average value of the Tevatron Electroweak 
      Working Group~\cite{TevEWWG}, $m_{t}=(172.4 \pm 1.2)~{\rm GeV}$, combining 
      published and also preliminary results from \Dzero and CDF. This mass, 
      interpreted as a pole mass, is translated into 
      $\ov m_{t}(\ov m_{t})=(165.017 \pm 1.156 \pm 0.11)~{\rm GeV}$ in the $\overline{\rm MS}$-scheme
      at one-loop order. It should be noted, however, that the identification of
      the measured mass value to the pole mass is under debate (with potential new systematics coming from this identification),
      see e.g.~\cite{Langenfeld:2010aj}.

\item In the Standard Model, the predicted decay width difference
  $\Delta\Gamma_d$ is small: 
      $\Delta\Gamma_d=(38.1^{+7.2}_{-14.1}) \times 10^{-4}\psinv$~\cite{ln} (with our inputs).
      The average between DELPHI and \babar\ measurements of the ratio 
      $\Delta\Gamma_d/\Gamma_d$ calculated by HFAG~\cite{DeltaGammad,BabarDeltaGammad,HFAG08} 
      is $0.009\pm0.037$.
      The experimental uncertainty is much larger than the size of the Standard Model prediction,
      $\Delta\Gamma_d/\Gamma_d=(58^{+11}_{-22}) \times 10^{-4}$, and the 
      measured value is in good agreement with the Standard Model prediction within experimental 
      uncertainties so that stringent constraints on New Physics contributions cannot be derived 
      at the present stage of precision. Since a huge amount of statistics will be needed to
      measure $\Delta\Gamma_d$ at the level predicted by the Standard Model,
      this situation will probably
      not change for quite a long time. Even then if a deviation from the Standard Model
      value were observed due to New Physics contributions in mixing it would show up
      beforehand in other observables like
      $a_\text{SL}^{d}$ or $\sin{2\beta}$. As a consequence, $\Delta\Gamma_d$ has no 
      visible impact in our discussion 
      and it is not used as an input to the fits presented here.

\item \CP violation in $B_{d}$ mixing (i.e., $|q/p| \ne 1$, with $q/p$ defined in
       Eq.~\eqref{eq:qp}) can be measured from the untagged dilepton rate asymmetry
      \begin{equation}
      a_\text{SL}^{d}=\frac{N_{\ell^{+}\ell^{+}}-N_{\ell^{-}\ell^{-}} }{N_{\ell^{+}\ell^{+}}+N_{\ell^{-}\ell^{-}}}
          = 2(1-|q/p|)~.
      \end{equation}
      With a tagged time-dependent decay asymmetry, one measures equivalently
        \begin{align}
              A_\text{SL}^q(t)& \:\equiv\:
              \frac{\Gamma(\bar B^0_q(t)\to l^{+}X)-\Gamma(B^0_q(t)\to l^{-}X)}
                   {\Gamma(\bar B^0_q(t)\to l^{+}X)+\Gamma(B^0_q(t)\to
                     l^{-}X)}\\ &=\frac{1-|q/p|^4}{1+|q/p|^4}= 2 (1-|q/p|) +
                   \Order\big((|q/p| - 1)^2\big),
        \end{align}
        with the time-dependence dropping out.
      A weighted average of \babar, Belle and CLEO measurements~\cite{ASLdBfactories} 
      results in $A_{\text{SL}}^d=-(47\pm 46)\times 10^{-4}$~\cite{HFAG08}, 
      which is a bit less than one standard deviation below the Standard Model prediction
      of $a_\text{SL}^d= (-7.58\epmuli{2.11}{0.64})\times 10^{-4}$
      \cite{ln,bbln,dega} (with our inputs).\\

\item Within the Standard Model the measurement of the $S=\sin 2\phi_d^{\psi K}$ coefficient in the time-dependent 
      \CP asymmetry 
      $A_{\CP}(t)=S\, \sin{(\Delta m_{d} \cdot t)} + C \, \cos{(\Delta m_{d} \cdot t)}$ 
      in decays of neutral $B_{d}$ mesons to final states $(c \bar{c}) K^{0}$ provides 
      a measurement of the parameter $\sin{2\beta}$,
      where $\beta =\arg [ -V_{td}^{\phantom{*}}V_{tb}^* / (V_{cd}^{\phantom{*}}V_{cb}^*) ]$,  
      to a very good approximation. 
      The current uncertainty of $0.023$ on $\sin 2\phi_d^{\psi K}$
       is still dominated by statistics~\cite{HFAG08}. 
      The difference between the measured $\sin 2\phi_d^{\psi K}$ coefficient and $\sin{2\beta}$ due to
      penguin contributions has been theoretically estimated in Ref.~\cite{BoosMannelReuter} to be below 
      the $10^{-3}$ level, while phenomenologically less stringent constraints on this difference are quoted in 
      Refs.~\cite{GrossmanKaganLigeti,LiMishima,Ciuchini1,Faller:2008zc}.\\
      When interpreting the measured $\sin 2\phi_d^{\psi K}$ coefficient as $\sin{(2\beta+\phi_{d}^{\Delta})}$
      the \emph{Standard Model4FC} hypothesis does not rigourously apply. However, the gluonic penguin 
      is \emph{OZI}-suppressed and the  $Z$-penguin is estimated to be small so that New Physics 
      in decay is assumed to be negligible with respect to the leading tree amplitude. 
      We neglect the effect from possible New Physics in $K-\bar{K}$ mixing on $\sin{(2\beta+\phi_{d}^{\Delta})}$, 
      which is justified given the small value of the well-measured \CP-violating parameter 
      $\epsilon_{K}$.

\item The measurement of $\sin{2\phi_d^{\psi K}}$ results in two solutions
      for $2\beta+\phi_{d}^{\Delta}$ (in $[0,\pi]$). 
      One of these solutions can be excluded by measuring 
      the sign of $\cos{2\phi_d^{\psi K}}$. 
      For a recent review of \babar\ and 
      Belle measurements see e.g. Ref.~\cite{lackerbeauty06}. 
      The current experimental 
      results from \babar\ and Belle using a time- and 
      angular-dependent analysis 
      of $B^0\to \jpsi K^{*0}$ decays, a time-dependent 
      Dalitz-plot analysis in 
      $\Bz/\Bzb \to D^{(*)0}/\bar{D}^{(*)0} h^{0}$ with 
      $\Bz/\Bzb \to D^{(*)0}/\bar{D}^{(*)0} h^{0}$, and 
      $\Bz/\Bzb \rightarrow D^{*+} D^{*-} K^{0}_{S}$,
      disfavour negative
            $\cos{2\phi_d^{\psi K}}$ values but
      HFAG deems it difficult to average the different measurements
      or to determine a reliable confidence level as a function of 
      $\cos{2\phi_d^{\psi K}}$~\cite{HFAG08}. Here, as a simplification, it 
      is only assumed that $\cos{2\phi_d^{\psi K}}>0$.
      
\item The constraint on the CKM angle $\alpha=\pi-\beta-\gamma$ 
      is obtained from time-dependent and
      time-independent measurements in the decays $B \to \pi\pi$, $B \to
      \rho\rho$, and $B \to \rho\pi$.  The time-dependent \CP
      asymmetries measured in $B \to \pi\pi$ provide information on the
      effective parameter $\sin{(2\alpha_\text{eff})}$ (which is a function of
      $\alpha$ and the penguin-to-tree ratio~\cite{Charles:1998qx}).  It is possible to
      translate this measurement into a constraint on $\alpha$ by
      exploiting isospin symmetry which allows to pin down 
      the penguin-to-tree ratio and thus to  determine the
      difference $\alpha-\alpha_\text{eff}$ from data~\cite{GronauLondon}.
      Under the assumption of exact isospin symmetry the amplitudes
      $A^{+-} \equiv A(B^{0} \to \pi^{+}\pi^{-})$, $A^{00} \equiv
      A(B^{0} \to \pi^{0}\pi^{0})$, and $A^{+0} \equiv A(B^{+} \to
      \pi^{+}\pi^{0})$ satisfy a triangular relationship:
      $\sqrt{2}A^{+0} - \sqrt{2}A^{00} = A^{+-}$.  A corresponding
      relationship holds for the \CP conjugated decays:
      $\sqrt{2}{\bar{A}}^{+0} - \sqrt{2}{\bar{A}}^{00} =
      {\bar{A}}^{+-}$.  These isospin triangles can be reconstructed by
      measuring the branching fractions and direct \CP asymmetries for
      the final states $B^{0} \to \pi^{+}\pi^{-}$, $B^{0} \to
      \pi^{0}\pi^{0}$, and $B^{+} \to \pi^{+}\pi^{0}$. Since one
      measures $\sin{(2\alpha_\text{eff})}$ and since the triangle has a twofold
      ambiguity for its apex in the complex
      plane, there is an eightfold ambiguity for $\alpha$ in $[0,\pi]$.
      The extraction of $\alpha$ from the isospin analysis is independent 
      of any possible New Physics contributions in the $\Delta I=1/2$ decay amplitude 
      except for the singular point $\alpha=0$~\cite{isospinprior}. If there
      is no New Physics contribution in the $\Delta I=3/2$ decay amplitude (as assumed here), 
      the
      extraction provides $\alpha=\pi-\gamma-\beta-\phi_{d}^{\Delta}/2$. As a consequence, 
      $\alpha$ is equivalent to $\gamma$ if $\beta+\phi_{d}^{\Delta}$ is measured e.g. 
      from $B \to \jpsi K_{S}$  as already pointed out above.
      
      Similar in line an isospin analysis can be performed for the $B
      \to \rho\rho$ system. In this case the analysis needs to take into
      account the measured longitudinal polarisation of the $\rho$
      mesons in the different final states $B^{0} \to \rho^{+}\rho^{-}$,
      $B^{0} \to \rho^{0}\rho^{0}$,
      and $B^{+} \to \rho^{+}\rho^{0}$. Finally, the $\rho\pi$ modes provides
      another crucial input for $\alpha$, by using a model for the Dalitz decay 
      into three pions in addition to the isospin symmetry~\cite{snyderquinn}.
      The results of these analyses that are based on the world averages of
      the various \babar\ and Belle measurements for the CP asymmetries and
      branching fractions determined by HFAG~\cite{HFAG08} 
      are displayed in Figure~\ref{fig-alpha}.
      The combined analysis results in $\alpha=(89.0^{+4.4}_{-4.2})^{\circ}$ at $68.3\percent$ CL.

      The most stringent constraint on $\alpha$ comes currently from the
      $B^{0} \to \rho \rho$ channel. The uncertainty is driven by the
      rather large branching fraction ${\mathcal{B}}(B^{+} \to \rho^{+}\rho^{0})$.
      The input value for ${\mathcal{B}}(B^{+} \to \rho^{+}\rho^{0})$ has changed
      recently when the \babar collaboration presented a new analysis on the 
      final data set~\cite{babar_rho+rho0}. The large branching fraction value 
      leads to an isospin triangle that just closes. The measurement uncertainty 
      is smaller than the expected uncertainty. As a consequence, the current 
      uncertainty quoted for $\alpha$ might be on the optimistic side.
      It should also be stressed that at this level of precision so-far neglected
      uncertainties (electroweak penguins, $\pi-\eta^{(')}$ mixing, $\rho-\omega$ 
      mixing, other isospin violations, finite $\rho$-width, etc.) should be 
      considered in more detail. 
\end{itemize}

\begin{nfigure}{Htb}
 \includegraphics[width=10cm]{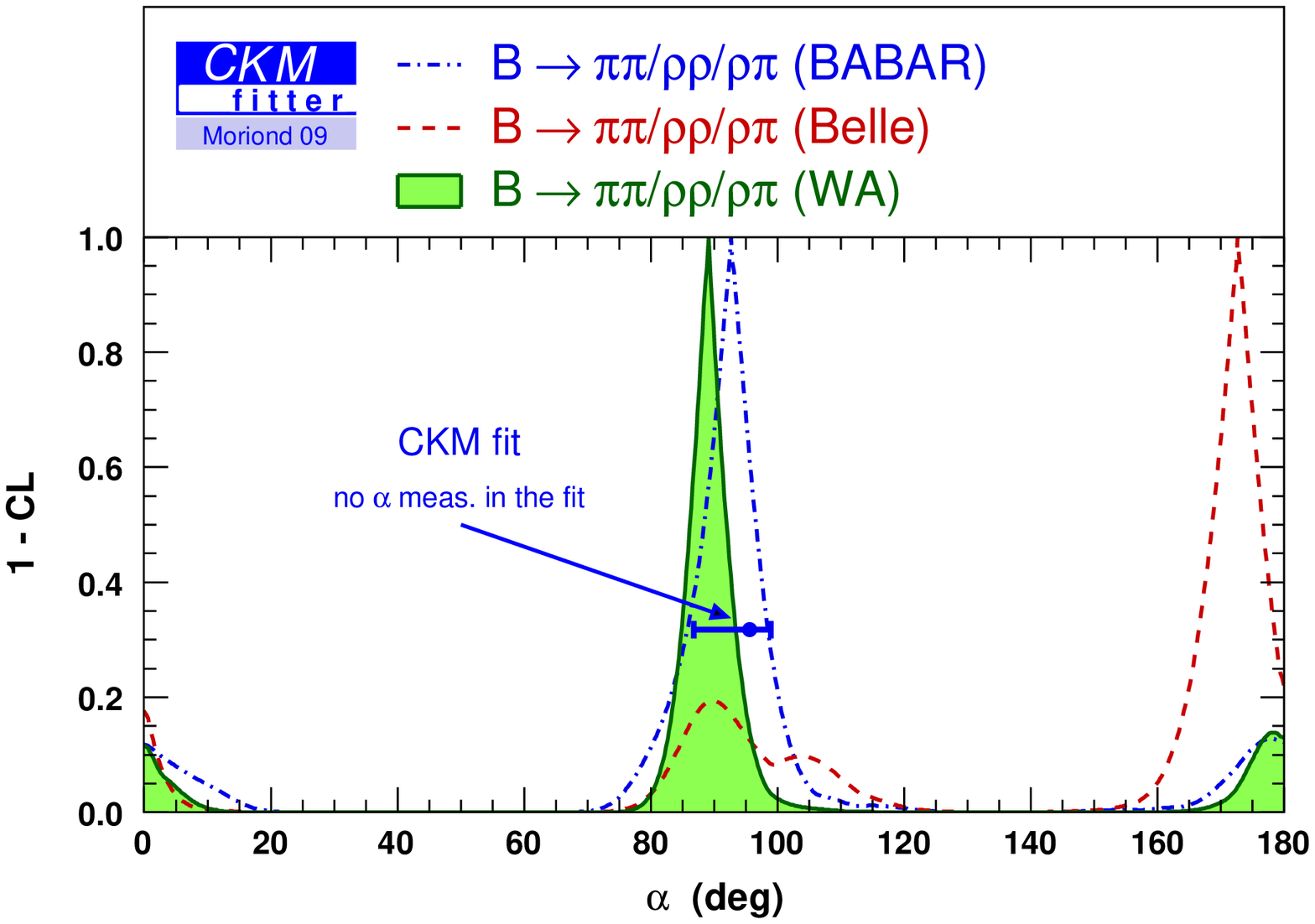}
 \caption{\small Constraint on the angle $\alpha$ from the isospin analyses of 
   $B\to \pi\pi$ and $B\to\rho\rho$ decays and a Dalitz plot analysis of 
    $B\to\rho\pi$ decays.  In the presence of New Physics in $B_d$ mixing, the
   quantity shown is $\pi-\gamma-\beta-\phi_{d}^{\Delta}/2$.}\label{fig-alpha}
\end{nfigure}

\begin{table}[Htp]
\renewcommand{\arraystretch}{1.3}
\centering
\begin{tabular}{|c|c|c|}\hline
Observable                                      & Value and uncertainties                      & Reference              \\
\hline
$|V_{ud}|$                                      & $0.97425 \pm 0.00022$                        &\cite{Hardy:2008gy}     \\
$|V_{us}|$                                      & $0.2254  \pm 0.0013$                         &\cite{flavianet}        \\
$m_{K}$                                         & $(497.614\pm 0.024)\mev$                     &\cite{PDG2008}          \\
$\GF$                                           & $(1.16637\pm 0.00001) \times 10^{-5}~\gev^2$ &\cite{PDG2008}          \\
$m_{W}$                                         & $(80.398 \pm 0.025)\gev$                     &\cite{PDG2008}          \\
$m_{B_{d}}$                                     & $(5.27917\pm 0.00029)\gev$                   &\cite{PDG2008}          \\
$m_{B^{+}}$                                     & $(5.27953\pm 0.00033)\gev$                   &\cite{PDG2008}          \\
$m_{B_{s}}$                                     & $(5.3663 \pm 0.0006)\gev$                    &\cite{PDG2008}          \\
$|V_{cb}|$                                      & $(40.89  \pm 0.38 \pm 0.59) \times 10^{-3}$  & see text               \\
$|V_{ub}|$                                      & $(3.92\pm0.09\pm0.45) \times 10^{-3}$        & see text               \\
$\mathcal{B}(B \rightarrow \tau \nu_{\tau})$    & $(1.68 \pm 0.31) \times 10^{-4}$             & see text               \\
$\gamma$                                        & $71{^{\,+21}_{\,-25}}^{\circ}$              & see text               \\
\hline
$\Delta m_{d}$                                  & $(0.507  \pm 0.005)\psinv$                   &\cite{PDG2008}          \\
$a_\text{SL}^{d}$                               & $(-47 \pm 46)\times 10^{-4}$                         &\cite{HFAG08}           \\
$\sin (2\phi_d^{\psi K})$                                   & $0.673   \pm 0.023$                          &\cite{HFAG09}           \\
$\cos (2\phi_d^{\psi K})$                                   & positive                                     &see text                \\
$\alpha$                                        & $89.0{^{\,+4.4}_{\,-4.2}}^{\circ}$           &see text                \\
\hline
$\Delta m_{s}$                                  & $(17.77  \pm 0.12)\psinv$                    &\cite{dmsexpvalue}      \\
$A_\text{SL}$                                   & $(-85 \pm 28)\times 10^{-4}$                       &see text and \cite{dimuon_evidence_d0,ASLCDF} \\
$a_\text{SL}^{s}$                               & $(-17 \pm 93)\times 10^{-4}$                         &\cite{ASLsD0_3}         \\
%$\Delta \Gamma^{\CP\prime}_{s}$                & $(0.090  \pm 0.050)\psinv$                   &\cite{D0DeltaGammasCP',CDFDeltaGammasCP'} \\
$\phi_s^{\psi\phi}$ vs. $\Delta \Gamma_{s}$ & see text                                     &\cite{taggedphaseCDF_2,taggedphaseD0,Punzi}                        \\ 
%Was: \phi_{s}
\hline
$|\epsilon_{K}|$                            & $(2.229  \pm 0.010) \times 10^{-3}$          &\cite{PDG2008}          \\
$\Delta m_{K}$                                  & $(3.483  \pm 0.006) \times 10^{-12}\mev$     &\cite{PDG2008}          \\
\hline
\end{tabular}
\caption[ExperimentalInputs]
{Experimental inputs used in the fits.}
\label{tab:ExperimentalInputs}
\end{table}

\begin{table}[htp]
\renewcommand{\arraystretch}{1.3}
\centering
\begin{tabular}{|c|c|c|}\hline
Theoretical Parameter                          & Value and uncertainties                 & Reference                                 \\
\hline
$f_{B_{s}}$                                    & $(231    \pm 3     \pm 15)\mev$         & see Section~\ref{ssec:MethodofAveraging}  \\
$\Bag_{B_{s}}\big(m_{b}\big)$                  & $0.841   \pm 0.013 \pm 0.020$           & see Section~\ref{ssec:MethodofAveraging}  \\
$f_{B_{s}}/f_{B_{d}}$                          & $1.209   \pm 0.007 \pm 0.023$           & see Section~\ref{ssec:MethodofAveraging}  \\
$\Bag_{B_{s}}/\Bag_{B_{d}}$                    & $1.01    \pm 0.01  \pm 0.03$            & see Section~\ref{ssec:MethodofAveraging}  \\
$\hat{\eta}_{B}$                               & $0.8393   \pm 0.0034$                   & see text                                  \\
$\ov m_{t}(\ov m_{t})$                         & $(165.017 \pm 1.156 \pm 0.11)\gev$      & see text and \cite{TevEWWG}               \\
${\hat{\Bag}}_{K}$                             & $(0.724  \pm 0.004 \pm 0.067)$          & see Section~\ref{ssec:MethodofAveraging}  \\
$f_{K}$                                        & $156.1\mev$                             & see text                                  \\
$\kappa_\epsilon$                              & $0.940 \pm 0.013 \pm 0.023$                      & see text                                  \\
$\eta_{tt}$                                    & $0.5765  \pm 0.0065$                    &{\rd \cite{bjw}}                           \\
$\eta_{ct}$                                    & $0.47    \pm 0.04$                      &{\rd \cite{Nierste2}}                      \\
$\eta_{cc}$                                    & $\rd (1.39 \pm 0.35) \lt( \frac{1.29\, \gev}{\ov m_c}\rt)^{1.1}$ &\cite{Nierste1}   \\
$\lambda_{K}$                                  & $(1.25   \pm 0.00 \pm 0.75)\gev$        & see text                                  \\
$\ov m_{c}(\ov m_{c})$                         & $(1.286  \pm 0.013 \pm 0.040)\gev$      & see text                                  \\
$\widetilde{\Bag}_{S,B_s}/\widetilde{\Bag}_{S,B_d}$ & $1.01 \pm 0 \pm 0.03$              &\cite{BGMPR}                               \\
$\widetilde{\Bag}_{S,B_s}(m_b)$                & $0.91 \pm 0.03 \pm 0.12$                &\cite{BGMPR}                               \\
$\rd \Lambda_{\ov{\rm MS}}^{(5)}$              & $(0.222 \pm 0.027)\gev$                 & from $\alpha_s(M_{Z})$ in \cite{PDG2008}  \\
$\overline{m}_{s}(\overline{m}_{b})$           & $(0.085 \pm 0.017)\gev$                 &\cite{ln}                                  \\
$\overline{m}_{b}(\overline{m}_{b})$           & $(4.248 \pm 0.051)\gev$                 &\cite{HFAG10}                              \\
$m_{b}^{pow}$                                  & $(4.7\pm 0 \pm 0.1)\gev$                &\cite{ln}                                  \\
$\Bag_{R_{0}}$                                 & $1.0 \pm 0.5$                           &\cite{ln}                                  \\
$\Bag_{\tilde{R}_{1}}$                         & $1.0 \pm 0.5$                           &\cite{ln}                                  \\
$\Bag_{R_{1}}$                                 & $1.0 \pm 0.5$                           &\cite{ln}                                  \\
%$\Bag_{R_{2}}$                                 & $1.0 \pm 0.5$                           &\cite{ln}                                  \\
$\Bag_{\tilde{R}_{2}}$                         & $1.0 \pm 0 \pm 0.5$                     &\cite{ln}                                  \\
%$\Bag_{R_{3}}$                                 & $1.0 \pm 0.5$                           &\cite{ln}                                  \\
$\Bag_{\tilde{R}_{3}}$                         & $1.0 \pm 0 \pm 0.5$                     &\cite{ln}                                  \\
\hline
\end{tabular}
\caption[TheoreticalInputs]
{Theoretical inputs used in the fits.}
\label{tab:TheoreticalInputs}
\end{table}

\boldmath
\subsection
{Observables in the \bbsuli system affected by New Physics in mixing}\label{ssec:InputsBsWithNP}
\unboldmath

\noindent 
We now discuss observables which can be possibly affected by $\vert \Delta F\vert =2$ New Physics 
contributions in the \Bs-\Bsb system.
\begin{itemize}
\item  The experimental input for the mass difference $\Delta m_s$ taken from HFAG is dominated by 
       the measurement of CDF~\cite{dmsexpvalue}. The dependence of the Standard Model prediction
       $\Delta m_s^{\text{SM}}$ on $\bar \rho-\bar \eta$ coordinates appears very 
       weak through the relevant CKM matrix elements term $\vert V^*_{ts}V^{\phantom{*}}_{tb}\vert^2$, 
       but the value of $\Delta m_s$ gives a direct constraint in the $\Delta_s$ plane, 
       by computing  $f_{B_{s}} \sqrt{\hat{\Bag}_{B_{s}}}$
       from $f_{B_{s}}$ and $\Bag_{B_{s}}$, given in Table~\ref{tab:TheoreticalInputs}.
       The hadronic matrix element for the $B_s$ system       
        can be related to the $B_d$ one via the 
       flavour-${\mathit SU}(3)$ breaking correction parameter $\xi$ defined through 
       $f_{B_s}^2 \Bag_{B_{s}} = \xi^2 f_{B_d}^2 \Bag_{B_{d}}$ 
       Measurements of $\Delta m_s$ thus reduce the uncertainties on $f_{B_d}^2 \Bag_{B_{d}}$
       since $\xi$ is better known from lattice QCD than $f_{B_d}^2 \Bag_{B_{d}}$.
       This relation also generates a strong correlation between the New Physics parameters 
       $|\Delta_d|$ and $|\Delta_s|$ when allowing for New Physics contributions to mixing.

\item  The Standard Model prediction \cite{ln} for $a_\text{SL}^{s}$ is at least one order of magnitude 
       smaller than the one for $a_\text{SL}^{d}$, see Table \ref{tab:fitResults_SM1}.
       Compared to the Standard Model prediction the \Dzero\ measurement has a quite large uncertainty, 
       $a_\text{SL}^{s}=(-17 \pm 91{^{+12}_{-23}}) \times 10^{-4}$~\cite{ASLsD0_3}, 
       and hence does not have a strong impact on the New Physics constraints in the $B_{s}$ sector. 
       It is nevertheless included in our analysis.

\item  \Dzero~\cite{dimuonexp} using $1~fb^{-1}$ and CDF~\cite{ASLCDF} have measured inclusive 
       dimuon CP asymmetries. The \Dzero result corresponding to the measurement quoted in 
       Ref.~\cite{dimuonexp} presented as 
       \(
           a_\text{SL}^{d} + \frac{f_{s} Z_{s}}{f_{d} Z_{d}} a_\text{SL}^{s} = -0.0028\pm0.0013\pm0.0009
       \)
       is provided in~\cite{D0CombinedCPasymmetry} whereas CDF~\cite{ASLCDF} quotes the result as
       \begin{equation}
       A_\text{SL}= \frac{f_{d} Z_{d} a_\text{SL}^{d} + f_{s} Z_{s} a_\text{SL}^{s}}{f_{d} Z_{d} + f_{s} Z_{s}}
                  =0.0080\pm0.0090\pm0.0068, 
       \end{equation}
       where $f_{d(s)}$ is the fraction of neutral $B_{d(s)}$ mesons 
       in the fragmentation and $Z_{d(s)}$ is given by 
       \begin{equation}
          Z_{q}= \frac{1}{1-y_{q}^{2}}-\frac{1}{1+x_{q}^{2}}
       \end{equation}
       with $y_{q}=\Delta \Gamma_{q}/2\Gamma_{q}$ and $x_{q}=\Delta m_{q}/\Gamma_{q}$
       (see also Ref.~\cite{nir2006}).
       Very recently \Dzero has presented a new measurement of $A_\text{SL}$~\cite{dimuon_evidence_d0}
       using $6.1~fb^{-1}$ integrated luminosity which shows a $3.2\sigma$  deviation from the 
       (almost zero) Standard Model prediction, and is the first direct evidence against the Standard Model in $B$ meson observables:
       $A_\text{SL}=-0.00957 \pm 0.00251 \pm 0.00146$. This result supersedes the former 
       result in Ref.~\cite{D0CombinedCPasymmetry}. The average between the new \Dzero result and the
       CDF result reads 
       \begin{equation}
       A_\text{SL}=-0.0085 \pm 0.0028,
       \end{equation}
        which is $2.9$ standard deviations away
       from the Standard Model prediction.\\
       For the interpretation of the measured observables we use the following values and 
       uncertainties for $f_{d(s)}$, $y_{q}$ and $x_{q}$: 
       $f_{d}=0.333 \pm 0.030$, $f_{s}=0.121 \pm 0.015$ with a correlation coefficient of
       $+0.439$ are taken from Ref.~\cite{HFAG10}, $x_{d}$ is calculated from 
       $\Delta m_{d} = (0.507 \pm 0.005)\cdot 10^{12} s^{-1}$~\cite{PDG2008} and $\tau_{B_{d}}=(1.525 \pm 0.009) \cdot 10^{-12} s$~\cite{HFAGforPDG2009},
       $y_{d}$ is set to zero since $\Delta \Gamma_{d}$ is expected to be very small
       in the Standard Model and is not affected by New Physics in our scenarios, $x_{s}$ is calculated from 
       $\Delta m_{s} = (17.77 \pm 0.12)\cdot 10^{12} s^{-1}$~\cite{PDG2008} and $y_{s}$ from 
       $\tau_{B_{s}}=(1.515 \pm 0.034) \cdot 10^{-12} s$~\cite{HFAGforPDG2009}, the measurement of
       $\Delta \Gamma_{s}$ in $B_s\to J/\psi\, \phi$ (see below)
       and the flavour-specific $B_{s}$
       lifetime $\tau_{B_{s}}^{FS}=(1.417 \pm 0.042) \cdot 10^{-12} s$~\cite{HFAGforPDG2009}. This results in
       $Z_{d}= 0.3741 \pm 0.0054$ and $Z_{s}= 1.0044^{+0.0058}_{-0.0032}$\, hence actually the error on $Z_{d,s}$
       has negligible impact on our global fits.
       
\item  CDF and \Dzero have presented time-dependent tagged analyses\cite{taggedphaseCDF,taggedphaseD0}
       of $B_s\to J/\psi\, \phi$ decays  which provide information on $\Delta_s$ through 
       $\Delta \Gamma_{s}$ and $\phi_s^{\psi\phi}$. 
       These analyses supersede the previous untagged studies of Refs.~\cite{dgexpnew, CDFdg} 
       (which use the same data sample, but without the tagging information)
       and we will not use the untagged results in our analysis. 
       These time-dependent tagged analyses have raised a lot of attention recently, 
       in particular when the UTfit collaboration claimed an evidence of New Physics 
       of at least $3 \sigma$ based on a global fit where these measurements played a 
       central role~\cite{Bona:2008jn}. It has been later argued, though, that this 
       conclusion came from an overinterpretation of the data~\cite{capri}~\footnote{In particular Ref.~\cite{Bona:2008jn} assumes
       that the only effect of the SU(3) assumption on the strong phases in the \Dzero analysis amounts to the suppression of 
       a mirror solution, without any impact on the accuracy and the location of the main solution.}. 
       In the framework of the HFAG~\cite{HFAG08} the CDF and \Dzero have determined 
       a combined constraint based on the measurements in Refs.~\cite{taggedphaseCDF,taggedphaseD0}. 
       In 2009 CDF has updated the analysis on a larger dataset using $2.8 fb^{-1}$ 
       of data~\cite{taggedphaseCDF_2}. The new average between \Dzero~\cite{taggedphaseD0} 
       and  CDF~\cite{taggedphaseCDF_2} has been presented in Summer 2009~\cite{Punzi}.
       Using this new average the deviation of the measured value $\phi_s^{\psi \phi}$  with 
       respect to the Standard Model value $\beta_s$ is essentially unchanged and reads $2.3$ standard deviations. 
       This average is our default input for the corresponding observables, supplemented by the 
       constraint on the flavour-specific $B_{s}$
       lifetime $\tau_{B_{s}}^{FS}=(1.417 \pm 0.042) \cdot 10^{-12}$~\cite{HFAGforPDG2009} which can be viewed as an independent measurement
       of $\Delta \Gamma_{s}$.
       New results for $B_s\to J/\psi\, \phi$ have been presented in Summer 2010 by
       CDF (with $5.2 fb^{-1}$)~\cite{taggedphaseCDF_3} and D\O\ (with $6.1 fb^{-1}$)~\cite{taggedphaseD0} 
       collaborations, in closer agreement to the Standard Model expectations, but these measurements    
       have not been combined together yet. They have not be included in the present analysis, 
       but are briefly discussed together with the Standard Model significance tests below.
\end{itemize}

\subsection{The neutral kaon system\label{Sec:InputsKaonWithNP}}

The master formula for $\epsilon_K$ has been given in \eq{cons}, from the relation 
between $\epsilon_K$ and $M_{12}^s$. The translation of $\epsilon_{K}$ into a constraint 
on $\rhobar$ and $\etabar$ suffers from sizeable uncertainties in the Wolfenstein 
parameter $A$ (the determination of which being driven by $|V_{cb}|^{4}$), ${\hat{\Bag}}_{K}$ 
(see Table~\ref{tab:TheoreticalInputs} and Section~\ref{ssec:MethodofAveraging}), 
from the long-distance corrections to the relation between $M_{12}^K$ and $\epsilon_K$ encoded in $\kappa_\epsilon$,  
and, though of less importance, from uncertainties in the QCD corrections coming from 
$\eta_{cc}$~\cite{Nierste1}, from the charm quark mass $\ov m_{c}(\ov m_{c})$ in 
the $\ov {\text{MS}}$ scheme, from $m_{t}$ and the perturbative QCD corrections 
$\eta_{tt}$~\cite{Nierste1} and $\eta_{ct}$~\cite{Nierste1}.

\begin{itemize}
\item From the experimental point of view, the number on the 
 LHS of Eq.~\eqref{cons} has shifted substantially over time. For instance in 1995 the 
 corresponding number was $1.21\cdot 10^{-7}$~\cite{Nierste2}. More recently, the 
 numerical value for $\epsilon_{K}$ has shifted by about $2.3~\%$ (a $3.7~\sigma$ effect) 
 between the 2004 and 2006 edition of the  Particle Data Group (PDG) from 
 $(2.284\pm0.014) \times 10^{-3}$~\cite{PDG2004} down to 
 $(2.232\pm0.007) \times 10^{-3}$~\cite{PDG2008}. 
 This shift has been mainly driven by improved measurements of the branching fraction 
 ${\mathcal{B}}(K_{L} \to \pi^{+}\pi^{-})$ performed by the KTeV, KLOE and NA48 collaborations 
 leading to a reduction of $5.5~\%$ of the semileptonic branching fraction values.
\item As discussed in sec.~\ref{ssec:kkmbasics}, 
 the relation between $\epsilon_K$ and \kkm\ is affected by several corrections
 encoded in $\kappa_\epsilon$. We have combined them with the experimental result for $\epsilon_K$ as indicated in this section.
\item
 The kaon decay constant $f_{K}$ is taken from the review on pseudoscalar decay 
 constants in Ref.~\cite{PDG2008} which is calculated from the measured branching 
 fraction ${\mathcal{B}}(K^{+} \rightarrow \mu^{+} \nu_{\mu}(\gamma))$ and the measured 
 charged kaon decay time using an external input for $|V_{us}|$. In Ref.~\cite{PDG2008}
 $|V_{us}|=0.2255 \pm 0.0019$ from $K_{\ell 3}$ decays is used as external input
 which leads to $f_{K}=155.5 \pm 0.2 \pm 0.8 \pm 0.2$ MeV where the first error is due
 to the experimental uncertainties, the second due to the uncertainty from $|V_{us}|$
 and the third due to higher order corrections~\cite{PDG2008}.
 With our input of $|V_{us}|=0.2246 \pm 0.0012$ this translates into
 $f_{K}=156.1 \pm 0.2 \pm 0.6 \pm 0.2$ MeV. In the fit, we do not consider the uncertainties
 on $f_{K}$ at this point since they are currently negligible with respect to the other 
 uncertainties. Since the $f_{K}$ value obtained in this way is anticorrelated with
 our $|V_{us}|$ input a consistent treatment would require including the leptonic kaon 
 decay in the fit and constrain $f_{K}$ simultaneously. This would then lead to an even
 smaller uncertainty on $f_{K}$ given the improved uncertainty on $|V_{us}|$ imposed by 
 the global fit. Such a fit is technically possible but has not been performed here since the 
 error from $f_{K}$ on $\epsilon_{K}$ does not play an important role. 
\item
 Various determinations of the charm quark mass are available. For instance,
 the charm quark mass in the kinetic mass scheme can be obtained from fits 
 to data from lepton energy and hadronic mass moments in $B \to X_{c}\ell\nu$ decays 
 combined with photon energy moments measured in $B \to X_{s}\gamma$ decays, see
 e.g. Refs.~\cite{BuchmuellerFlaecher,HFAG08}. 
 The most recent value for the kinetic mass quoted by HFAG is 
 $m_{c}^{kin}=(1.165\pm 0.050)~{\rm GeV}$~\cite{HFAG08}, corresponding to a
 value in the $\overline{\rm MS}$ scheme:
 $\ov m_{c}(\ov m_{c})=(1.265\pm0.060\pm0.050)~{\rm GeV}$.
 The first uncertainty on the charm quark mass is correlated with the b-quark 
 mass uncertainty obtained from the same fits quoted in Table~\ref{tab:ExperimentalInputs}
 with a linear correlation coefficient of order $98~\%$.
 A second uncertainty of $50$ MeV has been added following the discussion in 
 Ref.~\cite{BuchmuellerFlaecher}, to take into account 
 the low renormalisation scale and the size of higher-order perturbative corrections
 when translating the mass from one scheme to another.
 \\
 As an alternative, the charm quark mass can also be determined from $e^{+}e^{-}$
 annihilation data into hadrons created from quark-antiquark pairs. 
 The OPE-based method consists in writing sum rules for moments of the cross 
 section $\sigma(e^{+} e^{-} \rightarrow c \bar{c})$, which are dominated by 
 the perturbative term and the contribution proportional to the gluon condensate. 
 An older analysis of Steinhauser and K\"uhn based on a three-loop calculation
 finds $\ov m_{c}(\ov m_{c})=(1.304\pm0.027)~{\rm GeV}$ with a small uncertainty~\cite{SteinhauserKuehn}. 
 A similar analysis by Jamin and Hoang~\cite{JaminHoang} obtains a consistent 
 result, $\ov m_{c}(\ov m_{c})=(1.290\pm0.070)~{\rm GeV}$, but quotes a 
 significantly larger uncertainty. This can be ascribed to the choice of the 
 OPE scale separating short- and long-distance physics and it can be viewed 
 as the impact of (neglected) higher order terms in perturbation theory on 
 the determination of $m_c$ through such moments. 
 In a more recent calculation at four loops for the perturbative contribution, 
 the uncertainty has been further reduced:
 $\ov m_{c}(\ov m_{c})=(1.286\pm0.013)~{\rm GeV}$~\cite{Kuhn:2007vp}~\footnote{Using an 
 improved routine for the renormalisation group evolution of the computed 
 value $m_c(3 GeV)=(0.986 \pm 0.013)$ GeV the same group finds a slightly shifted central value 
 $\ov m_{c}(\ov m_{c})=1.268$ GeV~\cite{Steinhauser:2008pm}.}. However, in this reference, there 
 is only a limited discussion of the freedom in the choice of the renormalisation 
 scale for the perturbative series and the gluon condensate is varied only in a 
 limited range, even though these two effects were seen as bringing significant 
 systematics in Ref.~\cite{JaminHoang}.
 In the absence of further studies 
 on the systematics discussed above in the case of the four-loop analysis of 
 Ref.~\cite{Kuhn:2007vp}, we assign an additional theoretical uncertainty of 
 $0.040~{\rm GeV}$ and use as our input value: 
 $\ov m_{c}(\ov m_{c})=(1.286\pm0.013\pm0.040)~{\rm GeV}$ which is consistent 
 with the values from the moment fits but has smaller uncertainties.
\end{itemize}

\section{Quantitative Studies}\label{sec:quant} 

\subsection{Standard Model fit}\label{ssec:SMfit} 

In this section, we present the current status of the Standard Model CKM fit~\footnote{We stress an important difference
in our present definition of the Standard Model CKM fit, with respect to more `traditional' definitions, such as in Ref.~\cite{CKMfitter2}: we 
include here the full input list of Table~\ref{tab:ExperimentalInputs}, that is we also take into account $B_s\to J/\psi\phi$ and the semileptonic
asymetries. The latter observables have negligible impact on the determination of the CKM parameters, however, there is some visible effect on the fit predictions
for the subleading angles and asymetries in the $B_s$ system listed in Tables~\ref{tab:fitResults_SM1} and~\ref{tab:fitResults_SM2}.}.
Fig.~\ref{fig-rhoeta_smfit} shows the fit results in the $\bar\rho-\bar\eta$ 
\begin{nfigure}{Htb}
  \includegraphics[width=12cm]{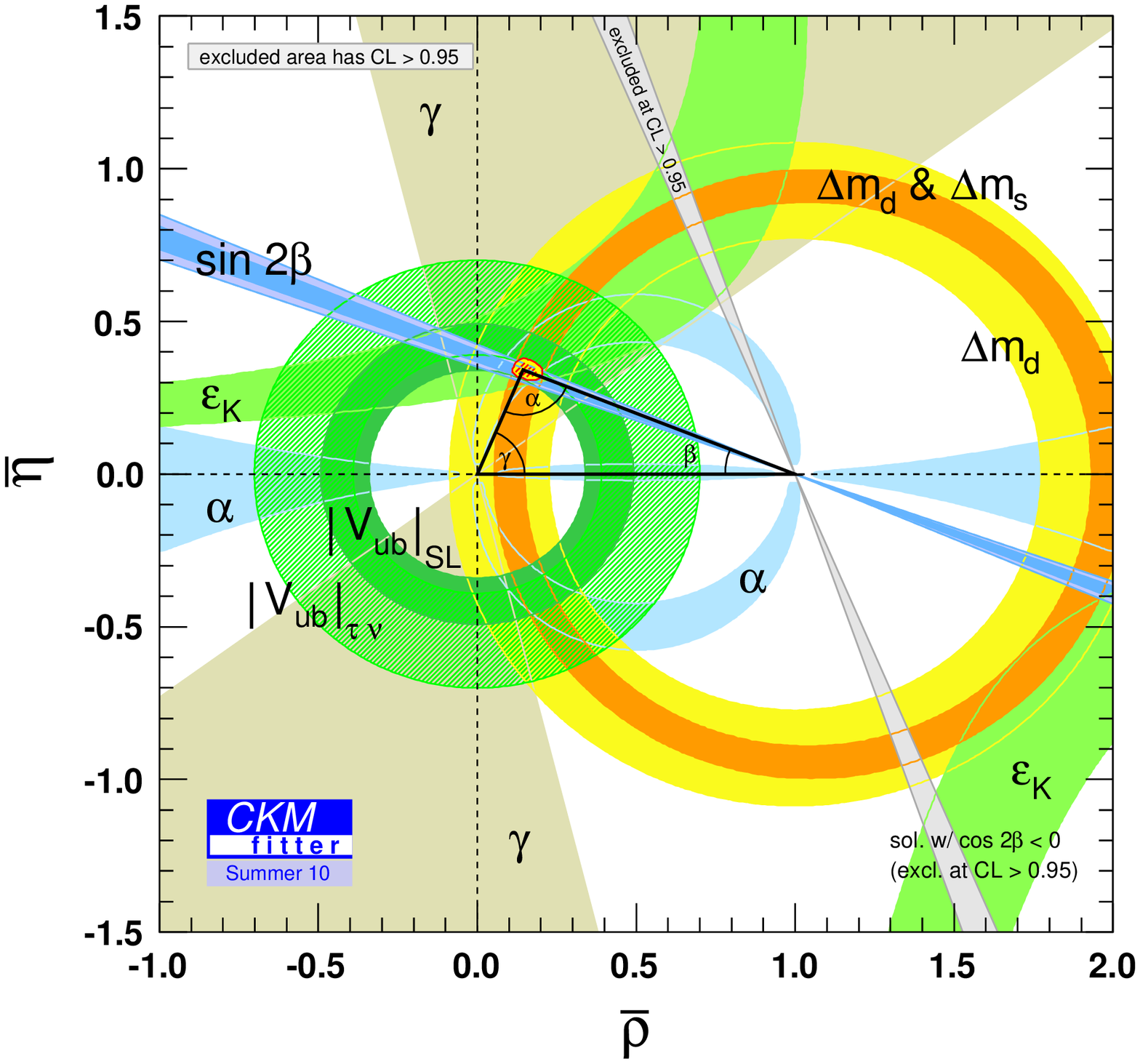}
  \caption{\small Constraint on the CKM $(\bar\rho,\bar\eta)$ coordinates from
                  the global Standard Model CKM-fit.
                  Regions outside the coloured areas have ${\rm CL} > 95.45~\%$. 
                  For the combined fit the yellow area inscribed by the contour 
                  line represents points with ${\rm CL} < 95.45~\%$. 
                  The shaded area inside this region represents points with ${\rm CL} < 68.3~\%$.
                  \label{fig-rhoeta_smfit}}
\end{nfigure}
plane. Table~\ref{tab:fitResults_SM1} shows the 
fit results for various parameters. 
\begin{table}[Htb]
\begin{center}\begin{tabular}{llll}  \hline
{ Quantity} & { central $\pm$ ${\rm CL}\equiv1\sigma$} & { $\pm$ ${\rm CL}\equiv2\sigma$  } & { $\pm$ ${\rm CL}\equiv3\sigma$ }  \\ 
 \hline  &&& \\[-0.3cm]
$A$ & $0.815^{\,+0.011}_{\,-0.029}$ & $0.815^{\,+0.020}_{\,-0.038}$ & $0.815^{\,+0.029}_{\,-0.046}$
\\[0.15cm]
$\lambda$ & $0.22543^{\,+0.00077}_{\,-0.00077}$ & $0.2254^{\,+0.0015}_{\,-0.0015}$ & $0.2254^{\,+0.0023}_{\,-0.0023}$
\\[0.15cm]
$\bar\rho$ & $0.144^{\,+0.029}_{\,-0.018}$ & $0.144^{\,+0.054}_{\,-0.028}$ & $0.144^{\,+0.068}_{\,-0.037}$
\\[0.15cm]
$\bar\eta$ & $0.342^{\,+0.016}_{\,-0.016}$ & $0.342^{\,+0.030}_{\,-0.028}$ & $0.342^{\,+0.045}_{\,-0.037}$
\\[0.15cm]
\hline &&& \\[-0.3cm]
$\widehat B_K$ (!) & $0.81^{\,+0.26}_{\,-0.12}$ & $0.81^{\,+0.34}_{\,-0.16}$ & $0.81^{\,+0.43}_{\,-0.19}$
\\[0.15cm]
$f_{B_s}$ [MeV] (!) & $234.9^{\,+9.3}_{\,-7.9}$ & $235^{\,+12}_{\,-12}$ & $235^{\,+16}_{\,-15}$
\\[0.15cm]
$\widehat\Bag_{B_{s}}$ (!)                   
                 & $ 1.115 ^{\,+ 0.090}_{\,-  0.036}$   &  1.115 $^{\,+  0.348}_{\,-  0.071}$   & 1.11 $^{\,+  0.54}_{\,-  0.10}$ \\[0.15cm]
$f_{B_s}/f_{B_d}$ (!) & $1.218^{\,+0.054}_{\,-0.046}$ & $1.22^{\,+0.11}_{\,-0.10}$ & $1.22^{\,+0.16}_{\,-0.15}$
\\[0.15cm]
$\Bag_{B_{s}}/\Bag_{B_{d}}$ (!)     
                 & $1.138 ^{\,+   0.076}_{\,- 0.094}$   & $ 1.14 ^{\,+  0.16}_{\,-  0.21}$     & $1.14 ^{\,+  0.24}_{\,-  0.30}$ \\[0.15cm]
$\widetilde{\Bag}_{S,B_s}(m_b)$ (!)     
                 & $-0.6 ^{\,+  1.4}_{\,- 2.3}$   & $-0.6 ^{\,+  2.8}_{\,-  3.9}$     & $-0.6 ^{\,+  4.3}_{\,-  5.5}$ \\[0.15cm]
\hline &&& \\[-0.3cm]
$J$~~$[10^{-5}]$ & $2.98^{\,+0.16}_{\,-0.18}$ & $2.98^{\,+0.30}_{\,-0.23}$ & $2.98^{\,+0.44}_{\,-0.28}$
\\[0.15cm]
\hline &&& \\[-0.3cm]
$\alpha\unit{deg}$ (!) & $97.6^{\,+2.4}_{\,-6.6}$ & $97.6^{\,+3.9}_{\,-10.7}$ & $97.6^{\,+5.5}_{\,-12.6}$
\\[0.15cm]
$\beta\unit{deg}$ (!) & $28.09^{\,+0.70}_{\,-1.49}$ & $28.1^{\,+1.4}_{\,-4.1}$ & $28.1^{\,+2.1}_{\,-7.2}$
\\[0.15cm]
$\gamma\unit{deg}$ (!) & $67.1^{\,+2.9}_{\,-4.5}$ & $67.1^{\,+4.4}_{\,-8.4}$ & $67.1^{\,+5.9}_{\,-10.4}$
\\[0.15cm]
$-2\beta_s$ [deg] (!) & $-2.083^{\,+0.097}_{\,-0.097}$ & $-2.08^{\,+0.18}_{\,-0.19}$ & $-2.08^{\,+0.27}_{\,-0.28}$
\\[0.15cm]
$-2\beta_s$ [deg] & $-2.084^{\,+0.097}_{\,-0.097}$ & $-2.08^{\,+0.17}_{\,-0.19}$ & $-2.08^{\,+0.23}_{\,-0.28}$
\\[0.15cm]
\hline\\
 \vspace{-0.5cm}
\end{tabular}\end{center}
 \caption[.]{\label{tab:fitResults_SM1}%\em
 Fit results of the Standard Model fit. The notation `(!)' means that the fit output represents the indirect constraint, \textit{i.e.}
              the corresponding direct input has been removed from the analysis.}
\end{table}
In Tables~\ref{tab:fitResults_SM1} and~\ref{tab:fitResults_SM2}
we also show the result of the fit for observables that have been individually 
excluded from the fit in order to quantify possible deviations between the 
individual input values and their fit predictions. The good overall agreement
of the combined Standard Model fit
mixes quantities that are in perfect agreement with their fit prediction, with others that
are individually at odds. Possible deviations
between a selection of measured observables and their Standard Model predictions are discussed
in more detail in the following.
\begin{table}[Htb]
\begin{center}\begin{tabular}{llll} \hline
{ Quantity} & { central $\pm$ ${\rm CL}\equiv1\sigma$} & { $\pm$ ${\rm CL}\equiv2\sigma$  } & { $\pm$ ${\rm CL}\equiv3\sigma$ }  \\
 \hline  &&& \\[-0.3cm]
$|V_{ud}|$ (!) & $0.97426^{\,+0.00030}_{\,-0.00030}$ & $0.97426^{\,+0.00060}_{\,-0.00061}$ & $0.97426^{\,+0.00089}_{\,-0.00091}$
\\[0.15cm]
$|V_{us}|$ (!) & $0.22545^{\,+0.00095}_{\,-0.00095}$ & $0.2254^{\,+0.0019}_{\,-0.0019}$ & $0.2254^{\,+0.0028}_{\,-0.0029}$
\\[0.15cm]
$|V_{ub}|$ (!) & $0.00356^{\,+0.00015}_{\,-0.00020}$ & $0.00356^{\,+0.00030}_{\,-0.00031}$ & $0.00356^{\,+0.00046}_{\,-0.00042}$
\\[0.15cm]
$|V_{cd}|$ & $0.22529^{\,+0.00077}_{\,-0.00077}$ & $0.2253^{\,+0.0015}_{\,-0.0015}$ & $0.2253^{\,+0.0023}_{\,-0.0023}$
\\[0.15cm]
$|V_{cs}|$ & $0.97341^{\,+0.00021}_{\,-0.00018}$ & $0.97341^{\,+0.00039}_{\,-0.00036}$ & $0.97341^{\,+0.00057}_{\,-0.00054}$
\\[0.15cm]
$|V_{cb}|$ (!) & $0.04508^{\,+0.00075}_{\,-0.00528}$ & $0.0451^{\,+0.0014}_{\,-0.0059}$ & $0.0451^{\,+0.0021}_{\,-0.0065}$
\\[0.15cm]
$|V_{td}|$ & $0.00861^{\,+0.00021}_{\,-0.00037}$ & $0.00861^{\,+0.00032}_{\,-0.00054}$ & $0.00861^{\,+0.00044}_{\,-0.00068}$
\\[0.15cm]
$|V_{ts}|$ & $0.04068^{\,+0.00043}_{\,-0.00138}$ & $0.04068^{\,+0.00081}_{\,-0.00169}$ & $0.0407^{\,+0.0012}_{\,-0.0020}$
\\[0.15cm]
$|V_{tb}|$ & $0.999135^{\,+0.000057}_{\,-0.000018}$ & $0.999135^{\,+0.000069}_{\,-0.000034}$ & $0.999135^{\,+0.000081}_{\,-0.000051}$
\\[0.15cm]
\hline &&& \\[-0.3cm]
$\phi_d$ [deg]  & $-5.8 ^{\,+ 2.0}_{\,-    3.6}$  & $-5.8 ^{\,+    2.7}_{\,-     4.6}$  & $-5.8 ^{\,+    2.9}_{\,-     5.7}$  \\[0.15cm]
$\phi_s$ [deg]  & $0.422 ^{\,+ 0.046}_{\,-    0.181}$  & $0.422 ^{\,+    0.098}_{\,-     0.248}$  & $0.42 ^{\,+    0.16}_{\,-     0.29}$  \\[0.15cm]
\hline &&& \\[-0.3cm]
$|\epsilon_K|$~~$[10^{-3}]$ (!) & $2.01^{\,+0.56}_{\,-0.65}$ & $2.01^{\,+0.70}_{\,-0.74}$ & $2.01^{\,+0.84}_{\,-0.83}$
\\[0.15cm]
$\Delta m_d \unit{ps^{-1}}$ (!) & $0.555^{\,+0.073}_{\,-0.046}$ & $0.55^{\,+0.11}_{\,-0.10}$ & $0.55^{\,+0.16}_{\,-0.15}$
\\[0.15cm]
$\Delta m_s \unit{ps^{-1}}$ (!) & $16.8^{\,+2.6}_{\,-1.5}$ & $16.8^{\,+4.1}_{\,-2.8}$ & $16.8^{\,+5.5}_{\,-3.4}$
\\[0.15cm]
$A_\text{SL}$~~$[10^{-4}]$ (!)            
                & $-3.67 ^{\,+1.09}_{\,-0.40}$  & $-3.67^{\,+ 1.52}_{\,-0.85}$   & $-3.7^{\,+ 1.7}_{\,- 1.3}$  \\[0.15cm]
$A_\text{SL}$~~$[10^{-4}]$           
                & $-3.68 ^{\,+1.03}_{\,-0.40}$  & $-3.68^{\,+ 1.52}_{\,-0.86}$   & $-3.7^{\,+ 1.7}_{\,- 1.3}$  \\[0.15cm]
$a_\text{SL}^{s}-a_\text{SL}^{d}$ $[10^{-4}]$             
                 & $7.93 ^{\,+ 0.66}_{\,- 2.14}$  & $7.9^{\,+ 1.3}_{\,- 3.0}$   & $7.9^{\,+ 2.0}_{\,- 3.4}$  \\[0.15cm]
$a_\text{SL}^{d}$~~$[10^{-4}]$ (!)            
                & $-7.58^{\,+ 2.11}_{\,-0.64}$ & $-7.6^{\,+2.9}_{\,- 1.3}$   & $-7.6^{\,+ 3.3}_{\,- 1.9}$  \\[0.15cm]
$a_\text{SL}^{d}$~~$[10^{-4}]$             
                & $-7.59^{\,+ 2.06}_{\,-0.64}$ & $-7.6^{\,+2.9}_{\,- 1.3}$   & $-7.6^{\,+ 3.3}_{\,- 1.9}$  \\[0.15cm]
$a_\text{SL}^{s}$~~$[10^{-4}]$ (!)            
                & $0.339^{\,+ 0.026}_{\,-0.090}$  & $0.339^{\,+0.052}_{\,- 0.130}$   & $0.339^{\,+ 0.079}_{\,- 0.147}$  \\[0.15cm]
$a_\text{SL}^{s}$~~$[10^{-4}]$             
                & $0.339^{\,+ 0.026}_{\,-0.090}$  & $0.339^{\,+0.052}_{\,- 0.130}$   & $0.339^{\,+ 0.079}_{\,- 0.147}$  \\[0.15cm]
$\Delta\Gamma_d \unit{ps^{-1}}$             
                & $0.00381^{\,+ 0.00072}_{\,-0.00141}$  & $0.0038^{\,+ 0.0011}_{\,- 0.0016}$   & $0.0038^{\,+ 0.0013}_{\,- 0.0018}$  \\[0.15cm]
$\Delta\Gamma_s \unit{ps^{-1}}$ (!)            
                & $0.104^{\,+ 0.060}_{\,-0.027}$  & $0.104^{\,+ 0.066}_{\,- 0.033}$   & $0.104^{\,+ 0.071}_{\,- 0.039}$  \\[0.15cm]
$\Delta\Gamma_s \unit{ps^{-1}}$            
                & $0.0818^{\,+ 0.0274}_{\,-0.0061}$  & $0.082^{\,+ 0.057}_{\,- 0.012}$   & $0.082^{\,+ 0.085}_{\,- 0.018}$  \\[0.15cm]
 \hline &&&      \\[-0.3cm]
$\BRB{\tau\nu}$~~$[10^{-4}]$ (!) & $0.764^{\,+0.087}_{\,-0.072}$ & $0.76^{\,+0.19}_{\,-0.11}$ & $0.76^{\,+0.29}_{\,-0.14}$
\\[0.15cm]
$\BRB{\tau\nu}$~~$[10^{-4}]$ & $0.833^{\,+0.109}_{\,-0.089}$ & $0.83^{\,+0.20}_{\,-0.15}$ & $0.83^{\,+0.28}_{\,-0.19}$
\\[0.15cm]
\hline
 \vspace{-0.5cm}
\end{tabular}\end{center}
\caption[.]{\label{tab:fitResults_SM2} %\em 
Fit results of the Standard Model fit. The notation `(!)' means that the fit output represents the indirect constraint, \textit{i.e.}
              the corresponding direct input has been removed from the analysis.}
\end{table}

One observes a sizeable discrepancy between the input value of
$\BRB{\tau\nu}$ (see Table~\ref{tab:ExperimentalInputs}) and its fit
prediction (see Table~\ref{tab:fitResults_SM2}) which is mainly driven
by the measured value of $\sin{2\beta}$, and was first discussed in
Ref.~\cite{CKMfitter08}. Removing either $\BRB{\tau\nu}$
or $\sin{2\beta}$ from the list of inputs results in a $\chi^{2}$ change
that corresponds to $2.9$ standard deviations.  This discrepancy could
arise either from a statistical fluctuation in the measured
$\BRB{\tau\nu}$ value, from too small (large) a value of $f_{B_{d}}$
($\HatBag_{B_{d}}$), or from New Physics in the $B \rightarrow \tau \nu$ and/or $\sin{2\beta}$
measurements. 
There is a specific correlation between $\sin{2\beta}$ and
$\BRB{\tau\nu}$ in the global fit that is a bit at odds with the direct
experimental determination.  This is best viewed in the
$(\sin{2\beta}$, $\BRB{\tau\nu})$ plane (see
Fig.~\ref{fig-sinbeta-Btotaunu}), regarding the prediction from the
global fit without using these measurements.
\begin{nfigure}{Htb}
  \includegraphics[width=10cm]{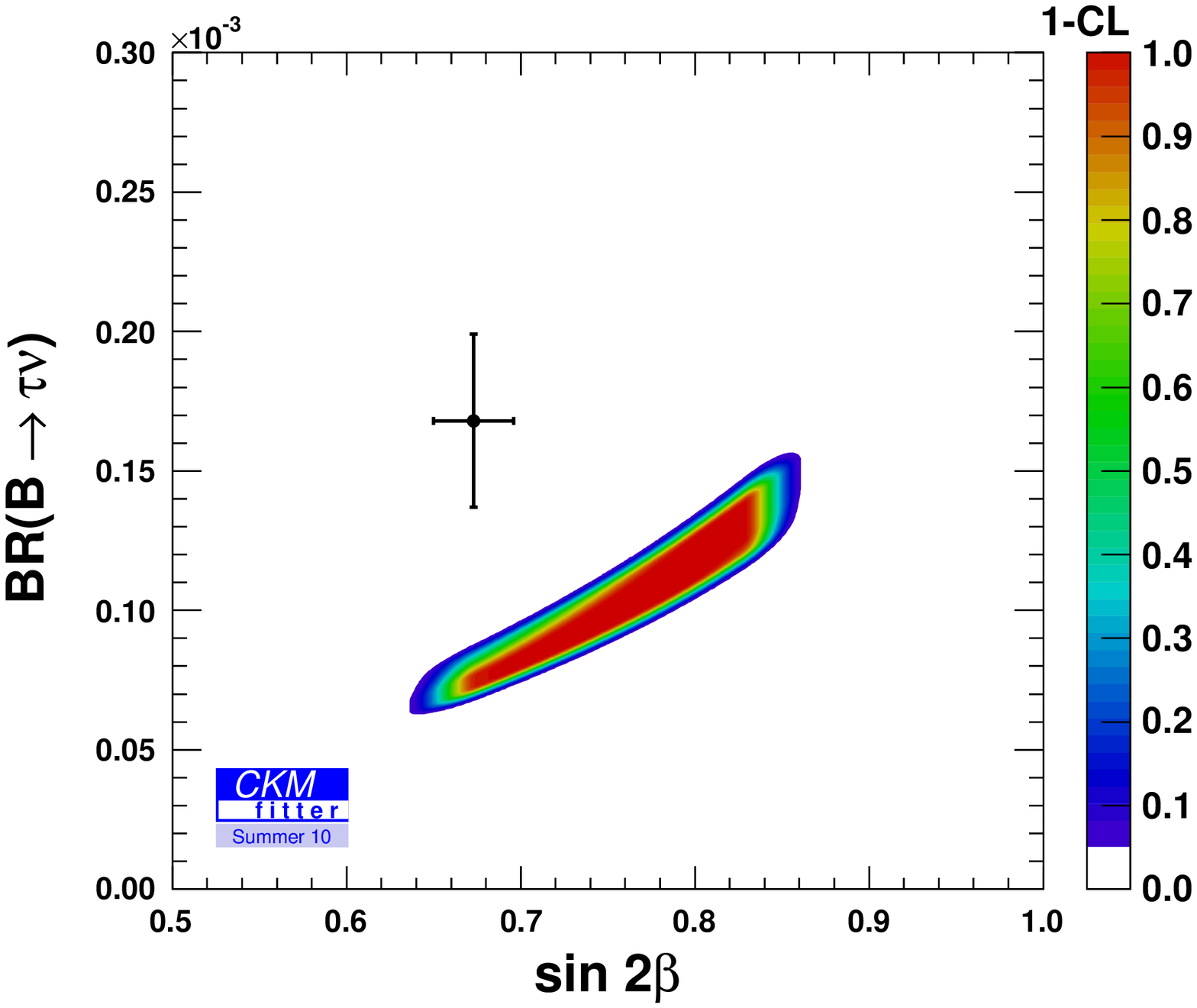}
  \caption{\small Constraint in the $(\sin{2\beta}$, $\BRB{\tau\nu})$
    plane. The coloured constraint represents the prediction for these
    quantities from the global fit when these inputs are removed
    while the cross represents the measurements with a  $1~\sigma$ uncertainty.\label{fig-sinbeta-Btotaunu}}
\end{nfigure}
The shape of the correlation can be understood more deeply by considering the ratio 
$\BRB{\tau\nu}/\Delta m_d$, where the decay constant $f_{B_{d}}$ cancels, leaving limited 
theoretical uncertainties restricted to the bag parameter $\HatBag_{B_{d}}$. 
The formula for the ratio displays explicitly 
that the correlation between $\BRB{\tau\nu}$ and the angle $\phi_d^{\psi K}$ is controlled by 
the values of $\HatBag_{B_{d}}$, and the angles $\gamma$ and  $\alpha=\pi-\beta-\gamma$:
\begin{eqnarray}
\frac{\BRB{\tau\nu}}{\Delta m_{d}} = \frac{3 \pi}{4} \frac{m_{\tau}^{2}}{m_{W}^{2} S(x_{t})} \left( 1 - \frac{m_{\tau}^{2}}{m_{B^{+}}^{2}} \right)^{2} \tau_{B^{+}} \frac{1}{\HatBag_{B_{d}} {\eta}_B} \frac{1}{{|V_{ud}|}^{2}} \left( \frac{\sin{\beta}}{\sin{\gamma}} \right)^{2}. 
 \label{BRBoverDeltamd}
\end{eqnarray}
The comparison of the indirect prediction of
$\HatBag_{B_{d}}$ from the above analytical formula (having only
$\BRB{\tau\nu}$, $\Delta m_{d}$, $\sin{2\beta}$, $\alpha$, $\gamma$ and
$|V_{ud}|$ as inputs, that is an almost completely theory-free
determination of $\HatBag_{B_{d}}$) with the current direct lattice
determination $\HatBag_{B_{d}} = 1.269^{+0.092}_{-0.090}$ is given in
Fig.~\ref{fig-BBd}. For this test the deviation is $2.9\sigma$,
dominated first by the experimental error on $\BRB{\tau\nu}$, $\alpha$, $\gamma$ and
second by the theoretical uncertainty on $\HatBag_{B_{d}}$. This tests clearly shows that
the semileptonic extraction of $|V_{ub}|$ has little to do with
the $\BRB{\tau\nu}$ anomaly.
Further insight is provided by Fig.~\ref{fig-BBd_fBd} where the constraints
on the decay constant $f_{B_{d}}$ and $f_{B_{d}} \sqrt{\HatBag_{B_{d}}}$ are
shown. We compare the fit inputs $f_{B_{d}}$ and $f_{B_{d}} \sqrt{\HatBag_{B_{d}}}$ 
taken from LQCD calculations with their predictions from the fit. The measured
$\BRB{\tau\nu}$ value leads to the constraint on $f_{B_{d}}$ represented by the 
green band. The orange band represents the constraint on $f_{B_{d}} \sqrt{\HatBag_{B_{d}}}$
thanks to the $\Delta m_{d}$ measurement. The combined prediction for both
quantities (red and yellow regions) reveals that the predicted value for
$f_{B_{d}} \sqrt{\HatBag_{B_{d}}}$ is in very good agreement with the LQCD 
input. Therefore, if the discrepancy is driven by too small a
$f_{B_{d}}$ value, the lattice-QCD artefact responsible for this underestimation
should not affect the more complicated determination
of the $\Delta B=2$ matrix element proportional to $f_{B_{d}} \sqrt{\HatBag_{B_{d}}}$,
as already demonstrated in Fig.~\ref{fig-BBd} in order 
to preserve the  good agreement between the predicted and calculated 
values for $f_{B_{d}} \sqrt{\HatBag_{B_{d}}}$.

\begin{nfigure}{H}
  \includegraphics[width=10cm]{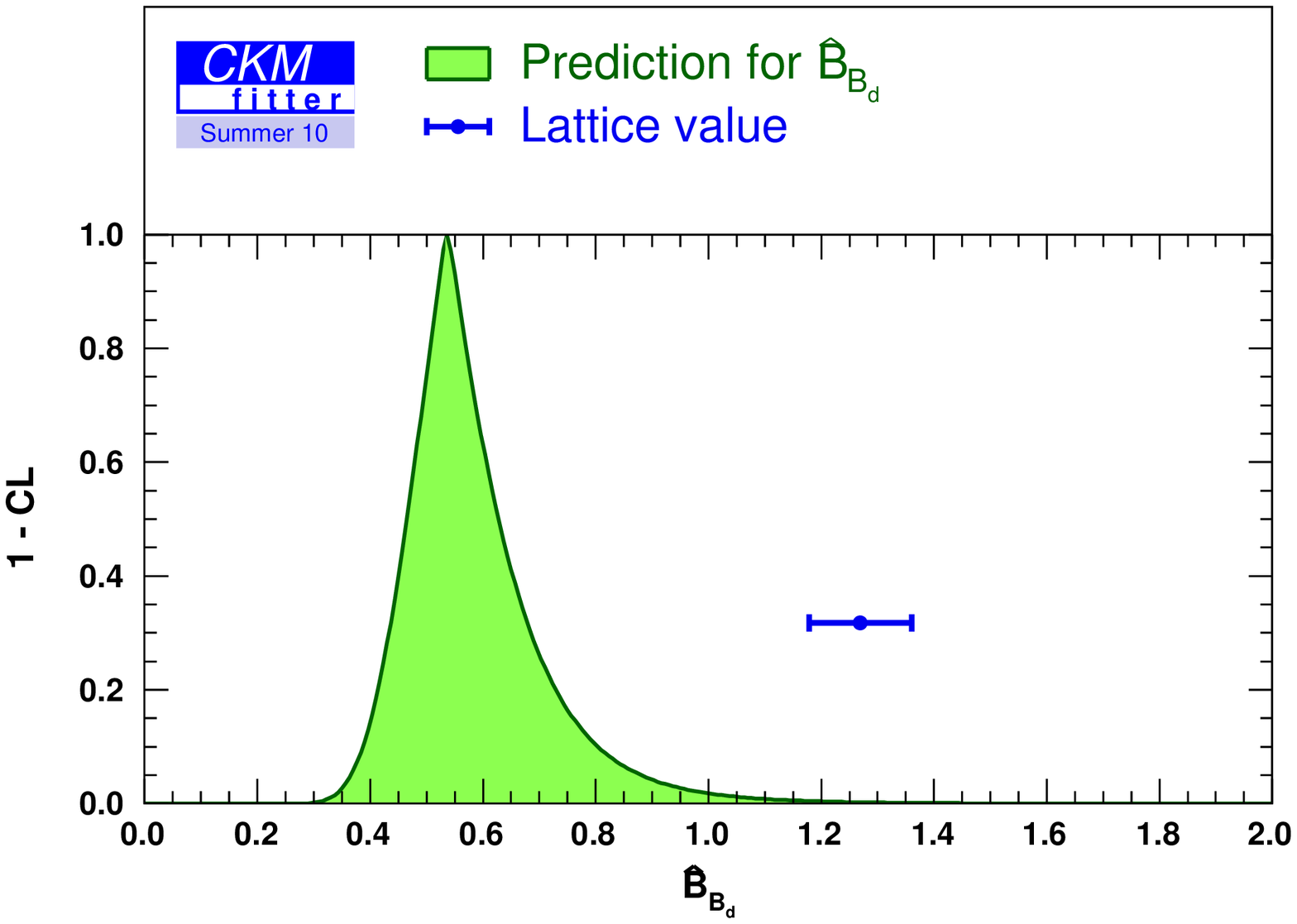}
  \caption{\small Comparison between the $\HatBag_{B_{d}}$ parameter used in the global fit
                  and its prediction.
                  The coloured constraint represents the prediction of $\HatBag_{B_{d}}$
                  quantities from the global fit while the blue point represents the 
                  input value with the corresponding
                  $1~\sigma$ uncertainty as used in the global fit.\label{fig-BBd}}
  \includegraphics[width=10cm]{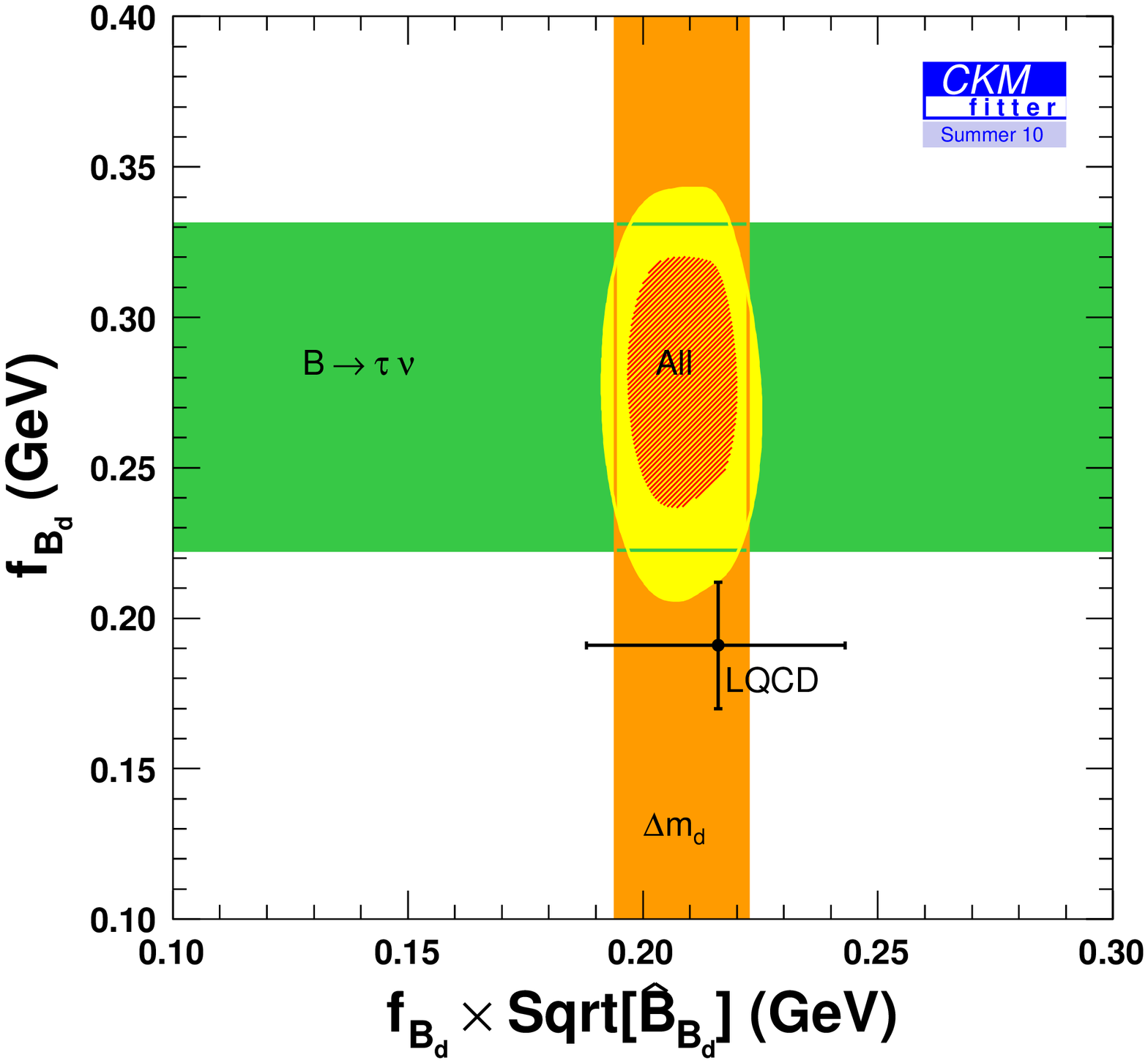}
  \caption{\small Constraint on $f_{B_{d}}$ and $f_{B_{d}} \sqrt{\HatBag_{B_{d}}}$. 
                  The green band shows the $95.45\%$ CL constraint on $f_{B_{d}}$ when 
                  $\BRB{\tau\nu}$ is included in the fit. The orange band represents 
                  the $95.45\%$ CL constraint on $f_{B_{d}} \sqrt{\HatBag_{B_{d}}}$
                  thanks to the $\Delta m_{d}$ measurement.
                  The combined $68.3\%$ CL and $95.45\%$ CL constraint is shown in red, 
                  respectively, in yellow. The black point shows the $1~\sigma$ 
                  uncertainty on $f_{B_{d}}$ and $f_{B_{d}} \sqrt{\HatBag_{B_{d}}}$ 
                  as used in the fit.\label{fig-BBd_fBd}}
\end{nfigure}

Another potential anomaly related to $|\epsilon_K|$ has been widely discussed 
in the literature~\cite{bg,smtensions}, but does not show up with our choice of 
inputs and statistical treatment. More details
can be found in the Appendix. Other interesting outcomes of the Standard Model global fit concern the prediction
of the recently measured CP-asymmetries  by the TeVatron experiments, namely in $B_s\to J/\psi\phi$ 
and in dimuonic inclusive decays (see Sect.~\ref{ssec:inputs}). The discrepancy of these measurements with respect to their 
Standard Model fit prediction, together with the $\BRB{\tau\nu}$ anomaly, are summarized in Table~\ref{pulls}.
\begin{table}[Htb]
\begin{center}\begin{tabular}{lcccc} \hline
Quantity & & Deviation & & \\ & wrt  SM fit & wrt Sc. I & wrt Sc. II & wrt Sc. III \\
 \hline  & \\[-0.3cm]
$\widehat B_K$ & $0.0~\sigma$ & - & $0.0~\sigma$ & - \\[0.15cm]
$f_{B_s}$ [MeV] & $0.0~\sigma$ & $0.9~\sigma$ & $0.8~\sigma$ & $1.2~\sigma$ \\[0.15cm]                  
$\widehat\Bag_{B_{s}}$ & $1.2~\sigma$ & $0.8~\sigma$ & $0.9~\sigma$ & $0.3~\sigma$ \\[0.15cm]
$f_{B_s}/f_{B_d}$ & $0.0~\sigma$ & $0.9~\sigma$ & $0.0~\sigma$ & $0.0~\sigma$ \\[0.15cm]
$\Bag_{B_{s}}/\Bag_{B_{d}}$ & $1.0~\sigma$ & $0.9~\sigma$ & $1.0~\sigma$ & $0.9~\sigma$ \\[0.15cm] 
$\widetilde{\Bag}_{S,B_s}(m_b)$ & $1.0~\sigma$ & $0.7~\sigma$ & $1.1~\sigma$ & $0.2~\sigma$ \\[0.15cm]
\hline &&& \\[-0.3cm]
$\alpha$ & $1.1~\sigma$ & $0.2~\sigma$ & $0.7~\sigma$ & $1.0~\sigma$ \\[0.15cm]
$\phi_d^\Delta+2\beta$  & $2.8~\sigma$ & $0.8~\sigma$ & $2.6~\sigma$ & $1.3~\sigma$ \\[0.15cm]
$\gamma$ & $0.0~\sigma$ & $0.0~\sigma$ & $0.0~\sigma$ & $0.0~\sigma$ \\[0.15cm]
$\phi_s^\Delta-2\beta_s$ & $2.3~\sigma$ & $0.5~\sigma$ & $2.4~\sigma$ & $1.6~\sigma$  \\[0.15cm]
\hline &&& \\[-0.3cm]
$|V_{ud}|$ & $0.0~\sigma$ & $0.0~\sigma$ & $0.0~\sigma$ & $0.1~\sigma$ \\[0.15cm]
$|V_{us}|$ & $0.0~\sigma$ & $0.0~\sigma$ & $0.0~\sigma$ & $0.0~\sigma$ \\[0.15cm]
$|V_{ub}|$ & $0.0~\sigma$ & $1.0~\sigma$ & $0.0~\sigma$ & $2.3~\sigma$ \\[0.15cm]
$|V_{cb}|$ & $0.0~\sigma$ & $0.0~\sigma$ & $1.6~\sigma$ & $1.8~\sigma$ \\[0.15cm]
\hline &&&& \\[-0.3cm]
$|\epsilon_K|$ & $0.0~\sigma$ & - & $0.0~\sigma$ & - \\[0.15cm]
$\Delta m_d$ & $1.0~\sigma$ & $0.9~\sigma$ & $1.0~\sigma$ & $0.8~\sigma$ \\[0.15cm]
$\Delta m_s$ & $0.3~\sigma$ & $0.7~\sigma$ & $0.9~\sigma$ & $1.2~\sigma$ \\[0.15cm]
$A_\text{SL}$ & $2.9~\sigma$ & $1.2~\sigma$ & $2.9~\sigma$ & $2.2~\sigma$ \\[0.15cm]
$a_\text{SL}^{d}$ & $0.9~\sigma$ & $0.2~\sigma$ & $0.8~\sigma$ & $0.3~\sigma$ \\[0.15cm]
$a_\text{SL}^{s}$ & $0.2~\sigma$ & $0.7~\sigma$ & $0.2~\sigma$ & $0.0~\sigma$ \\[0.15cm]
$\Delta\Gamma_s$ & $1.0~\sigma$ & $0.2~\sigma$ & $1.1~\sigma$ & $0.9~\sigma$ \\[0.15cm]
\hline &&&& \\[-0.3cm]
$\BRB{\tau\nu}$ & $2.9~\sigma$ & $0.7~\sigma$ & $2.6~\sigma$ & $1.0~\sigma$ \\[0.15cm]
\hline &&&& \\[-0.3cm]
$\BRB{\tau\nu}$ and $A_\text{SL}$ & $3.7~\sigma$ & $0.9~\sigma$ & $3.5~\sigma$ & $2.0~\sigma$ \\[0.15cm]
$\phi_s^\Delta-2\beta_s$ and $A_\text{SL}$ & $3.3~\sigma$ & $0.8~\sigma$ & $3.3~\sigma$ & $2.3~\sigma$ \\[0.15cm]
$\BRB{\tau\nu}$, $\phi_s^\Delta-2\beta_s$ and $A_\text{SL}$ & $4.0~\sigma$ & $0.6~\sigma$ & $3.8~\sigma$ & $2.1~\sigma$ \\[0.15cm]
\hline \end{tabular} \end{center}
\caption{Pull values for selected parameters and observables in the Standard Model (SM) and Scenarios I, II, III global fits, in terms of the number of 
equivalent standard deviations between the direct measurement (Tables~\ref{tab:ExperimentalInputs}, \ref{tab:TheoreticalInputs}) and the full indirect
fit predictions (Tables~\ref{tab:fitResults_SM1}, \ref{tab:fitResults_SM2}, \ref{tab:fitResults_NPBDBS}, \ref{tab:fitResults_NPBDBS2},
\ref{tab:fitResults_MFV_II}, \ref{tab:fitResults_MFV_II2}, \ref{tab:fitResults_MFV_III}, \ref{tab:fitResults_MFV_III2}). These numbers are computed from the $\chi^2$ difference with and without 
the input, interpreted with the appropriate number of degrees of freedom. The zero entries are due to the existence of
flat likelihoods in the Rfit model for theoretical uncertainties~\cite{CKMfitter1,CKMfitter2}.
The last three lines show the pulls for specific correlated
combinations of
the three most anomalous observables.}\label{pulls}
\end{table}
It is worth noting that the Standard Model does not correlate these anomalies between each other, because the standard prediction
for CP-violation ($-2\phi_s^{\psi\phi}$ and $A_\text{SL}$) in the $B_s$ system is essentially zero, and hence at leading 
order has no common parameter with the $\BRB{\tau\nu}$ anomaly.

\clearpage

\boldmath
\subsection[Scenario I]{Scenario I: New Physics in \bbmd\ and \bbms}\label{ssec:bdbs} 
\unboldmath

In this section, we present the CKM fit for scenario I where New Physics in mixing is 
independently allowed in the $B_d$ and $B_s$ systems (\textit{i.e.} $\Delta_d$ and $\Delta_s$ 
are independent). 
These fits exclude the constraint from $\epsilon_K$ since it is not possible to obtain 
non-trivial constraints for the three New Physics parameters in the $K$ sector. The first study 
of this kind using only $B$-factory data has been performed in~\cite{Laplace:2002ik} 
followed by a complete quantitative analysis~\cite{CKMfitter2} profiting from the large 
data sets of \babar and Belle. Analyses taking also into account the $B_s$ system have
been performed by the UTfit collaboration~\cite{UTfit2,UTfitDeltaF2}.

\begin{figure}[!Htb]
  \begin{center}
  \mbox{\epsfig{file=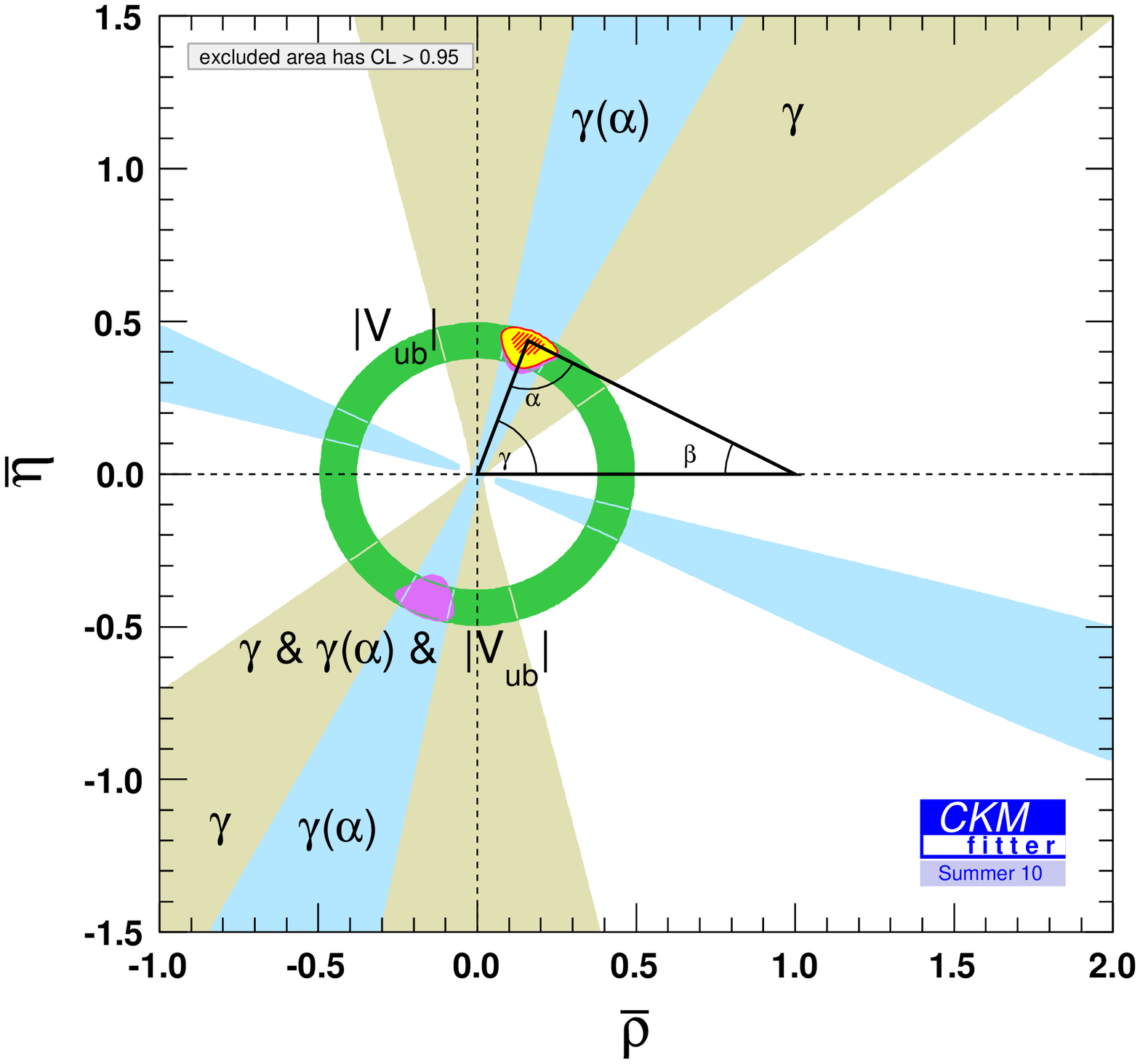,width=12cm}}
  \caption{\small 
Constraint on the CKM $(\bar\rho,\bar\eta)$ coordinates, using only the 
    inputs which are not affected by new physics in mixing: $|V_{ud}|$, $|V_{us}|$, 
    $|V_{cb}|$, $|V_{ub}|$ from semileptonic decays and from $B \rightarrow \tau \nu$, 
    $\gamma$ (directly and from the combination $\gamma(\alpha)$ of $\alpha-\phi_d^{\Delta}/2$ and $\beta+\phi_d^{\Delta}/2$).
    Regions outside the coloured areas have ${\rm CL} > 95.45~\%$. For the combined fit,
    two solutions are available: the usual solution corresponds to
    the yellow area (points with ${\rm CL} < 95.45~\%$, the 
    shaded region corresponding to points with ${\rm CL} < 68.3~\%$), and the second solution
    corresponds to the purple region.}\label{rhobar-etabar-NP}.
  \end{center}
\end{figure}

In Fig.~\ref{rhobar-etabar-NP} we show the $\bar\rho-\bar\eta$ plane for this fit, 
allowing us to constrain the parameters of the CKM matrix using observables not affected by New Physics
according to our hypothesis.
There are two allowed solutions in $\bar\rho-\bar\eta$ which cannot be distinguished 
when using only $|V_{ud}|$, $|V_{us}|$, $|V_{cb}|$, $|V_{ub}|$ from semileptonic decays 
and from $B \rightarrow \tau \nu$, $\gamma$ and $\alpha-\phi_d^\Delta/2$ and $\beta+\phi_d^\Delta /2$.
Once $a_\text{SL}^{d}$ is added, the second solution at negative $\bar\rho$ and 
$\bar\eta$ values is clearly disfavoured leaving as the only solution the one
with positive $\bar\rho$ and $\bar\eta$ values.
Figs.~\ref{fig-Deltad_scenario1} and~\ref{fig-Deltas_scenario1} show
\begin{nfigure}{!Htb}
  \includegraphics[width=12cm]{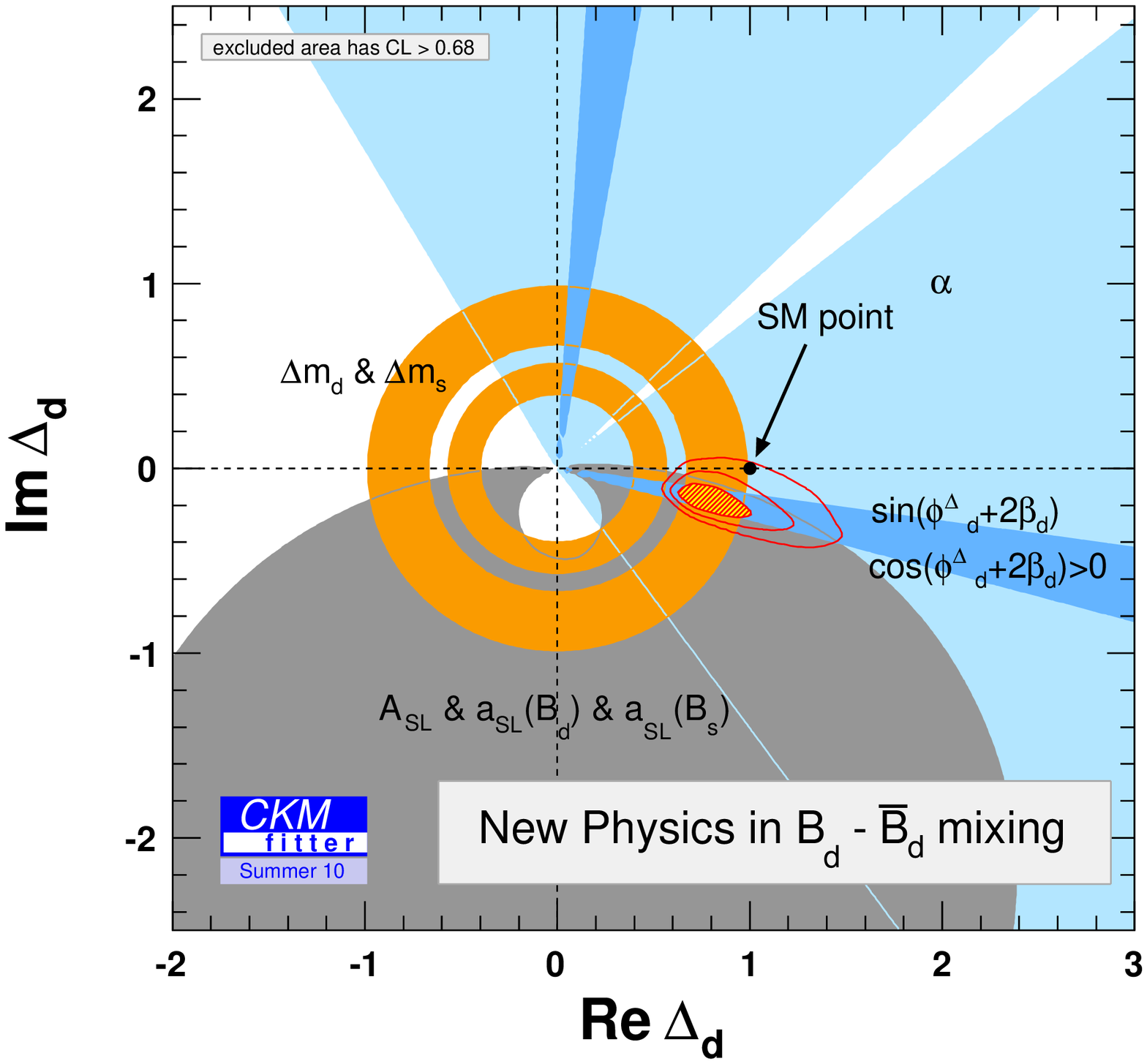}
  \caption{\small Constraint on the complex parameter $\Delta_d$ 
                  from the fit in scenario I.
                  For the individual constraints the coloured areas represent regions
                  with ${\rm CL} < 68.3~\%$. For the combined fit the red area shows
                  the region with ${\rm CL} < 68.3~\%$ while the two additional contour 
                  lines inscribe the regions with ${\rm CL} < 95.45~\%$, 
                  and  ${\rm CL} < 99.73~\%$, respectively.
                  \label{fig-Deltad_scenario1}}
\end{nfigure}
the results of the global fit for scenario I in the
complex $\Delta_d$ and $\Delta_s$ planes, respectively.
We emphasize that we assume that $\Delta_d$ and $\Delta_s$ are
taken as independent in this scenario, but that some of the
constraints correlate them (such that $a_{\rm fs}$ from the inclusive
dimuon asymmetry, and the ratio $\Delta m_d/\Delta m_s$). Therefore
the figures should be understood as two-dimensional projections of a
single multidimensional fit, and not as independent computations.
The constraint from $\Delta m_{d}$ in the
$\mbox{Re}{\Delta_d}-\mbox{Im}{\Delta_d}$ plane shows two allowed
ring-like regions. They correspond to the two allowed solutions in the
$\bar\rho-\bar\eta$ plane when $a_\text{SL}^{d}$ is excluded from the
list of inputs. Indeed, in this New Physics scenario, $\Delta m_{d}$ is
proportional to the product $|\Delta_d|^2 \cdot |V_{td}V_{tb}^*|^2$,
where the second factor is different for the two allowed solutions
since it is the side of the unitarity triangle connecting $(1,0)$ and
$(\ov{\rho},\ov{\eta})$.  The impact of $a_\text{SL}^{d}$ highlights
the power of this measurement to exclude a large region of the possible
New Physics parameter space even with a measurement precision of $O(5 \cdot
10^{-3})$. In the combined fit the inner ring (which corresponds to
the solution for $\phi_d^\Delta$ in the first quadrant near the 
$\imag \Delta_d$ axis) in the complex $\Delta_d$ plane is disfavoured. 
This  leaves us with an allowed region for $|\Delta_d|$ which is in 
agreement with the Standard Model value $\Delta_d=1$, albeit with possible deviations 
up to $40~\%$. The New Physics phase $\phi^\Delta_d$, mainly driven by 
the $\BRB{\tau\nu}$ vs. $\sin2\phi_d^{\psi K}$ correlation, has the best fit value
at $-12.9^{\circ}$. It can be as large as $-21.8^{\circ}$ ($-27.9^{\circ}$) 
at the $2~\sigma$ ($3~\sigma$) level and shows currently a deviation 
from the Standard Model of $2.7~\sigma$. 
It is interesting to note that the combined individual constraint from 
$a_\text{SL}^{d}$, $a_\text{SL}^{s}$ and $A_\text{SL}$ \textit{also} 
favours a negative New Physics phase $\phi_{\Delta_d}$, mainly due to 
the measured negative $a_\text{SL}^{d}$ value. When $\BRB{\tau\nu}$ 
is excluded from the inputs $\mbox{Im}{\Delta_d}$ and hence 
$\phi_{\Delta_d}$ is in good agreement with the Standard Model value 
(see Fig.~\ref{fig-Deltad_woBtaunu_scenario1}). At the same time the
allowed range for $|\Delta_d|$ is significantly enlarged since
$\BRB{\tau\nu}$ helps to reduce the uncertainty on $\Delta m_{d}$:
the two rings are enlarged and merge into a single one.

\begin{nfigure}{!Htb}
  \includegraphics[width=12cm]{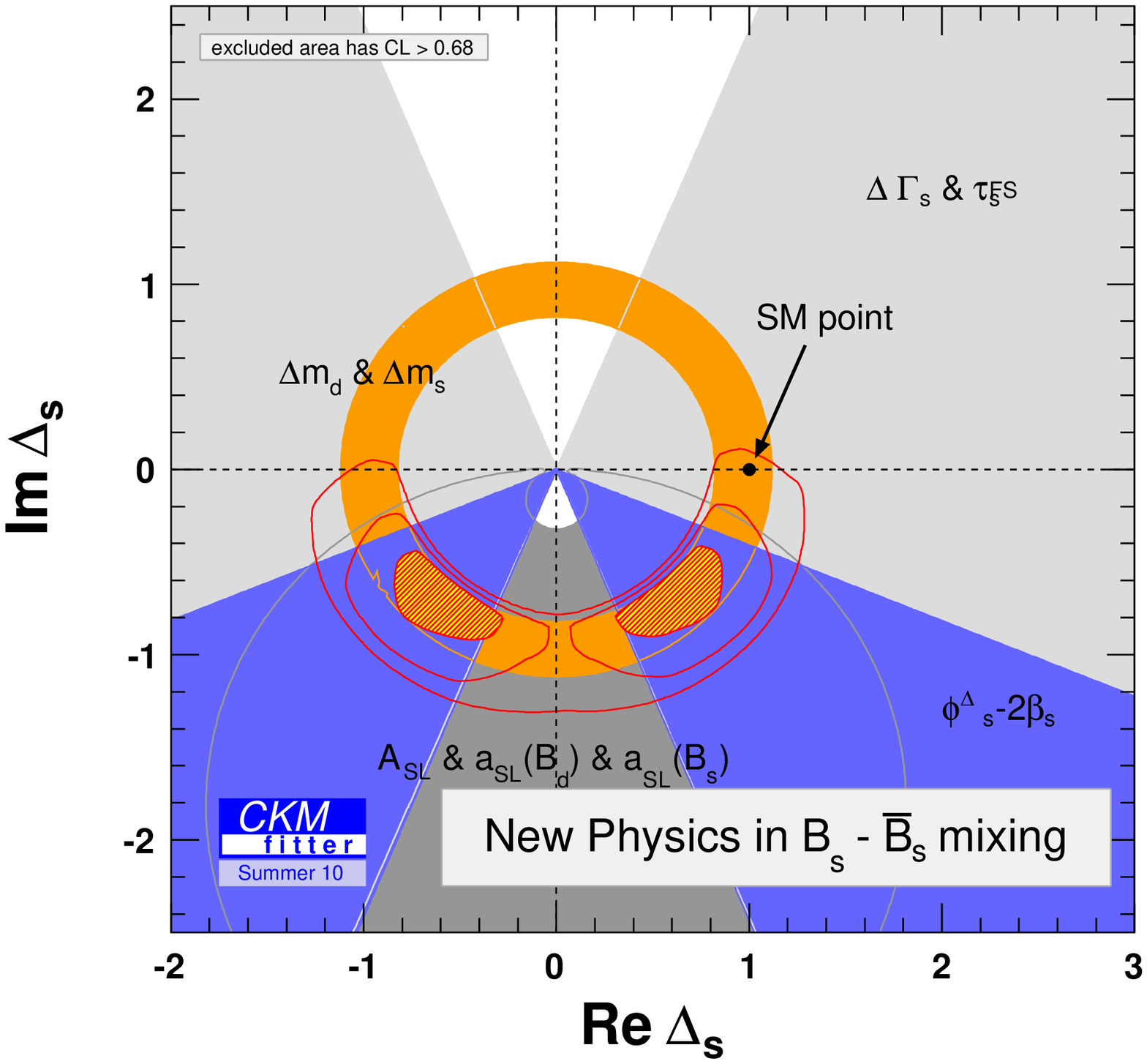}
  \caption{\small Constraint on the complex parameter $\Delta_s$ 
                  from the fit in scenario I.
                  For the individual constraints the coloured areas represent regions
                  with ${\rm CL} < 68.3~\%$. For the combined fit the red area shows
                  the region with ${\rm CL} < 68.3~\%$ while the two additional contour 
                  lines inscribe the regions with ${\rm CL} < 95.45~\%$, and 
                  ${\rm CL} < 99.73~\%$, respectively. 
                  \label{fig-Deltas_scenario1}}
\end{nfigure}
The constraint on $|\Delta_s|$ from $\Delta m_{s}$ is more stringent
than that for $|\Delta_d|$ - thanks to the smaller theoretical
uncertainty in its prediction compared to $\Delta m_{d}$ - and in good
agreement with the Standard Model point.  It is interesting to note that
also for the $B_{s}$ system the constraint from $\BRB{\tau\nu}$ plays a
non-negligible role: when removing this measurement from the list of
inputs the constraint on $|\Delta_s|$ becomes weaker since this
measurement improves the input on the decay constant $f_{B_{s}}$ through
the SU(3)-breaking parameter $\xi$ (compare
Fig.~\ref{fig-Deltas_woBtaunu_scenario1} with
Fig.~\ref{fig-Deltas_scenario1}).  There is evidence for a non-zero New Physics
phase $\phi^\Delta_s$ at the $3.1~\sigma$ level. This discrepancy is
driven by the $A_\text{SL}$ from D\O\ and by the $\phi_s^{\psi\phi}$
analyses from Tevatron, but is expected to be somewhat relaxed by the
updated measurements of $\phi_s^{\psi\phi}$~\cite{taggedphaseCDF_3,taggedphaseD0_2}.

We also note that there is an interesting difference in the allowed size of the 
New Physics contribution when comparing the $B_{d}$ and the $B_{s}$ systems. While in the 
$B_{s}$ system the size of the New Physics contribution is essentially constrained by the 
$\Delta m_{s}$ measurement alone this is not the case in the $B_{d}$ system. 
Indeed, 
%This is due to the fact that 
the theoretical prediction of $\Delta m_{d}$ strongly 
depends on the Wolfenstein parameters $\rhobar$ and $\etabar$ whereas this 
dependence is very weak for  $\Delta m_{s}$. The constraint on New Physics in $B_{d}$-mixing
relies thus on $|V_{ub}|$ on one hand and $\gamma$ on the other hand, the latter 
being currently dominated by the combination of the $\sin{2\phi_d^{\psi K}}$ and $\alpha$ 
measurements which is independent of New Physics contributions in $B$-mixing.
The theory prediction for the oscillation frequency $\Delta m_{d}$
depends on the quantity
$|\Delta_d|\equiv\sqrt{(\real \Delta_d)^{2}+(\imag\Delta_d)^{2}}$. Without
a good constraint on $|\Delta_d|$ from other observables it can
only be predicted with a very large uncertainty as observed in
Table~\ref{tab:fitResults_SM2}. The only other observables that are
sensitive to the modulus of $\Delta_d$
are $a_\text{SL}^{d}$ and 
$A_\text{SL}$ but those are measured with a precision that is significantly above 
the Standard Model prediction and thus do no constrain very much the range of $\Delta m_{d}$ 
(even though they proved powerful in eliminating the negative $(\ov\rho,\ov\eta)$ 
solution) (the same statement holds for $\Delta m_{s}$).

Tables~\ref{tab:fitResults_NPBDBS} and~\ref{tab:fitResults_NPBDBS2} show the fit results  for various parameters and observables. We
also show the result of the fit for quantities  that have been individually excluded from the fit in order to quantify possible 
deviations between the individual input values and their fit predictions. The corresponding pull values are listed in
Table~\ref{pulls}. Among other things it is interesting to note that the
indirect fit prediction for the dimuonic asymmetry $A_\text{SL}=(-42 ^{\,+ 20}_{\,- 19})\times 10^{-4}$ is consistent at 1.2 standard
deviations with the \Dzero/CDF average $(-85\pm 28)\times 10^{-4}$ used here, and  remains more precise in spite of the uncertainties on
the theoretical and New Physics parameters. Hence future improvements of this measurement are  expected to give crucial information on
the underlying physics.

Another important output of our global analysis is the prediction of the difference $a_\text{SL}^{s}-a_\text{SL}^{d}$, that will be measured
by the LHCb experiment in a close future~\cite{RavenLHCb}. It reads $a_\text{SL}^{s}-a_\text{SL}^{d}=(-39 ^{\,+ 31}_{\,- 24})\times 10^{-4}$ 
($-93\times 10^{-4}<a_\text{SL}^{s}-a_\text{SL}^{d}<36\times 10^{-4}$ at the $3~\sigma$ level), to be compared to the Standard Model result 
$a_\text{SL}^{s}-a_\text{SL}^{d}=(7.93 ^{\,+ 0.66}_{\,- 2.14})\times 10^{-4}$ ($4.5\times 10^{-4}<a_\text{SL}^{s}-a_\text{SL}^{d}<9.9\times 10^{-4}$ at $3~\sigma$).
\begin{table}[!Htb]
 \setlength{\tabcolsep}{0.8pc}
 \begin{center}\begin{tabular}{llll} \hline
 Quantity & central  ${\pm {\rm CL}}\equiv1\sigma$       &  ${\pm {\rm CL}}\equiv2\sigma$    &  ${\pm {\rm CL}}\equiv3\sigma$  \\
 \hline  &&& \\[-0.3cm]
$A$        & $    0.801 ^{\,+    0.024}_{\,-    0.017}$  & $0.801^{\,+    0.034}_{\,-    0.026}$ & $0.801^{\,+    0.043}_{\,-    0.036}$  \\[0.15cm]
$\lambda$  & $  0.22542 ^{\,+  0.00077}_{\,-  0.00077}$  & $0.2254^{\,+  0.0015}_{\,-  0.0015}$   & $0.2254^{\,+  0.0023}_{\,-  0.0023}$  \\[0.15cm]
$\bar\rho$ & $    0.159 ^{\,+    0.036}_{\,-    0.035}$  & $0.159^{\,+    0.070}_{\,-    0.067}$ & $0.16^{\,+    0.14}_{\,-    0.10}$  \\[0.15cm]
$\bar\eta$ & $    0.438 ^{\,+    0.019}_{\,-    0.029}$  & $0.438^{\,+    0.033}_{\,-    0.069}$ & $0.438^{\,+    0.047}_{\,-    0.113}$  \\[0.15cm]
 \hline &&      \\[-0.3cm]
$\mbox{Re}{(\Delta_d)}$  & $ 0.735 ^{\,+ 0.182}_{\,-    0.082}$  & $0.74^{\,+    0.38}_{\,-     0.13}$  & $0.74^{\,+    0.63}_{\,-    0.17}$  \\[0.15cm]
$\mbox{Im}{(\Delta_d)}$  & $-0.168 ^{\,+ 0.055}_{\,-    0.066}$  & $-0.17^{\,+    0.12}_{\,-     0.13}$  & $-0.17^{\,+    0.20}_{\,-     0.22}$ \\[0.15cm]
$|\Delta_d|$ & $0.747 ^{\,+ 0.195}_{\,-    0.079}$  & $0.75 ^{\,+    0.40}_{\,-     0.13}$  & $0.75 ^{\,+    0.66}_{\,-     0.17}$  \\[0.15cm]
$\phi^\Delta_d$ [deg]  & $-12.9 ^{\,+ 3.8}_{\,-    2.7}$  & $-12.9 ^{\,+    8.9}_{\,-     4.9}$  & $-12.9 ^{\,+    14.9}_{\,-     7.0}$  \\[0.15cm]
$\mbox{Re}{(\Delta_s)}$  & $ -0.57  ^{\,+ 0.18}_{\,-     0.17}$   & $-0.57^{\,+    0.39}_{\,-    0.39}$ & $-0.57^{\,+    1.80}_{\,-    -0.64}$  \\[0.15cm]
                       & \textit{or} $ 0.56  ^{\,+ 0.19}_{\,-     0.15}$   & \textit{or} $0.56^{\,+    0.42}_{\,-    0.36}$ &   \\[0.15cm]
$\mbox{Im}{(\Delta_s)}$  & $-0.69  ^{\,+ 0.16}_{\,-     0.14}$   & $-0.69^{\,+    0.39}_{\,-    0.34}$ & $-0.69^{\,+    0.66}_{\,-    0.56}$  \\[0.15cm]
$|\Delta_s|$ & $0.887 ^{\,+ 0.143}_{\,-    0.064}$  & $0.887 ^{\,+    0.338}_{\,-     0.093}$  & $0.89 ^{\,+    0.46}_{\,-     0.12}$  \\[0.15cm]
$\phi^\Delta_s$ [deg]  & $-130 ^{\,+ 13}_{\,-    12}$  & $-130 ^{\,+    28}_{\,-     28}$  &   \\[0.15cm]
                       & \textit{or} $ -51.6  ^{\,+ 14.2}_{\,-     9.7}$   & \textit{or} $-52^{\,+    32}_{\,-    25}$ &  $-52 ^{\,+    50}_{\,-    123}$ \\[0.15cm]
 \hline &&      \\[-0.3cm]
$f_{B_s}$ [MeV] (!)                   
                 & $ 278 ^{\,+   83}_{\,-      34}$     & $278^{\,+     125}_{\,-  58}$     & $278^{\,+    155}_{\,-     82}$ \\[0.15cm]
$f_{B_s}/f_{B_d}$ (!)     
                 & $1.09 ^{\,+    0.11}_{\,-   0.23}$   & $1.09^{\,+    0.27}_{\,- 0.41}$    & $1.09^{\,+    0.50}_{\,-    0.47}$ \\[0.15cm]
$\widehat\Bag_{B_{s}}$ (!)                   
                 & $ 2.21 ^{\,+   0.70}_{\,-   1.06}$   & $2.2^{\,+    1.7}_{\,-  1.6}$     & $2.2^{\,+    2.9}_{\,-     2.1}$ \\[0.15cm]
$\Bag_{B_{s}}/\Bag_{B_{d}}$ (!)     
                 & $0.48 ^{\,+    0.69}_{\,-   0.24}$   & $>0.12$     & $>0.01$ \\[0.15cm]
$\widetilde{\Bag}_{S,B_s}(m_b)$ (!)     
                 & $3.4 ^{\,+  2.2}_{\,- 3.0}$   & $ 3.4 ^{\,+  5.2}_{\,-  4.5}$     & $ 3.4 ^{\,+  8.7}_{\,-  7.7}$ \\[0.15cm]
 \hline &&      \\[-0.3cm]
$J$~~$[10^{-5}]$ 
           & $  3.69 ^{\,+ 0.19 }_{\,-    0.22}$         & $3.69^{\,+     0.31}_{\,-     0.55}$ & $3.69^{\,+     0.43}_{\,-     0.93}$  \\[0.15cm]
 \hline &&      \\[-0.3cm]
$\alpha\unit{deg}$  (!) 
           & $  79 ^{\,+ 22}_{\,-  15}$            & $79^{\,+     37}_{\,-     24}$ & $79^{\,+     50}_{\,-     31}$ \\[0.15cm]
$\beta\unit{deg}$    (!) 
           & $  27.2 ^{\,+ 1.1}_{\,-   3.1}$        & $27.2^{\,+    2.0}_{\,-    6.0}$ & $27.2^{\,+    2.8}_{\,-    9.1}$  \\[0.15cm]
$\gamma\unit{deg}$ (!) & $70.0^{\,+4.3}_{\,-4.5}$ & $70.0^{\,+8.5}_{\,-9.2}$ & $70^{\,+13}_{\,-20}$ \\[0.15cm]
\hline && \\[-0.3cm]
$\phi^\Delta_d+2\beta$ [deg]  (!)& $28 ^{\,+ 17}_{\,-    32}$  & $28 ^{\,+    38}_{\,-     55}$  & $28 ^{\,+    74}_{\,-     79}$ \\[0.15cm]
%                &  & \textit{or}  $-165 ^{\,+    86}_{\,-     15}$  & \textit{or} $-165 ^{\,+    105}_{\,-     15}$ \\[0.15cm]
$\phi_d$ [deg]  & $-17.9 ^{\,+ 4.9}_{\,-    5.8}$  & $-17.9 ^{\,+    10.2}_{\,-     9.5}$  & $-18 ^{\,+    17}_{\,-     12}$ \\[0.15cm]
$\phi^\Delta_s-2\beta_s$ [deg]  (!)& $-127 ^{\,+ 13}_{\,-    17}$  & $<-94$  & $<40$ \\[0.15cm]
                & \textit{or} $-58 ^{\,+ 17}_{\,-    13}$  & \textit{or} $-58 ^{\,+    52}_{\,-     34}$  &  $>135$ \\[0.15cm]
$\phi^\Delta_s-2\beta_s$ [deg]  & $-132 ^{\,+ 13}_{\,-    12}$  & $-132 ^{\,+    28}_{\,-     28}$  &  \\[0.15cm]
                & \textit{or} $-54.6 ^{\,+ 14.5}_{\,-    9.3}$  & \textit{or} $-55 ^{\,+    32}_{\,-     25}$  & $-55 ^{\,+    51}_{\,-     122}$ \\[0.15cm]
$\phi_s$ [deg]  & $-129 ^{\,+ 13}_{\,-    11}$  & $-129 ^{\,+    28}_{\,-     27}$  &   \\[0.15cm]
                & \textit{or} $-51 ^{\,+ 14}_{\,-   10}$  & \textit{or} $-51 ^{\,+    32}_{\,-     25}$  & $-51 ^{\,+    50}_{\,-    123}$ \\[0.15cm]
 \hline
 \end{tabular}
 \end{center}
 \vspace{-0.5cm}
 \caption[.]{\label{tab:fitResults_NPBDBS}%\em
              Fit results in the New Physics scenario I. The notation `(!)' means that the fit output represents the indirect constraint, \textit{i.e.}
              the corresponding direct input has been removed from the analysis.}
\end{table}
\begin{table}[Htb]
 \setlength{\tabcolsep}{0.8pc}
\begin{center}\begin{tabular}{llll} \hline
 Quantity & central  ${\pm \mathrm{CL}}\equiv1\sigma$       &  ${\pm \mathrm{CL}}\equiv2\sigma$    &  ${\pm \mathrm{CL}}\equiv3\sigma$  \\[0.15cm]
 \hline  && \\[-0.3cm]
$|V_{ud}|$(!) 
           & $  0.97426 ^{\,+  0.00030}_{\,-  0.00030}$  & $0.97426^{\,+  0.00060}_{\,- 0.00061}$  & $0.97426^{\,+  0.00089}_{\,- 0.00091}$ \\[0.15cm]
$|V_{us}|$(!) 
           & $  0.22542 ^{\,+   0.00095}_{\,-  0.00095}$    & $0.2254^{\,+  0.0019}_{\,-  0.0019}$   & $0.2254^{\,+  0.0028}_{\,-  0.0029}$ \\[0.15cm]
$|V_{ub}|$(!) 
           & $  0.00501 ^{\,+0.00104}_{\,-  0.00064}$  & $0.0050^{\,+0.0015}_{\,- 0.0011}$  & $0.0050^{\,+0.0020}_{\,- 0.0016}$ \\[0.15cm]
$|V_{cd}|$ & $  0.22529 ^{\,+  0.00077}_{\,-  0.00077}$  & $0.2253^{\,+  0.0015}_{\,-  0.0015}$   & $0.2253^{\,+  0.0023}_{\,-  0.0023}$  \\[0.15cm]
$|V_{cs}|$ & $  0.97344 ^{\,+  0.00019}_{\,-  0.00021}$  & $0.97344^{\,+  0.00037}_{\,- 0.00039}$  & $0.97344^{\,+  0.00055}_{\,- 0.00057}$  \\[0.15cm]
$|V_{cb}|$(!) 
            & $  0.0407 ^{\,+  0.0121}_{\,-  0.0075}$  & $0.041^{\,+  0.032}_{\,- 0.018}$  & $0.041^{\,+  0.047}_{\,-  0.032}$ \\[0.15cm]
$|V_{td}|$ & $  0.00871 ^{\,+  0.00047}_{\,-  0.00042}$  & $0.00871^{\,+  0.00079}_{\,- 0.00078}$  & $0.0087^{\,+  0.0011}_{\,- 0.0015}$  \\[0.15cm]
$|V_{ts}|$ & $  0.04001 ^{\,+  0.00112}_{\,-  0.00078}$  & $0.0400^{\,+  0.0015}_{\,- 0.0012}$  & $0.0400^{\,+  0.0019}_{\,-  0.0015}$  \\[0.15cm]
$|V_{tb}|$ & $  0.999161 ^{\,+ 0.000032}_{\,- 0.000047}$ & $0.999161^{\,+ 0.000047}_{\,- 0.000063}$ & $0.999161^{\,+ 0.000062}_{\,- 0.000079}$  \\[0.15cm]
 \hline &&      \\[-0.3cm]
$\Delta m_d$ $[{\rm ps}^{-1}]$ (!)            
                & $ 0.25 ^{\,+     0.34}_{\,- 0.16}$    &  $>0.04$                              &  \\[0.15cm]
$\Delta m_s$ $[{\rm ps}^{-1}]$ (!)            
                & $ 8.6 ^{\,+14.7}_{\,- 5.2}$    &  $>0.6$          &  \\[0.15cm]
$A_\text{SL}$ $[10^{-4}]$ (!)            
                & $-42 ^{\,+ 20}_{\,- 19}$ & $-42^{\,+ 29}_{\,- 27}$   & $-42^{\,+ 40}_{\,- 33}$ \\[0.15cm]
$A_\text{SL}$ $[10^{-4}]$             
                & $-55.7 ^{\,+ 14.9}_{\,- 8.7}$ & $-56^{\,+ 31}_{\,- 16}$   & $-56^{\,+ 42}_{\,- 21}$ \\[0.15cm]
$a_\text{SL}^{s}-a_\text{SL}^{d}$ $[10^{-4}]$             
                 & $-39 ^{\,+ 31}_{\,- 24}$  & $-39^{\,+ 58}_{\,- 39}$   & $-39^{\,+ 75}_{\,- 54}$ \\[0.15cm]
$a_\text{SL}^{d}$ $[10^{-4}]$ (!)            
                & $-36.2^{\,+ 13.9}_{\,-5.9}$ & $-36^{\,+ 23}_{\,- 11}$ & $-36^{\,+ 35}_{\,- 16}$ \\[0.15cm]
$a_\text{SL}^{d}$ $[10^{-4}]$             
                & $-36.7^{\,+ 13.4}_{\,-5.5}$ & $-37^{\,+ 23}_{\,- 11}$ & $-37^{\,+ 34}_{\,- 16}$ \\[0.15cm]
$a_\text{SL}^{s}$ $[10^{-4}]$ (!)            
                 & $-84.9 ^{\,+ 33.4}_{\,- 9.8}$  & $-85^{\,+ 64}_{\,- 20}$   & $-85^{\,+ 83}_{\,- 27}$ \\[0.15cm]
$a_\text{SL}^{s}$ $[10^{-4}]$            
                 & $-75 ^{\,+ 30}_{\,- 18}$  & $-75^{\,+ 56}_{\,- 29}$   & $-75^{\,+ 74}_{\,- 36}$ \\[0.15cm]
$\Delta\Gamma_d$ $[{\rm ps}^{-1}]$ (!)            
                & $0.00577^{\,+ 0.00095}_{\,-0.00257}$  & $0.0058^{\,+ 0.0015}_{\,- 0.0035}$   & $0.0058^{\,+ 0.0022}_{\,- 0.0040}$ \\[0.15cm]
$\Delta\Gamma_s$ $[{\rm ps}^{-1}]$ (!)            
                & $-0.118^{\,+ 0.068}_{\,-0.034}$  &    &   \\[0.15cm]
                & \textit{or} $0.128^{\,+ 0.029}_{\,-0.062}$ & $0.128^{\,+ 0.053}_{\,- 0.305}$   & $0.128^{\,+ 0.068}_{\,- 0.324}$  \\[0.15cm]
$\Delta\Gamma_s$ $[{\rm ps}^{-1}]$             
                & $-0.109^{\,+ 0.029}_{\,-0.025}$  & $-0.109^{\,+ 0.074}_{\,-0.049}$   &  \\[0.15cm]
                & \textit{or} $0.106^{\,+ 0.035}_{\,-0.020}$ & $0.106^{\,+ 0.057}_{\,- 0.065}$   & $0.106^{\,+ 0.076}_{\,- 0.284}$  \\[0.15cm]
 \hline &&      \\[-0.3cm]
$\mathcal{B}(B\to\tau\nu)$ $[10^{-4}]$ (!) & $ 1.457 ^{\,+ 0.073}_{\,- 0.405}$ & $ 1.46 ^{\,+ 0.15}_{\,- 0.84}$ & $ 1.46 ^{\,+ 0.23}_{\,- 0.89}$ \\[0.15cm]
$\mathcal{B}(B\to\tau\nu)$ $[10^{-4}]$  & $ 1.468 ^{\,+ 0.072}_{\,- 0.143}$ & $ 1.47 ^{\,+ 0.15}_{\,- 0.38}$ & $ 1.47 ^{\,+ 0.22}_{\,- 0.64}$ \\[0.15cm]
 \hline
 \end{tabular}
 \end{center}
 \vspace{-0.5cm}
 \caption[.]{\label{tab:fitResults_NPBDBS2}%\em
              Fit results in the New Physics scenario I. The notation `(!)' means that the fit output represents the indirect constraint, \textit{i.e.}
              the corresponding direct input has been removed from the analysis. }
\end{table}

\begin{table}[!Htb]
\begin{center}\begin{tabular}{ll} \hline
Hypothesis & p-value \\ 
 \hline  & \\[-0.3cm]
$\mathrm{Im}(\Delta_d)=0$ (1D) & $2.7~\sigma$ \\[0.15cm]
$\mathrm{Im}(\Delta_s)=0$ (1D) & $3.1~\sigma$ \\[0.15cm]
 \hline  & \\[-0.3cm]
$\Delta_d=1$ (2D) & $2.7~\sigma$ \\[0.15cm]
$\Delta_s=1$ (2D) & $2.7~\sigma$ \\[0.15cm]
$\mathrm{Im}(\Delta_d)=\mathrm{Im}(\Delta_s)=0$ (2D) & $3.8~\sigma$ \\[0.15cm]
 \hline  & \\[-0.3cm]
$\Delta_d=\Delta_s$ (2D) & $2.1~\sigma$ \\[0.15cm]
 \hline  & \\[-0.3cm]
$\Delta_d=\Delta_s=1$ (4D) & $3.6~\sigma$ \\[0.15cm]
\hline \end{tabular} \end{center}
\caption{p-values for various Standard Model hypotheses in the framework of New Physics scenario I, in terms of the number of 
equivalent standard deviations. These numbers are computed from the $\chi^2$ difference with and without 
the hypothesis constraint, interpreted with the appropriate number of degrees of freedom.}\label{ScIpvalues}
\end{table}
In contrast to the Standard Model fit, our Scenario I relates the $B_d$ and $B_s$ anomalies through the correlated determination of the $\Delta$
parameters. Hence it is particularly interesting to compute the p-values associated  with the hypothesis that some specific combination
of the $\Delta$ parameters take their Standard Model value. This is shown in Table~\ref{ScIpvalues}. We have listed several hypotheses 
because the `Standard Model' null hypothesis is \textit{composite}, \textit{i.e.} it does not allow to compute the 
expected distribution of  measurements in a
given set of experiments, because the Standard Model does not predict the value of  its fundamental parameters. Hence one is \textit{a priori} free
to choose among the numerical hypotheses tested in Table~\ref{ScIpvalues} the one that models the Standard Model hypothesis.
This choice introduces some
arbitrariness, and thus slightly different answers to the same underlying question.
In the present context, one may view the hypothesis
$\Delta_d=\Delta_s=1$ as the most natural choice to represent the Standard Model: this hypothesis is excluded
at $3.6\sigma$ by our global analysis.

However, one should be aware that this hypothesis somewhat dilutes the
most anomalous pieces of information that are related to CP-violation,
by including in the test CP-conserving components (corresponding to the
real part of $\Delta_d$ and $\Delta_s$). Let us imagine that we consider
a more general class of models allowing for CP-violation in several
different processes that are well compatible with the Standard Model. We would have
introduced a different $\Delta$ parameter for each process. The test
corresponding to all $\Delta$ parameters being equal to 1 would then
receive a small contribution from the three anomalies that we have
discussed, but this would be hidden by the other measurements in
agreement with the Standard Model expectations. This illustates why it is sometimes
worth testing reduced hypotheses, such as
$\mathrm{Im}(\Delta_d)=\mathrm{Im}(\Delta_s)=0$, in order to single out
specific deviations from the Standard Model. The latter CP hypothesis is excluded at
$3.8\sigma$ by our global analysis.

We also learn from this table that scenario III to be discussed
below, is slightly disfavoured by the data when one considers it as a
subcase of scenario I ($\Delta_d=\Delta_s$), in agreement with the third
paper of Ref.~\cite{paperafterD0}. Finally as already stressed above,
the various evidences against the Standard Model shown in Table~\ref{ScIpvalues}
will be relaxed when the new Tevatron average of the $B_s\to J/\psi\phi$
tagged analysis is
available~\cite{taggedphaseCDF_3,taggedphaseD0_2}. However, a very rough
estimate allows us to predict that at least the
$\mathrm{Im}(\Delta_d)=\mathrm{Im}(\Delta_s)=0$ hypothesis
(\textit{i.e.} no CP-violating phase in either $B_d$ or $B_s$ mixing
amplitudes) will remain disfavoured by more than three standard
deviations. Indeed although the mixing CP-phase is expected to come back
closer to the Standard Model value~\cite{taggedphaseCDF_3,taggedphaseD0_2}, it
remains well compatible with the indirect constraint from the dimuonic
asymmetry, as shown by Fig.~\ref{fig-phis_comparison}. In particular,
the best value for $\phi_s^\Delta-2\beta_s$ is only about one standard
deviation below the most recent CDF and \Dzero updates for $B_s\to
J/\psi \phi$~\cite{taggedphaseCDF_3,taggedphaseD0_2}, not included in
the present analysis. Again it would be interesting to know the precise
form of the combined CDF and \Dzero likelihoods in order to quantify by
how much the difference of the indirect global fit prediction of
$A_\text{SL}$ with its direct measurement is increased with respect to
our estimate of $1.2~\sigma$ in Table~\ref{pulls}.
\begin{nfigure}{htb}
 \includegraphics[width=10cm]{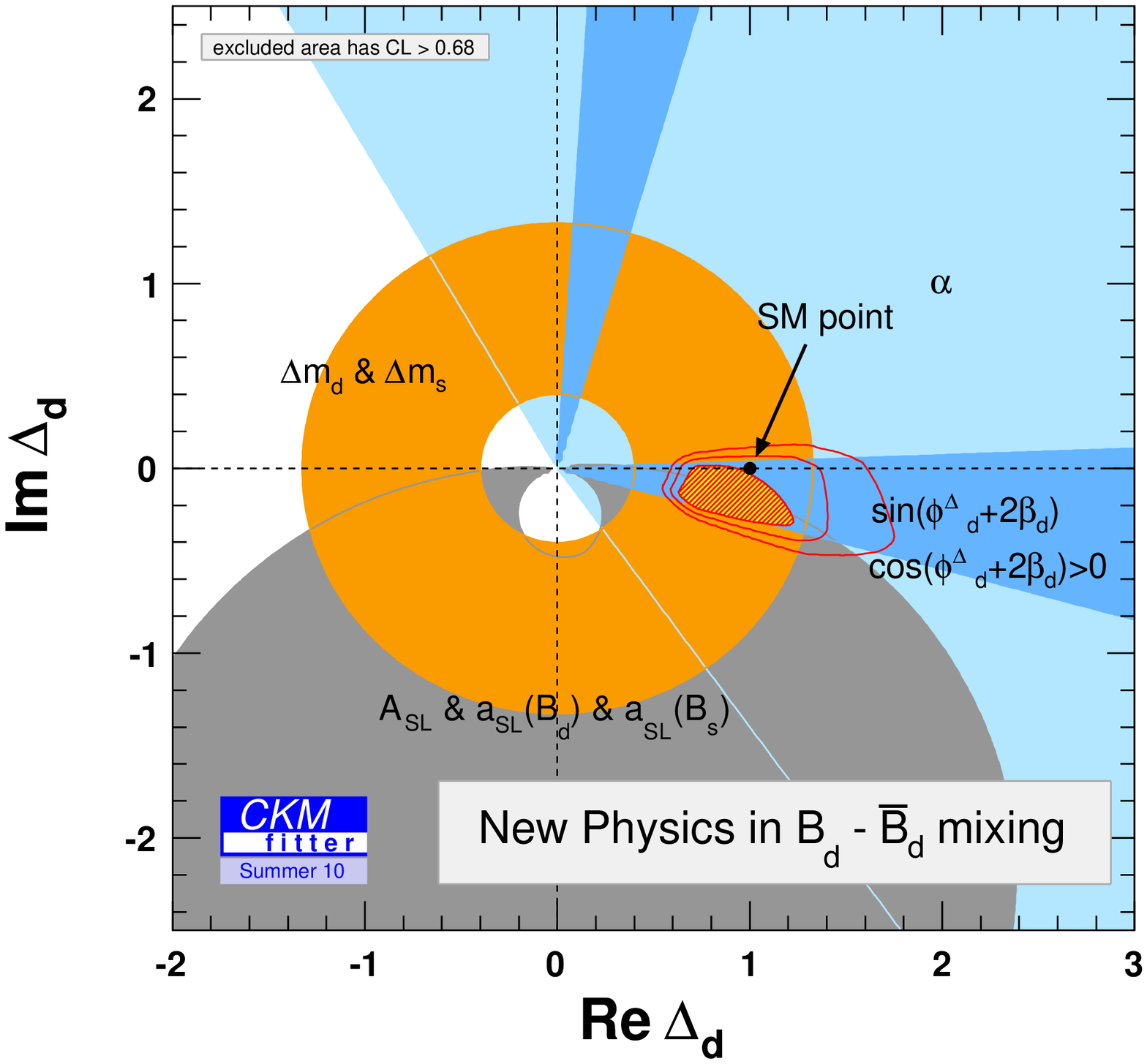}
   \caption{\small Fit result for the complex parameter $\Delta_d$ 
                   in scenario I when excluding $\BRB{\tau\nu}$
                   from the list of inputs.
                   For the combined fit the red area shows
                   the region with ${\rm CL} < 68.3~\%$ while the two additional contour
                   lines inscribe the regions with ${\rm CL} < 95.45~\%$,
                   and  ${\rm CL} < 99.73~\%$, respectively.
               \label{fig-Deltad_woBtaunu_scenario1}}
%\end{nfigure}
%\begin{nfigure}{htb}
 \includegraphics[width=10cm]{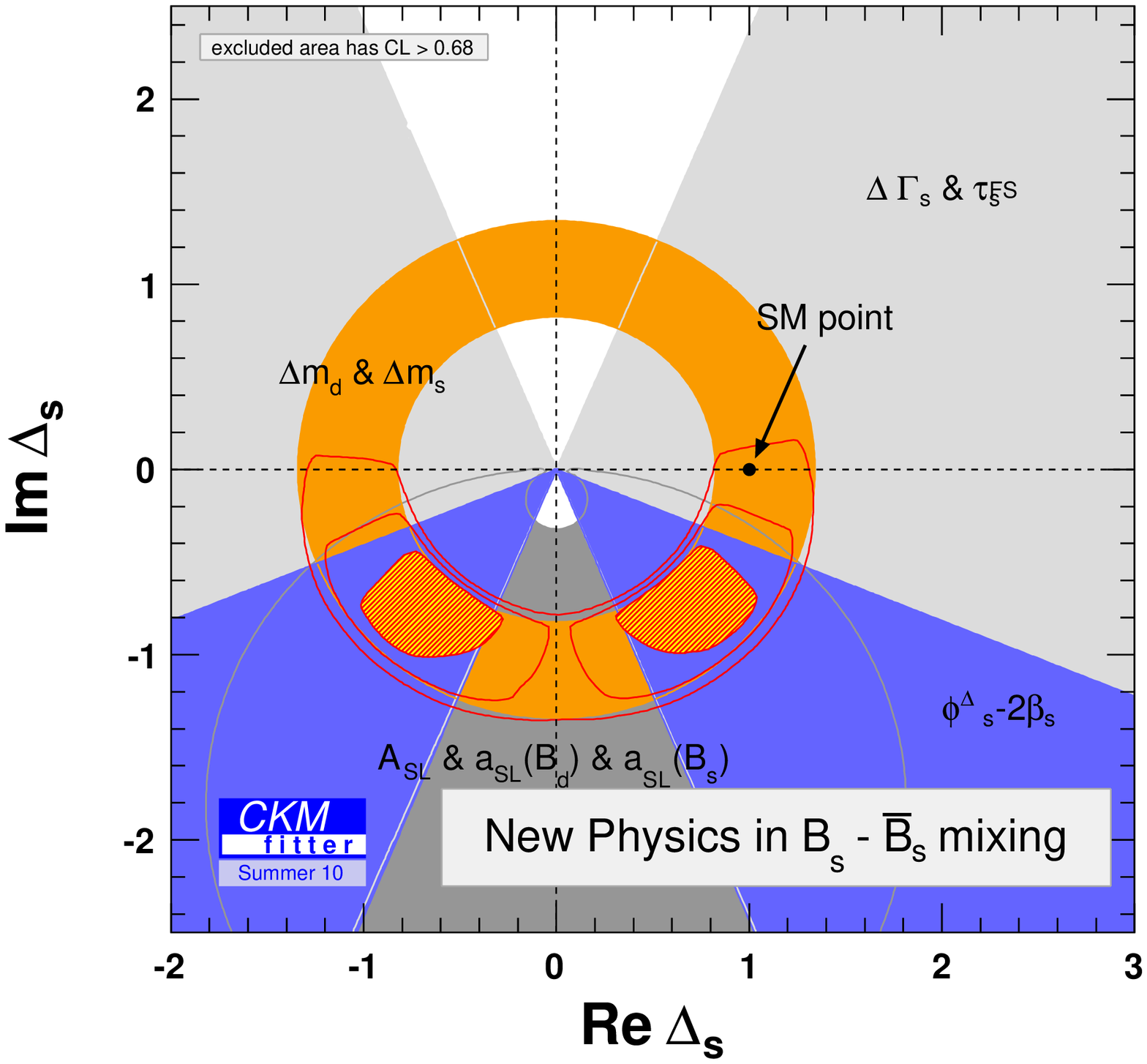}
   \caption{\small  Fit result for the complex parameter $\Delta_s$ 
                   in scenario I when excluding $\BRB{\tau\nu}$
                   from the list of inputs.
                   For the combined fit the red area shows
                   the region with ${\rm CL} < 68.3~\%$ while the two additional contour
                   lines inscribe the regions with ${\rm CL} < 95.45~\%$,
                   and  ${\rm CL} < 99.73~\%$, respectively.
\label{fig-Deltas_woBtaunu_scenario1}}
\end{nfigure}

\begin{nfigure}{Htb}
 \includegraphics[width=10cm]{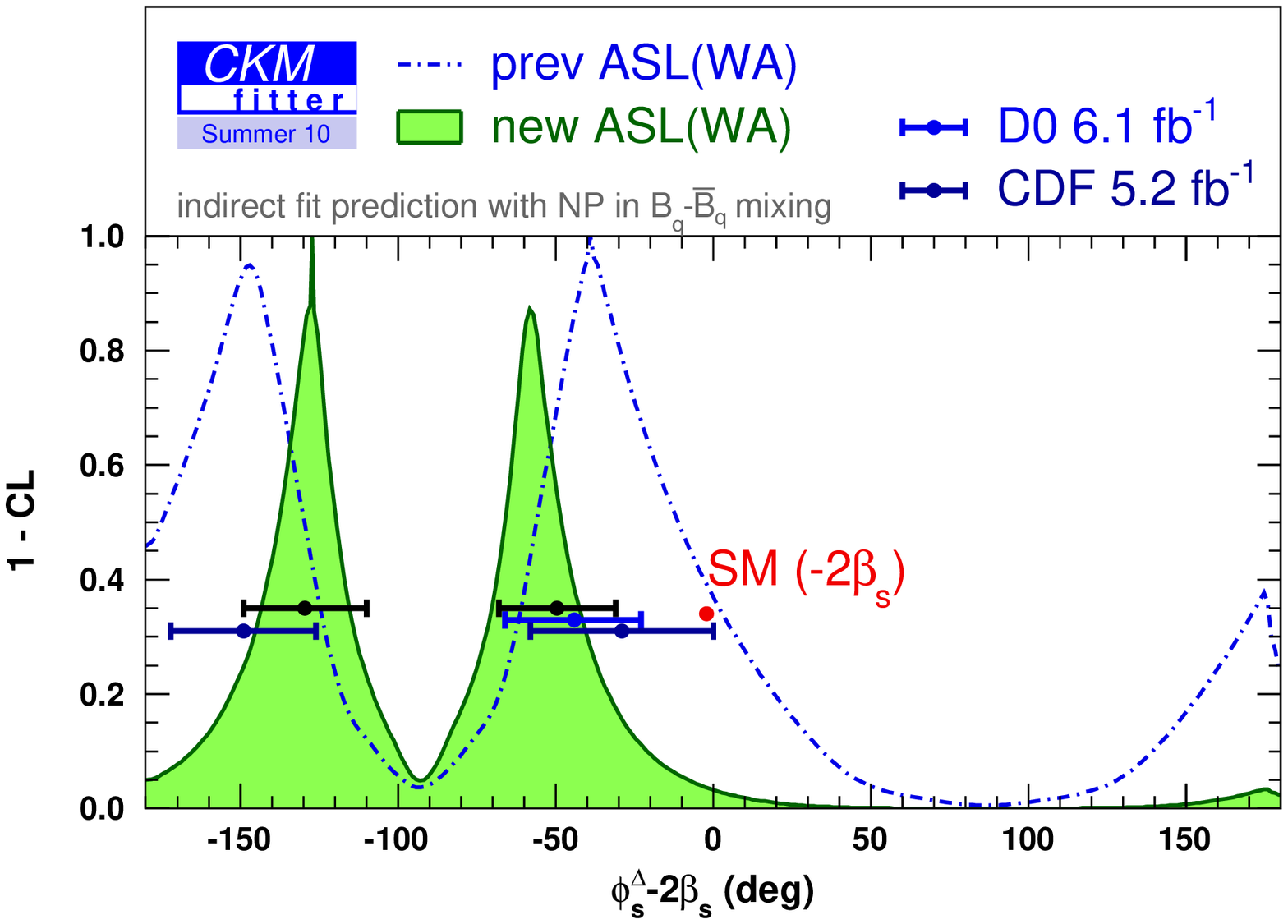}
   \caption{\small New Physics scenario I. 
   Indirect constraint on the CP phase  $\phi_s^{\psi\phi}$,
   compared with the direct measurement by the TeVatron: previous world average~\cite{Punzi} (in black) and
   the Summer 2010 CDF and \Dzero\ updates~\cite{taggedphaseCDF_3,taggedphaseD0_2} (in blue). The dotted line represents the full
   scenario I fit with the previous world average for $a_\mathrm{fs}$, while the green curve is the update after the \Dzero\ evidence
   for a non zero dimuonic asymmetry.
                  \label{fig-phis_comparison}}
\end{nfigure}

\clearpage
\subsection[Scenario II]{Scenario II: MFV with small bottom--Yukawa
coupling}\label{ssec:mfv1} 

In this section, we discuss the MFV scenario which allows to connect the $B$
and kaon sectors. Such a kind of numerical analysis has been first presented 
in~\cite{UTfit1}. In this scenario, there is only one additional real New Physics parameter
$\Delta$, see \eq{defsc2}. 
As a consequence, this scenario has difficulties to describe a situation 
where the data prefer a non-zero New Physics phase in $B$-mixing.
Indeed the scenario II hypothesis embedded in scenario I, that is $\Delta_d=\Delta_s=\Delta$
with Im$(\Delta)=0$, is disfavoured by 3.7 standard deviations.
The quality of the fit does not change when $\epsilon_{K}$ is removed 
from the list of inputs. 

Fig.~\ref{fig-Delta_scenario2} shows the impact of the various constraints
on the parameter $\Delta$. The constraint is only slightly changed when
adding to the $B$-meson observables - where $\BRB{\tau\nu}$
has been excluded - the $\epsilon_{K}$ constraint. On the other hand, 
when adding $\BRB{\tau\nu}$, the constraint gets significantly stronger
at the $1\sigma$ level while at $2\sigma$ the reduction in the allowed
parameter space is modest.
\begin{nfigure}{Htb}
  \includegraphics[width=10cm]{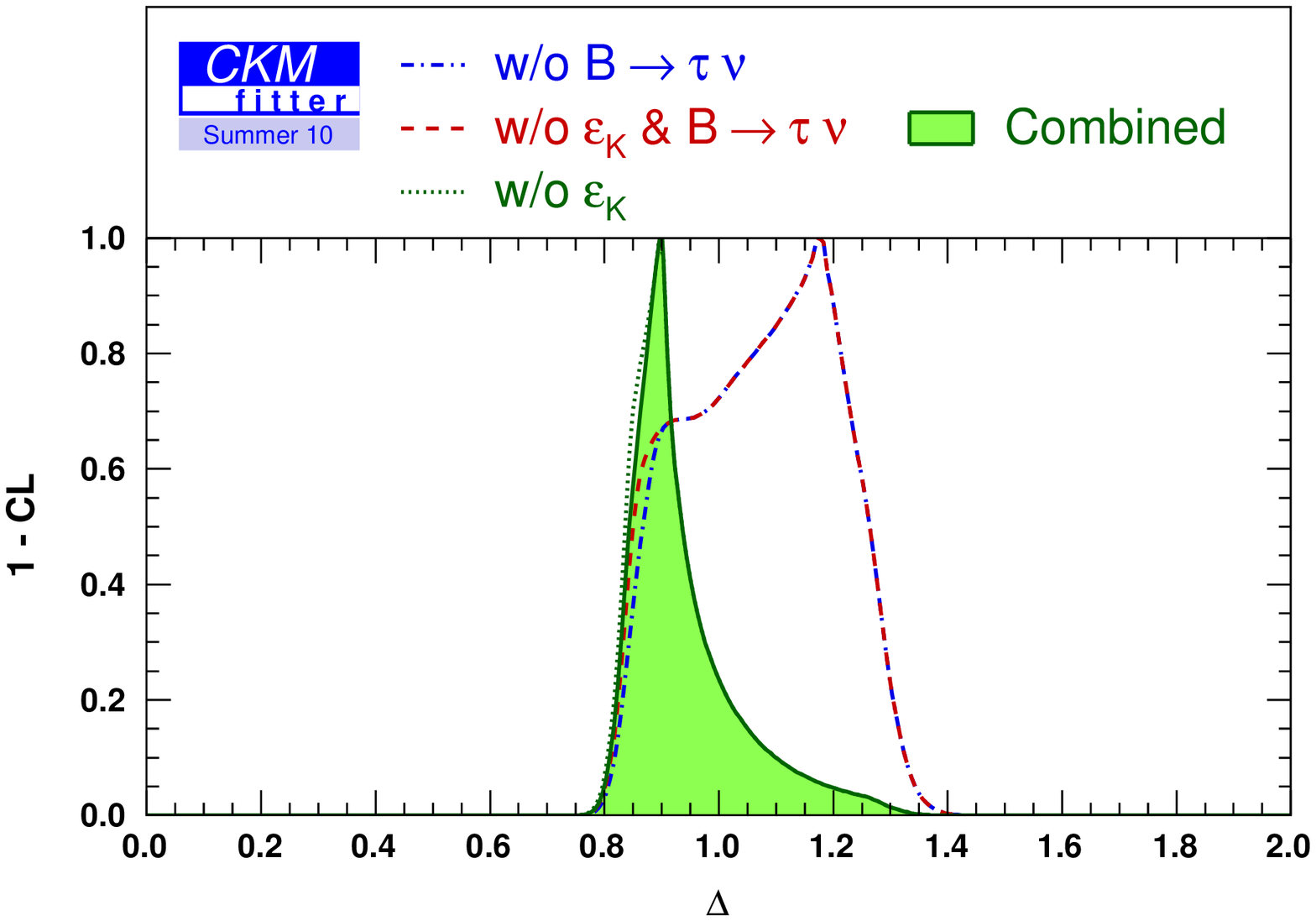}
  \caption{\small Constraint on the real parameter $\Delta$ from
                  the fit in scenario II. The red dashed line represents
                  the constraint if only $B_{d}$ and $B_{s}$ observables
                  are used excluding $\BRB{\tau\nu}$ and $\epsilon_{K}$.
                  When adding $\epsilon_{K}$ the constraint is essentially
                  unchanged (blue dotted-dashed line). A significantly 
                  stronger constraint is obtained when adding $\BRB{\tau\nu}$ 
                  (dotted green). The full constraint when adding both, 
                  $\BRB{\tau\nu}$ and $\epsilon_{K}$, is shown in green.
                  \label{fig-Delta_scenario2}}
\end{nfigure}
In Tables~\ref{tab:fitResults_MFV_II} and~\ref{tab:fitResults_MFV_II2}
we further provide
the constraints on individual parameters obtained from the combined fit in scenario II 
as well as predictions for the parameters not used as fit inputs. The compatibility of
the parameter $\Delta$ with 1 is good, meaning that this New Physics scenario does not describe
the data better than the Standard Model, as expected since all discussed anomalies seem to require new CP phases.

\begin{table}[Htb]
 \setlength{\tabcolsep}{0.8pc}
 \begin{center}\begin{tabular}{llll} \hline
 Quantity & central  ${\pm \mathrm{CL}}\equiv1\sigma$       &  ${\pm \mathrm{CL}}\equiv2\sigma$    &  ${\pm \mathrm{CL}}\equiv3\sigma$  \\[0.15cm]
 \hline  &&& \\[-0.3cm]
$A$        & $  0.8172 ^{\,+   0.0093}_{\,-   0.0199}$   & $0.817^{\,+ 
0.019}_{\,-    0.037}$ & $0.817^{\,+    0.028}_{\,-    0.047}$ \\[0.15cm]
$\lambda$  & $  0.22543 ^{\,+  0.00077}_{\,-  0.00077}$  & $0.2254^{\,+ 
0.0015}_{\,- 0.0015}$   & $0.2254^{\,+  0.0023}_{\,-  0.0023}$ \\[0.15cm]
$\bar\rho$ & $  0.145 ^{\,+    0.026}_{\,-    0.019}$    & $0.145^{\,+   
0.048}_{\,-   0.028}$ & $0.145^{\,+    0.066}_{\,-    0.037}$ \\[0.15cm]
$\bar\eta$ & $  0.342 ^{\,+    0.016}_{\,-    0.011}$    & $0.342^{\,+   
0.030}_{\,-   0.024}$ & $0.342^{\,+    0.045}_{\,-    0.038}$ \\[0.15cm]
 \hline &&&      \\[-0.3cm]
$\Delta$   & $ 0.899 ^{\,+ 0.072}_{\,-    0.069}$        & $0.90^{\,+   
0.31}_{\,-     0.10}$  & $0.90^{\,+    0.45}_{\,-    0.13}$ \\[0.15cm]
 \hline &&&      \\[-0.3cm]
$B_K$ (!)                              
                 & $ 0.88 ^{\,+    0.23}_{\,-     0.15}$ & $0.88^{\,+     0.34}_{\,-
    0.26}$ & $0.88^{\,+     0.44}_{\,-     0.31}$ \\[0.15cm]
$f_{B_s}$ [MeV] (!)                   
                 & $ 254.5 ^{\,+   8.4}_{\,-      11.9}$     & $254^{\,+    17}_{\,-37}$       & $254^{\,+    26}_{\,-     72}$ \\[0.15cm]
$\widehat\Bag_{B_{s}}$ (!)               
   
                 & $ 0.763 ^{\,+ 0.545}_{\,-  0.075}$   & $0.76^{\,+0.80}_{\,- 0.14}$   & $0.76^{\,+    1.12}_{\,-   0.19}$ \\[0.15cm]
$f_{B_s}/f_{B_d}$ (!)     
                 & $1.217 ^{\,+    0.053}_{\,-   0.035}$   & $1.217^{\,+0.113}_{\,- 0.076}$    & $1.22^{\,+    0.16}_{\,-    0.18}$ \\[0.15cm]
$\Bag_{B_{s}}/\Bag_{B_{d}}$ (!)     
                 & $1.136 ^{\,+   0.076}_{\,- 0.095}$   & $1.14^{\,+0.16}_{\,- 0.19}$     & $1.14^{\,+  0.24}_{\,-  0.28}$ \\[0.15cm]
$\widetilde{\Bag}_{S,B_s}(m_b)$ (!)     
                 & $-0.9 ^{\,+  1.3}_{\,- 2.2}$   & $-0.9 ^{\,+  2.6}_{\,-  3.7}$     & $-0.9 ^{\,+  4.0}_{\,-  5.1}$ \\[0.15cm]
 \hline &&&      \\[-0.3cm]
$J$~~$[10^{-5}]$ 
           & $  2.99 ^{\,+ 0.15 }_{\,-    0.12}$         & $2.99^{\,+     0.29}_{\,-     0.29}$ & $2.99^{\,+     0.44}_{\,-     0.42}$ \\[0.15cm]
 \hline &&&      \\[-0.3cm]
$\alpha$ [deg] (!)
           & $ 93.3 ^{\,+ 5.7}_{\,-  4.5}$            & $93.3^{\,+     8.5}_{\,-     6.7}$ & $93.3^{\,+     10.6}_{\,-     8.6}$ \\[0.15cm]
$\beta$  [deg]  (!)
           & $  28.05 ^{\,+ 0.70}_{\,-   2.27}$        & $28.1^{\,+    1.4}_{\,-    4.8}$ & $28.1^{\,+    2.1}_{\,-    7.4}$ \\[0.15cm]
$\gamma$ [deg] (!)   
           & $  67.1 ^{\,+ 2.9}_{\,-   3.8}$        & $67.1^{\,+    4.5}_{\,-
   7.2}$ & $67.1^{\,+    6.0}_{\,-    10.0}$ \\[0.15cm]
 \hline  &&& \\[-0.3cm]
$\phi_d$ [deg]  & $-5.7 ^{\,+ 1.5}_{\,-    3.3}$  & $-5.7 ^{\,+    2.5}_{\,-     4.2}$  & $-5.7 ^{\,+    2.8}_{\,-     5.2}$  \\[0.15cm]
$-2\beta_s$ [deg] (!) & $-2.085^{\,+0.071}_{\,-0.095}$ & $-2.08^{\,+0.15}_{\,-0.18}$ & $-2.08^{\,+0.23}_{\,-0.27}$
\\[0.15cm]
$-2\beta_s$ [deg] & $-2.083^{\,+0.070}_{\,-0.097}$ & $-2.08^{\,+0.15}_{\,-0.19}$ & $-2.08^{\,+0.24}_{\,-0.28}$
\\[0.15cm]
$\phi_s$ [deg] & $0.401 ^{\,+ 0.042}_{\,-    0.124}$  & $0.401 ^{\,+    0.090}_{\,-     0.211}$  & $0.40 ^{\,+    0.14}_{\,-     0.26}$  \\[0.15cm]
 \hline
 \end{tabular}
 \end{center}
 \vspace{-0.5cm}
 \caption[.]{\label{tab:fitResults_MFV_II}%\em
              Fit results in the New Physics scenario II. The notation `(!)' means that the fit output represents the indirect constraint, \textit{i.e.}
              the corresponding direct input has been removed from the analysis. }
\end{table}
\begin{table}[Htb]
 \setlength{\tabcolsep}{0.8pc}
\begin{center}\begin{tabular}{llll} \hline
 Quantity & central  ${\pm \mathrm{CL}}\equiv1\sigma$       &  ${\pm
\mathrm{CL}}\equiv2\sigma$   &  ${\pm \mathrm{CL}}\equiv3\sigma$  \\[0.15cm]
 \hline  &&& \\[-0.3cm]
$|V_{ud}|$(!) 
           & $  0.97426 ^{\,+  0.00030}_{\,-  0.00030}$  & $0.97426^{\,+ 
0.00060}_{\,- 0.00061}$  & $0.97426^{\,+  0.00089}_{\,- 0.00091}$ \\[0.15cm]
$|V_{us}|$(!) 
           & $  0.22545 ^{\,+   0.00095}_{\,-  0.00095}$    & $0.2254^{\,+ 
0.0019}_{\,-  0.0019}$   & $0.2254^{\,+  0.0028}_{\,-  0.0029}$ \\[0.15cm]
$|V_{ub}|$(!) 
           & $  0.00357 ^{\,+0.00015}_{\,-  0.00014}$  &
$0.00357^{\,+0.00030}_{\,- 0.00029}$  & $0.00357^{\,+0.00046}_{\,- 0.00044}$
\\[0.15cm]
$|V_{cd}|$ & $  0.22529 ^{\,+  0.00077}_{\,-  0.00077}$  & $0.2253^{\,+  0.0015}_{\,-  0.0015}$   & $0.2253^{\,+  0.0023}_{\,-  0.0023}$ \\[0.15cm]
$|V_{cs}|$ & $  0.97341 ^{\,+  0.00018}_{\,-  0.00018}$  & $0.97341^{\,+  0.00036}_{\,- 0.00036}$  & $0.97341^{\,+  0.00054}_{\,- 0.00054}$ \\[0.15cm]
$|V_{cb}|$(!) 
           & $  0.0493 ^{\,+  0.0028}_{\,-  0.0061}$  & $0.0493^{\,+ 
0.0048}_{\,- 0.0094}$  & $0.0493^{\,+  0.0066}_{\,-  0.0116}$ \\[0.15cm]
$|V_{td}|$ & $  0.00863 ^{\,+  0.00020}_{\,-  0.00023}$  & $0.00863^{\,+  0.00032}_{\,- 0.00055}$  & $0.00863^{\,+  0.00043}_{\,- 0.00082}$ \\[0.15cm]
$|V_{ts}|$ & $  0.04078 ^{\,+  0.00038}_{\,-  0.00097}$  & $0.04078^{\,+  0.00076}_{\,- 0.00171}$  & $0.0408^{\,+  0.0011}_{\,-  0.0021}$ \\[0.15cm]
$|V_{tb}|$ & $  0.999131 ^{\,+ 0.000040}_{\,- 0.000016}$ & $0.999131^{\,+ 0.000071}_{\,- 0.000032}$ & $0.999131^{\,+ 0.000088}_{\,- 0.000048}$ \\[0.15cm]
 \hline &&&      \\[-0.3cm]
$\epsilon_K$~~$[10^{-3}]$ (!)         & $ 1.87 ^{\,+    0.54}_{\,-     0.55}$ & $ 1.87 ^{\,+    0.95}_{\,-     0.67}$ & $ 1.87 ^{\,+    1.21}_{\,-     0.77}$  \\[0.15cm]
$\Delta m_d$ $[{\rm ps}^{-1}]$ (!)            
                & $ 0.554 ^{\,+     0.073}_{\,- 0.047}$  & $0.554^{\,+0.114}_{\,-    0.095}$  & $0.55^{\,+    0.16}_{\,-     0.14}$ \\[0.15cm]
$\Delta m_s$ $[{\rm ps}^{-1}]$ (!)            
                & $ 16.2 ^{\,+     1.7}_{\,- 1.4}$       & $16.2^{\,+3.7}_{\,-   2.6}$  & $16.2^{\,+    8.5}_{\,-     3.6}$ \\[0.15cm]
$A_\text{SL}$ $[10^{-4}]$ (!)            
                & $-4.04 ^{\,+1.01}_{\,-0.53}$  & $-4.0^{\,+1.9}_{\,-1.1}$   & $-4.0^{\,+ 2.4}_{\,- 1.6}$  \\[0.15cm]
$A_\text{SL}$ $[10^{-4}]$             
                & $-4.06 ^{\,+0.96}_{\,-0.55}$  & $-4.1^{\,+1.9}_{\,-1.1}$   & $-4.1^{\,+ 2.4}_{\,- 1.7}$  \\[0.15cm]
$a_\text{SL}^{s}-a_\text{SL}^{d}$ $[10^{-4}]$             
                & $8.74 ^{\,+0.99}_{\,-1.97}$  & $8.7^{\,+1.9}_{\,-4.0}$   & $8.7^{\,+ 2.8}_{\,- 4.9}$  \\[0.15cm]
$a_\text{SL}^{d}$ $[10^{-4}]$ (!)            
                & $-8.36^{\,+ 1.93}_{\,-0.94}$ & $-8.4^{\,+3.9}_{\,- 1.8}$   & $-8.4^{\,+ 4.7}_{\,- 2.7}$  \\[0.15cm]
$a_\text{SL}^{d}$ $[10^{-4}]$             
                & $-8.37^{\,+ 1.90}_{\,-0.95}$ & $-8.4^{\,+3.8}_{\,- 1.8}$   & $-8.4^{\,+ 4.7}_{\,- 2.7}$  \\[0.15cm]
$a_\text{SL}^{s}$ $[10^{-4}]$ (!)            
                & $0.373^{\,+ 0.045}_{\,-0.083}$  & $0.373^{\,+0.078}_{\,- 0.168}$   &    $0.37^{\,+ 0.11}_{\,- 0.21}$  \\[0.15cm]
$a_\text{SL}^{s}$ $[10^{-4}]$            
                & $0.373^{\,+ 0.045}_{\,-0.083}$  & $0.373^{\,+0.078}_{\,- 0.168}$   &    $0.37^{\,+ 0.11}_{\,- 0.21}$  \\[0.15cm]
$\Delta\Gamma_d$ $[{\rm ps}^{-1}]$ (!)            
                & $0.00426^{\,+ 0.00056}_{\,-0.00154}$  & $0.0043^{\,+ 0.0011}_{\,- 0.0021}$   & $0.0043^{\,+ 0.0016}_{\,- 0.0024}$  \\[0.15cm]
$\Delta\Gamma_s$ $[{\rm ps}^{-1}]$ (!)            
                & $0.110^{\,+ 0.079}_{\,-0.022}$  & $0.110^{\,+ 0.089}_{\,- 0.037}$   & $0.110^{\,+ 0.097}_{\,- 0.051}$  \\[0.15cm]
$\Delta\Gamma_s$ $[{\rm ps}^{-1}]$             
                & $0.0946^{\,+ 0.0174}_{\,-0.0082}$  & $0.095^{\,+ 0.047}_{\,- 0.025}$   & $0.095^{\,+ 0.088}_{\,- 0.037}$  \\[0.15cm]
 \hline &&&      \\[-0.3cm]
$\mathcal{B}(B\to\tau\nu)$ $[10^{-4}]$ (!) & $ 0.653 ^{\,+ 0.277}_{\,- 0.040}$ & $ 0.653 ^{\,+ 0.404}_{\,- 0.077}$ & $ 0.65 ^{\,+ 0.52}_{\,- 0.11}$ \\[0.15cm]
$\mathcal{B}(B\to\tau\nu)$ $[10^{-4}]$ & $ 0.92 ^{\,+ 0.12}_{\,- 0.10}$ & $ 0.92 ^{\,+ 0.23}_{\,- 0.26}$ & $ 0.92 ^{\,+ 0.33}_{\,- 0.34}$ \\[0.15cm]
 \hline
 \end{tabular}
 \end{center}
 \vspace{-0.5cm}
 \caption[.]{\label{tab:fitResults_MFV_II2}%\em
              Fit results in the New Physics scenario II. The notation `(!)' means that the fit output represents the indirect constraint, \textit{i.e.}
              the corresponding direct input has been removed from the analysis. }
 \end{table}

\clearpage
\subsection[Scenario III]{Scenario III: Generic MFV}\label{ssec:mfv2} 

In Fig.~\ref{fig-Delta_scenario3} we present the result from the
combined fit to $B_{d}$ and $B_{s}$ observables in the complex plane
$\Delta=\Delta_s=\Delta_d$ for the MFV hypothesis allowing for a
large bottom Yukawa coupling $y_b$ (scenario III).  In scenario
III, $\Delta$ can have a sizeable complex component proportional to
$y_b^2$. One expects that the constraint $\Delta_s=\Delta_d$
reduces the size of the allowed New Physics contributions significantly with
respect to the general case studied in scenario I.
However, our fit result in \fig{fig-Delta_scenario3} still allows for a New Physics 
contribution of order $-20~\%$ to $+40~\%$.
The New Physics phase in this scenario shows evidence for a deviation from the Standard Model
since both $a_\mathrm{fs}$  and the $\phi^\Delta_{s}$ measurement on the one hand
and  the discrepancy between $\sin 2\phi_d^{\psi K}$
 and $\BRB{\tau\nu}$ on the other hand, prefer a negative New Physics phase in $B_{d,s}$ mixing. 
\begin{nfigure}{Htb}
\includegraphics[width=12cm]{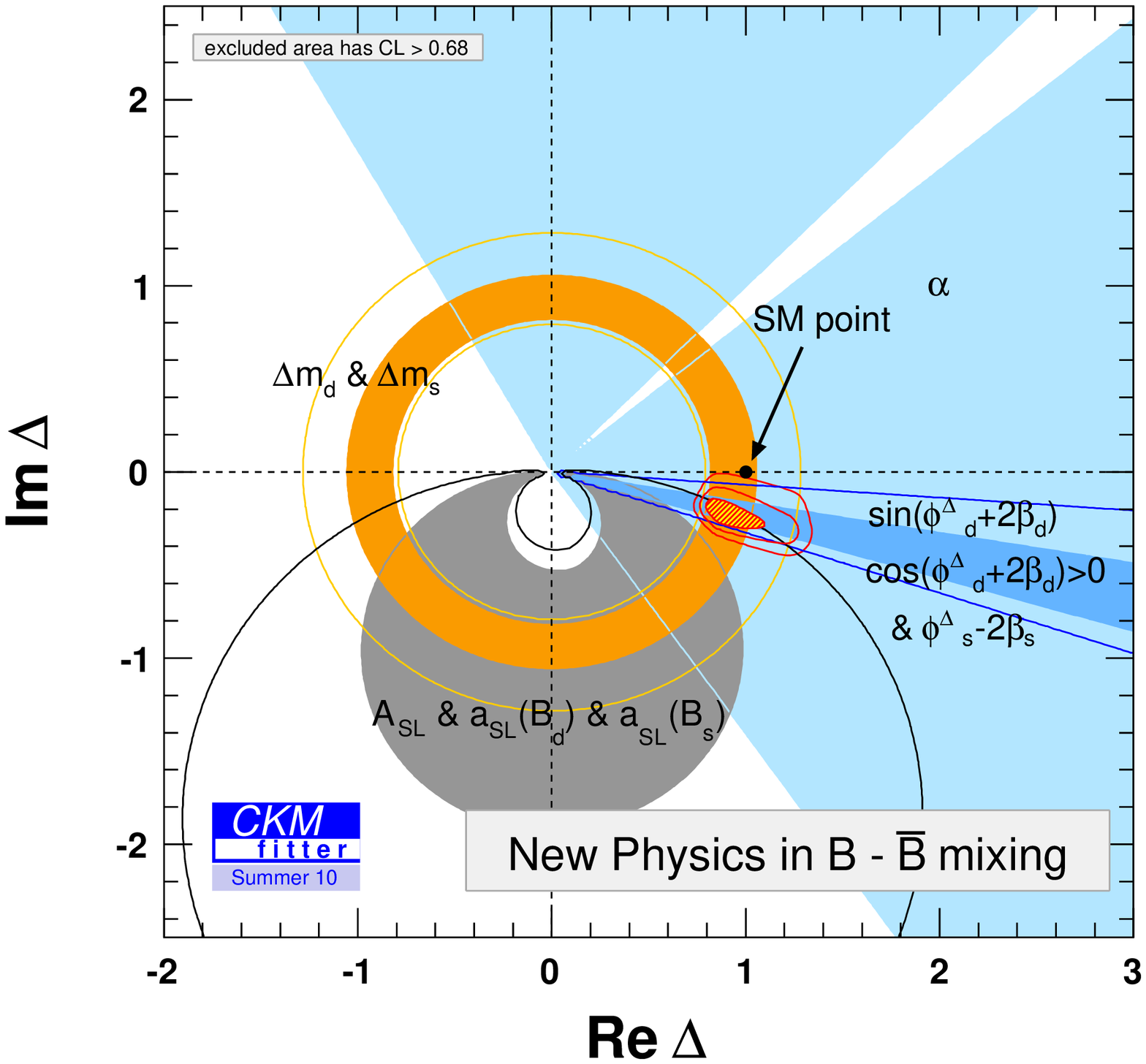}
 \caption{\small Constraint on the complex parameter $\Delta \equiv \Delta_d=\Delta_s$ from 
                  the fit in scenario III. 
                  For the individual constraints the coloured areas represent regions
                  with ${\rm CL} < 68.3~\%$. In addition, a ${\rm CL} < 95.45~\%$ contour
                  is shown for the individual constraints obtained from
                  $\Delta m_{d}$ 
                  and $\Delta m_{s}$,
                  from $A_{SL}$, $a_{SL}^{d}$, and $a_{SL}^{s}$, and from $\phi_{d}^{\psi K}$
                  and $\phi_{s}^{\psi \phi}$.
                  For the combined fit the red area shows
                  the region with ${\rm CL} < 68.3~\%$ while the two additional contour 
                  lines inscribe the regions with ${\rm CL} < 95.45~\%$ and 
                  ${\rm CL} < 99.73~\%$, respectively.\label{fig-Delta_scenario3}}
\end{nfigure}

%\begin{nfigure}{htb}
% \includegraphics[width=10cm]{globalNP_newMFV_BdBsEpsK_ChangeOfVariables_woBtaunu_all.eps}
%  \caption{\small Constraint on the $\mbox{Re}{\Delta}-\mbox{Im}{\Delta}$ coordinates from the fit
%                 in scenario III when excluding $\BRB{\tau\nu}$
%                  from the list of inputs.
%                  The shaded area shows the allowed region with ${\rm CL} < 68~\%$.
%                  The allowed region with ${\rm CL} < 95~\%$ is inscribed by the contour line
%                  shown.\label{fig-Delta_woBtaunu_scenario3}}
%\end{nfigure}

Compared to scenario II this model is in much better agreement with the
data regardless of whether one takes into account
$\BRB{\tau\nu}$ as an input or not. While the $B_{d}$ sector constraints do not
allow too large a New Physics phase, the $B_{s}$ sector prefers a large New Physics phase,
though with large errors. As a consequence and as already discussed above,
the quality of the fit in scenario III is
manifestly worse ($2.1~\sigma$) than in scenario I.

In this scenario, the $B_{d}$- and $B_{s}$-systems decouple from the kaon system 
and hence the constraint on $\mbox{Re}{\Delta}$ and $\mbox{Im}{\Delta}$ cannot be 
improved by adding $\epsilon_{K}$ as this introduces the additional NP parameter 
$\Delta_K^{tt}$ in the theoretical prediction for $\epsilon_{K}$. However, when 
including $\epsilon_{K}$ in the fit one is able to constrain $\Delta_K^{tt}$ 
from the fitted values of the CKM parameters.
The parameter is in agreement with the Standard Model expectation but could differ by more than 
$40~\%$ from $\Delta_K^{tt}=1$. Similarly to the previous scenarios,
one-dimensional results from the fit in scenario III are shown in Tables~\ref{tab:fitResults_MFV_III}
and~\ref{tab:fitResults_MFV_III2}. Again one sees that in particular the semileptonic asymmetries 
$a_\text{SL}^{d}$ and $A_\text{SL}$ are more precisely predicted than the measurements, so that
improvements of the data will be extremely instructive. The asymmetry difference is now predicted to be
$a_\text{SL}^{s}-a_\text{SL}^{d}=(18.6 ^{\,+ 2.0}_{\,- 10.2})\times 10^{-4}$
($-1\times 10^{-4}<a_\text{SL}^{s}-a_\text{SL}^{d}<25\times 10^{-4}$ at $3~\sigma$), hence a moderate positive value is preferred with respect to Scenario
I that predicts a larger negative value.

Finally, two tests of the Standard Model are interesting to study within scenario III, and are shown in Table~\ref{ScIIIpvalues}.
\begin{table}[!Htb]
\begin{center}\begin{tabular}{ll} \hline
Hypothesis & p-value \\ 
 \hline  & \\[-0.3cm]
$\mathrm{Im}(\Delta)=0$ (1D) & $3.5~\sigma$ \\[0.15cm]
 \hline  & \\[-0.3cm]
$\Delta=1$ (2D) & $3.3~\sigma$ \\[0.15cm]
\hline \end{tabular} \end{center}
\caption{p-values for various Standard Model hypotheses in the framework of New Physics scenario III, in terms of the number of 
equivalent standard deviations. These numbers are computed from the $\chi^2$ difference with and without 
the hypothesis constraint, interpreted with the appropriate number of degrees of freedom.}\label{ScIIIpvalues}
\end{table}
As in scenario I, they show evidence for New Physics, and that scenario III describes the data significantly better than either the Standard Model or
scenario II, assuming as before that the improved data on $B_s\to J/\psi\phi$ do not change the overall picture dramatically.

\clearpage

\begin{table}[Htb]
 \setlength{\tabcolsep}{0.8pc}
 \begin{center}\begin{tabular}{llll} \hline
 Quantity & central  ${\pm \mathrm{CL}}\equiv1\sigma$       &  ${\pm \mathrm{CL}}\equiv2\sigma$    &  ${\pm \mathrm{CL}}\equiv3\sigma$  \\[0.15cm]
 \hline  &&& \\[-0.3cm]
$A$        & $  0.7928 ^{\,+   0.0288}_{\,-   0.0092}$   & $ 0.793 ^{\,+    0.040}_{\,-    0.018}$ & $0.793 ^{\,+    0.050}_{\,-    0.027}$ \\[0.15cm]
$\lambda$  & $  0.22536 ^{\,+  0.00077}_{\,-  0.00077}$  & $ 0.2254 ^{\,+  0.0015}_{\,-  0.0015}$   & $ 0.2254 ^{\,+  0.0023}_{\,-  0.0023}$ \\[0.15cm]
$\bar\rho$ & $  0.177 ^{\,+    0.032}_{\,-    0.012}$    & $ 0.177 ^{\,+    0.059}_{\,-    0.026}$ & $ 0.177 ^{\,+    0.072}_{\,-    0.042}$ \\[0.15cm]
$\bar\eta$ & $  0.439 ^{\,+    0.011}_{\,-    0.022}$    & $ 0.439 ^{\,+    0.021}_{\,-    0.050}$ & $ 0.439^{\,+    0.031}_{\,-    0.086}$ \\[0.15cm]
 \hline &&&      \\[-0.3cm]
$\mbox{Re}{(\Delta)}$  & $ 0.876 ^{\,+ 0.126}_{\,-    0.061}$  & $ 0.876 ^{\,+    0.346}_{\,-     0.097}$  &  $0.88^{\,+    0.43}_{\,-    0.13}$ \\[0.15cm]
$\mbox{Im}{(\Delta)}$  & $-0.226 ^{\,+ 0.057}_{\,-    0.048}$  & $-0.23 ^{\,+    0.11}_{\,-     0.12}$  & $-0.23 ^{\,+    0.18}_{\,-     0.19}$ \\[0.15cm]
$|\Delta|$  & $0.907 ^{\,+ 0.128}_{\,-    0.070}$  & $0.91 ^{\,+    0.35}_{\,-     0.11}$  & $0.91 ^{\,+    0.45}_{\,-     0.14}$ \\[0.15cm]
$\phi^{\Delta}$ [deg]  & $-14.4 ^{\,+ 2.9}_{\,-    2.1}$  & $-14.4 ^{\,+    6.5}_{\,-     4.2}$  & $-14.4 ^{\,+    11.4}_{\,-     6.3}$ \\[0.15cm]
$\Delta_K^{tt}$ & $1.01 ^{\,+ 0.40}_{\,-    0.38}$  & $1.01^{\,+    0.53}_{\,-     0.43}$  & $1.01^{\,+    0.69}_{\,-     0.47}$ \\[0.15cm]
 \hline &&&      \\[-0.3cm]
$f_{B_s}$ [MeV] (!)                   
                 & $ 274 ^{\,+   20}_{\,-      24}$     & $ 274 ^{\,+    46}_{\,-  51}$       & $ 274 ^{\,+    77}_{\,-     83}$ \\[0.15cm]
$\widehat\Bag_{B_{s}}$ (!)                   
                 & $ 1.44 ^{\,+ 0.33}_{\,-  0.61}$   &  $1.44^{\,+  0.99}_{\,-  1.37}$   & $1.4^{\,+  1.6}_{\,-  1.5}$ \\[0.15cm]
$f_{B_s}/f_{B_d}$ (!)     
                 & $1.230 ^{\,+    0.066}_{\,-   0.048}$   & $1.23 ^{\,+    0.13}_{\,- 0.10}$    & $1.23 ^{\,+    0.18}_{\,-    0.21}$ \\[0.15cm]
$\Bag_{B_{s}}/\Bag_{B_{d}}$ (!)     
                 & $1.155 ^{\,+   0.090}_{\,- 0.138}$   & $ 1.16 ^{\,+  0.18}_{\,-  0.24}$     & $1.16 ^{\,+  0.28}_{\,-  0.40}$ \\[0.15cm]
$\widetilde{\Bag}_{S,B_s}(m_b)$ (!)     
                 & $1.14 ^{\,+  0.98}_{\,- 1.91}$   & $ 1.1 ^{\,+  2.8}_{\,-  4.0}$     & $ 1.1 ^{\,+  4.6}_{\,-  5.4}$ \\[0.15cm]
 \hline &&&      \\[-0.3cm]
$J$~~$[10^{-5}]$ 
           & $  3.61 ^{\,+ 0.16 }_{\,-    0.14}$  & $3.61^{\,+     0.27}_{\,-     0.37}$ & $3.61^{\,+     0.36}_{\,-     0.67}$ \\[0.15cm]
 \hline &&&      \\[-0.3cm]
$\alpha$   [deg] (!)
           & $  88.0 ^{\,+ 4.4}_{\,-  4.9}$         & $88.0^{\,+    6.9}_{\,-     6.6}$ & $88.0^{\,+     10.4}_{\,-     8.0}$ \\[0.15cm]
$\beta$    [deg] (!)
           & $ 28.01 ^{\,+ 0.66}_{\,-   1.47}$        & $28.0^{\,+    1.3}_{\,-    3.4}$ & $28.0^{\,+    2.0}_{\,-    8.5}$ \\[0.15cm]
$\gamma$ [deg] (!)& $ 68.1 ^{\,+ 1.3}_{\,- 4.3}$         & $ 68.1^{\,+    2.6}_{\,-     8.0}$ & $ 68.1 ^{\,+     3.9}_{\,-     9.8}$ \\[0.15cm]
 \hline \\[-0.3cm]
$\phi^{\Delta}+2\beta$ [deg]  (!)& $32.7 ^{\,+ 7.4}_{\,-    8.1}$  & $33 ^{\,+    16}_{\,-     29}$  & $33 ^{\,+    27}_{\,-     45}$  \\[0.15cm]
%&&&\textit{or} $-100 ^{\,+    27}_{\,-     21}$\\[0.15cm]
$\phi_d$ [deg]  & $-20.9 ^{\,+ 3.8}_{\,-    4.5}$  & $-20.9 ^{\,+    7.8}_{\,-     7.9}$  & $-21 ^{\,+    13}_{\,-     11}$  \\[0.15cm]
$\phi^{\Delta}-2\beta_s$ [deg]  (!) & $-16.6 ^{\,+ 3.7}_{\,-    2.2}$  & $-16.6 ^{\,+    8.1}_{\,-     4.3}$  & $-16.6 ^{\,+    13.9}_{\,-     6.5}$  \\[0.15cm]
$\phi^{\Delta}-2\beta_s$ [deg]  & $-17.1 ^{\,+ 3.0}_{\,-    2.1}$  & $-17.1 ^{\,+    6.8}_{\,-     4.3}$  & $-17.1 ^{\,+    11.8}_{\,-     6.4}$  \\[0.15cm]
$\phi_s$ [deg] & $-13.9 ^{\,+ 2.9}_{\,-    2.1}$  & $-13.9 ^{\,+    6.6}_{\,-     4.3}$  & $-13.9 ^{\,+    11.4}_{\,-     6.4}$  \\[0.15cm]
 \hline &&&      
 \end{tabular}
 \end{center}
 \vspace{-0.5cm}
 \caption[.]{\label{tab:fitResults_MFV_III} %\em
              Fit results in the New Physics scenario III. The notation `(!)' means that the fit output represents the indirect constraint, \textit{i.e.}
              the corresponding direct input has been removed from the analysis. }
 \end{table}

\begin{table}[Htb]
 \setlength{\tabcolsep}{0.8pc}
\begin{center}\begin{tabular}{llll} \hline
 Quantity & central  ${\pm \mathrm{CL}}\equiv1\sigma$       &  ${\pm \mathrm{CL}}\equiv2\sigma$    &  ${\pm \mathrm{CL}}\equiv3\sigma$  \\[0.15cm]
 \hline  &&& \\[-0.3cm]
$|V_{ud}|$(!) 
           & $  0.97430 ^{\,+  0.00030}_{\,-  0.00030}$  & $ 0.97430 ^{\,+  0.00060}_{\,- 0.00060}$  & $ 0.97430 ^{\,+  0.00089}_{\,- 0.00090}$ \\[0.15cm]
$|V_{us}|$(!) 
           & $  0.22534 ^{\,+   0.00095}_{\,-  0.00095}$    & $ 0.2253 ^{\,+  0.0019}_{\,-  0.0019}$   & $ 0.2253 ^{\,+  0.0028}_{\,-  0.0029}$ \\[0.15cm]
$|V_{ub}|$(!) 
           & $  0.00577 ^{\,+0.00045}_{\,-  0.00068}$  & $ 0.00577 ^{\,+0.00086}_{\,- 0.00124}$  & $ 0.0058 ^{\,+0.0012}_{\,- 0.0017}$ \\[0.15cm]
$|V_{cd}|$ & $  0.22524 ^{\,+  0.00077}_{\,-  0.00077}$  & $0.2252^{\,+  0.0015}_{\,-  0.0015}$   & $0.2252^{\,+  0.0023}_{\,-  0.0023}$ \\[0.15cm]
$|V_{cs}|$ & $  0.97347 ^{\,+  0.00018}_{\,-  0.00018}$  & $0.97347^{\,+  0.00036}_{\,- 0.00037}$  & $0.97347^{\,+  0.00053}_{\,- 0.00056}$ \\[0.15cm]
$|V_{cb}|$(!) 
           & $  0.0323 ^{\,+  0.0036}_{\,-  0.0035}$  & $ 0.0323 ^{\,+  0.0096}_{\,- 0.0076}$  & $ 0.032 ^{\,+  0.017}_{\,-  0.011}$ \\[0.15cm]
$|V_{td}|$ & $  0.00847 ^{\,+  0.00032}_{\,-  0.00032}$  & $ 0.00847 ^{\,+  0.00046}_{\,- 0.00061}$  & $ 0.00847 ^{\,+  0.00058}_{\,- 0.00076}$ \\[0.15cm]
$|V_{ts}|$ & $  0.03961 ^{\,+  0.00137}_{\,-  0.00037}$  & $0.03961 ^{\,+  0.00184}_{\,- 0.00074}$  & $0.0396^{\,+  0.0022}_{\,-  0.0011}$ \\[0.15cm]
$|V_{tb}|$ & $  0.999179 ^{\,+ 0.000016}_{\,- 0.000057}$ & $0.999179 ^{\,+ 0.000031}_{\,- 0.000077}$ & $0.999179 ^{\,+ 0.000046}_{\,- 0.000094}$ \\[0.15cm]
 \hline &&&      \\[-0.3cm]
$\Delta m_d$ $[{\rm ps}^{-1}]$ (!)            
                & $ 0.562 ^{\,+     0.081}_{\,- 0.068}$  & $ 0.56 ^{\,+    0.13}_{\,-     0.12}$  & $ 0.56 ^{\,+    0.18}_{\,-     0.20}$ \\[0.15cm]
$\Delta m_s$ $[{\rm ps}^{-1}]$ (!)            
                & $ 14.9 ^{\,+     2.3}_{\,- 1.1}$       & $ 14.9 ^{\,+    4.8}_{\,-     2.0}$  & $ 14.9 ^{\,+    7.9}_{\,-     2.9}$ \\[0.15cm]
$A_\text{SL}$ $[10^{-4}]$ (!)            
                & $-22.6 ^{\,+7.9}_{\,-3.4}$  & $-22.6 ^{\,+ 12.9}_{\,- 7.9}$   & $-23 ^{\,+ 19}_{\,- 14}$  \\[0.15cm]
$A_\text{SL}$ $[10^{-4}]$             
                & $-23.4 ^{\,+5.5}_{\,-3.6}$  & $-23.4 ^{\,+ 11.0}_{\,- 9.0}$   & $-23 ^{\,+ 17}_{\,- 14}$  \\[0.15cm]
$a_\text{SL}^{s}-a_\text{SL}^{d}$ $[10^{-4}]$           
                & $18.6^{\,+ 2.0}_{\,-10.2}$  & $18.6^{\,+ 4.1}_{\,- 17.2}$   & $18.6^{\,+ 6.3}_{\,- 19.4}$  \\[0.15cm]
$a_\text{SL}^{d}$ $[10^{-4}]$ (!)            
                & $-32.4^{\,+ 11.2}_{\,-3.8}$ & $-32.4^{\,+ 15.9}_{\,- 7.7}$   & $-32^{\,+ 22}_{\,-12}$  \\[0.15cm]
$a_\text{SL}^{d}$ $[10^{-4}]$            
                & $-32.5^{\,+ 10.2}_{\,-3.8}$ & $-32.5^{\,+ 15.5}_{\,- 7.7}$   & $-33^{\,+ 22}_{\,-12}$  \\[0.15cm]
$a_\text{SL}^{s}$ $[10^{-4}]$(!)            
                & $-13.9^{\,+ 3.4}_{\,-4.2}$  & $-13.9^{\,+ 7.5}_{\,- 13.2}$   & $-14^{\,+ 12}_{\,- 19}$  \\[0.15cm]
$a_\text{SL}^{s}$ $[10^{-4}]$           
                & $-13.9^{\,+ 3.4}_{\,-4.2}$  & $-13.9^{\,+ 7.5}_{\,- 13.2}$   & $-14^{\,+ 12}_{\,- 19}$  \\[0.15cm]
$\Delta\Gamma_d$ $[{\rm ps}^{-1}]$ (!)            
                & $0.00432^{\,+ 0.00053}_{\,-0.00180}$  & $0.0043^{\,+ 0.0012}_{\,- 0.0023}$   & $0.0043^{\,+ 0.0015}_{\,- 0.0027}$  \\[0.15cm]
$\Delta\Gamma_s$ $[{\rm ps}^{-1}]$ (!)            
                & $0.164^{\,+ 0.021}_{\,-0.074}$  & $0.164^{\,+ 0.031}_{\,- 0.097}$   & $0.164^{\,+ 0.040}_{\,- 0.109}$  \\[0.15cm]
$\Delta\Gamma_s$ $[{\rm ps}^{-1}]$            
                & $0.100^{\,+ 0.023}_{\,-0.019}$  & $0.100^{\,+ 0.072}_{\,- 0.034}$   & $0.100^{\,+ 0.090}_{\,- 0.046}$  \\[0.15cm]
 \hline &&&      \\[-0.3cm] 
$\mathcal{B}(B\to\tau\nu)$ $[10^{-4}]$ (!) & $ 1.36 ^{\,+ 0.10}_{\,- 0.35}$ & $ 1.36 ^{\,+ 0.23}_{\,- 0.59}$ & $ 1.36 ^{\,+ 0.31}_{\,- 0.77}$ \\[0.15cm]
$\mathcal{B}(B\to\tau\nu)$ $[10^{-4}]$  & $ 1.38 ^{\,+ 0.13}_{\,- 0.12}$ & $ 1.38 ^{\,+ 0.23}_{\,- 0.34}$ & $ 1.38 ^{\,+ 0.31}_{\,- 0.52}$ \\[0.15cm]
 \hline
 \end{tabular}
 \end{center}
 \vspace{-0.5cm}
 \caption[.]{\label{tab:fitResults_MFV_III2} %\em
              Fit results in the New Physics scenario III. The notation `(!)' means that the fit output represents the indirect constraint, \textit{i.e.}
              the corresponding direct input has been removed from the analysis. }
 \end{table}

\clearpage
\section{Conclusions}\label{sec:conclusions}

In this paper, we have studied three different 
scenarios of New Physics in $|\Delta F|=2$ transitions. 
The complex parameters quantifying the New Physics contributions to 
\bbmq\ are $\Delta_q\equiv |\Delta_q|\cdot e^{i\phi_d^\Delta} \equiv 
M_{12}^q/M_{12}^{{\rm SM},q}$. In \kkm\ three parameters 
$\Delta_K^{tt}$, $\Delta_K^{ct}$, and $\Delta_K^{cc}$ are needed.

We have first recalled the result of the Standard Model fit using the
current available data sets.  In the $B$ system an interesting effect
is observed: the inclusion of $\BRB{\tau\nu}$ obtained from a
combination of \babar\ and Belle's measurements deviates from its
prediction in the Standard Model fit by $2.9~\sigma$ which either
points to a large statistical fluctuation in the experimental numbers,
to a problem in the calculation of 
both the decay constant $f_{B_d}$ and the bag parameter
$\Bag_{B_{d}}$ on the Lattice, or to New Physics contributions in
$\sin{2\beta}$ and/or in $B \rightarrow \tau\nu$. If there were New
Physics contributions to $B_{d}$ mixing this discrepancy would point
to a negative non-vanishing New Physics phase $\phi^\Delta_d$.  A second hint
for a deviation is observed 
%% when comparing the measurement of the mixing phase $\phi_s$ in $B_{s}$
%% mixing with its theoretically rather clean prediction
in the \bbms\ phase $\phi_s^\Delta-2\beta_s$ measured in $B_s\to J/\psi \phi$
though the significance is here only around $2.3~\sigma$. The
largest discrepancy actually comes from the recent improved
measurement of the dimuonic asymmetry by the \Dzero\ collaboration,
which is at odds at the $3.2~\sigma$ level with respect to the
indirect fit prediction ($2.9~\sigma$ when the average with the CDF
measurement of the same quantity is made).  Furthermore, the
correction factor $\kappa_{\epsilon}$ \cite{bg} in the theoretical
prediction of $\epsilon_{K}$ decreases the quality of the Standard
Model fit. However, with our inputs (especially the range for
  ${\hat{\Bag}}_{K}$ in \tab{tab:TheoreticalInputs}) and with the
conservative Rfit error treatment of theoretical uncertainties used in
our fit we do not observe a significant deviation between the measured
$\epsilon_{K}$ value and its prediction from a Standard Model fit
excluding the $\epsilon_{K}$ measurement.

In our New Physics scenario I, we have considered uncorrelated New Physics
contributions to $B_{d}$, $B_{s}$ and $K$ mixing. That is, the 
complex parameters $\Delta_d$ and $\Delta_s$ are allowed to vary 
independently and the kaon sector is omitted, since the  
%% . We do not
%% include the kaon sector since the only relevant observable in this
%% case, $\epsilon_{K}$, would depend on three independent 
New Physics parameters $\Delta_K^{tt}$, $\Delta_K^{ct}$, and $\Delta_K^{cc}$
are unrelated to all other observables entering the fit.
%% that cannot be constrained in such an analysis (to our knowledge,
%% the parameters $\Delta_K^{ct}$ and $\Delta_K^{cc}$ have not been
%% discussed in the literature so far).
The experimental data are well described in this
scenario which can accomodate negative New Physics phases preferred by 
a)
  the discrepancy between $\BRB{\tau\nu}$ and $\sin{2\beta}$ both
  measured at the B-factories  \babar\ and Belle, b) the
  $2\beta_{s}$ measurements in $B_s\to J/\psi \phi$ at
 the Tevatron, and c) the
  dimuon asymmetry $a_\mathrm{fs}$ measured by \Dzero. The size of the
New Physics contribution both in $B_{s}$-mixing and $B_{d}$-mixing can be as
large as $40~\%$ and the hypothesis of zero new CP phases, 
  $\phi_d=\phi_s=0$,  is disfavoured by as much as $3.8$ standard
  deviations in this scenario (see \tab{ScIpvalues}). 
The large parameter region emphasises that, despite of the
success of the $B$-factories and the Tevatron, there is still
considerable room for New Physics in $B_{d}$ as well as in $B_{s}$ mixing.
%% possibly hidden by the still sizeable theoretical uncertainties in the
%% relevant hadronic parameters.

In addition, we have considered two Minimal Flavour Violation
scenarios. The first MFV scenario, scenario II, corresponds to small
bottom Yukawa couplings, leading to 
$\Delta_d=\Delta_s=\Delta_K^{tt}=\Delta$ with all New Physics phases 
identical to zero, $\phi_s^\Delta=\phi_d^\Delta=
\phi_K^{ij\,\Delta}=0$.  This scenario
$\Delta_d=\Delta_s=\Delta_K^{tt}=\Delta$ with vanishing New Physics phases has
been widely studied in the literature.  
In this scenario, the New Physics parameter $\Delta_K^{cc}$
is equal to one to a very good approximation and $\Delta_K^{ct}$ only
deviates by a few percent from one where the deviation can be
estimated in terms of $\Delta_K^{tt}$. Since in this scenario no New Physics
phases are allowed, the deviations seen in the Standard Model fit are still
present. As a consequence, this MFV scenario is currently 
disfavoured by 3.7 standard deviations, but not totally excluded yet. 
The New Physics parameter $\Delta$ can deviate from one by about $+40~\%$ at $95~\%$
C.L..  The constraint gets only slighly relaxed when removing either
$\BRB{\tau\nu}$ or $\epsilon_{K}$ from the inputs to the fit.

Our scenario III is a generic MFV scenario with large 
bottom Yukawa coupling and arbitrary new flavour-blind CP phases.
In this scenario, the kaon sector is unrelated to the $B$-sector.
As in scenario II, one has $\Delta_d=\Delta_s=\Delta$, however, this
time with a complex parameter $\Delta$. 
The New Physics parameters in $\epsilon_{K}$ are unrelated to $\Delta$ and
$\epsilon_{K}$ can be removed from the input list.
However, when including $\epsilon_{K}$ one is able
to constrain the New Physics parameter $\Delta_K^{tt}$ which is found to be
consistent with the Standard Model value of 1, but can deviate from unity by about
$\pm 40~\%$.  The fit describes the data significantly better than the
Standard Model fit and than the fit in scenario II since the data prefer a
negative New Physics phase in $B_{d}$ and in $B_{s}$ mixings.  
The hypothesis of a zero new CP phase, $\imag \Delta =0$, 
is disfavoured by 3.5 standard deviations (see \tab{ScIIIpvalues}).
As in the
other New Physics scenarios, the allowed size of the New Physics contribution in
$B_{d}$ and in $B_{s}$ mixing can be as large as $+40~\%$.

The several scenarios discussed here show that we have indeed
sensitivity to New Physics in the $|\Delta F|=2$ sector.  While the
  overall picture of current data reveals strong hints for New
  Physics, the current experimental uncertainties prevent us from
  excluding the Standard Model, as highlighted by the $p$-values of
each hypothesis. It has to be seen how this picture evolves with the
improvement of both experimental and theoretical results. When 
we completed this study, new results from the Tevatron
experiments were given for the measurement of $\phi_s^{\psi\phi}$, in better
agreement with the Standard Model, which will be included in our
analysis once the CDF and D\O\ collaborations have agreed on a
combination of their results.  Importantly, more precise
measurement of the CP asymmetries in flavour-specific decays,
  $a_\text{SL}^q$, from either the Tevatron or the LHC experiments,
may become crucial in the future: for the time being the
theory-and-data driven fit predictions for $a_\text{SL}^q$ are more
precise than the direct measurements, as can be verified by
  comparing e.g.\ Tabs.~\ref{tab:ExperimentalInputs} and
  \ref{tab:fitResults_NPBDBS2}.  Hence future more precise data
on $a_\text{SL}^q$ could help to discriminate between the Standard
  Model and different scenarios of New Physics. Meanwhile, our
  predictions of $a_\text{SL}^d$ in Tabs.~\ref{tab:fitResults_NPBDBS2},
  \ref{tab:fitResults_MFV_II} and \ref{tab:fitResults_MFV_III} are an
  important side result of our analyses: they permit a fast
  extraction of $a_\text{SL}^s$ from future measurements of $a_{\rm
    fs}$ (or $a_\text{SL}^s-a_\text{SL}^d$ considered by LHCb), which is
  more accurate than what is obtained using the experimental value of $a_{\rm
    fs}^d$ listed in \tab{tab:ExperimentalInputs}.
   
 As an illustration of this statement, we show  the indirect fit prediction for the difference 
$a_{\text{SL}}^s-a_{\text{SL}}^d$ as a function of $\phi_s^\Delta-2\beta_s$, in the Standard Model and 
the  New Physics Scenarios I and III in Fig.~\ref{fig-DeltaAsl-psiphi}. The prediction in Scenario I and III differs by about 
two standard deviations, hence a precise direct measurement of either 
observable could not only exclude the Standard Model, but also select one of 
the New Physics scenarios. Thanks to the specific two-dimensional correlation it could 
also invalidate all scenarios, which would then imply that there are other 
sources of New Physics than just contributions to the mixing amplitudes. The 
LHCb experiment is expected to measure both CP asymmetries, with an accuracy 
of about $\sigma({\phi_s^\Delta-2\beta_s})=4^\circ$ and 
$\sigma({a_{\text{SL}}^s-a_{\text{SL}}^d})=20\times 10^{-4}$ for 1~fb$^{-1}$ of integrated 
luminosity~\cite{RavenLHCb}.
   
   \begin{nfigure}{Htb}
  \includegraphics[width=12cm]{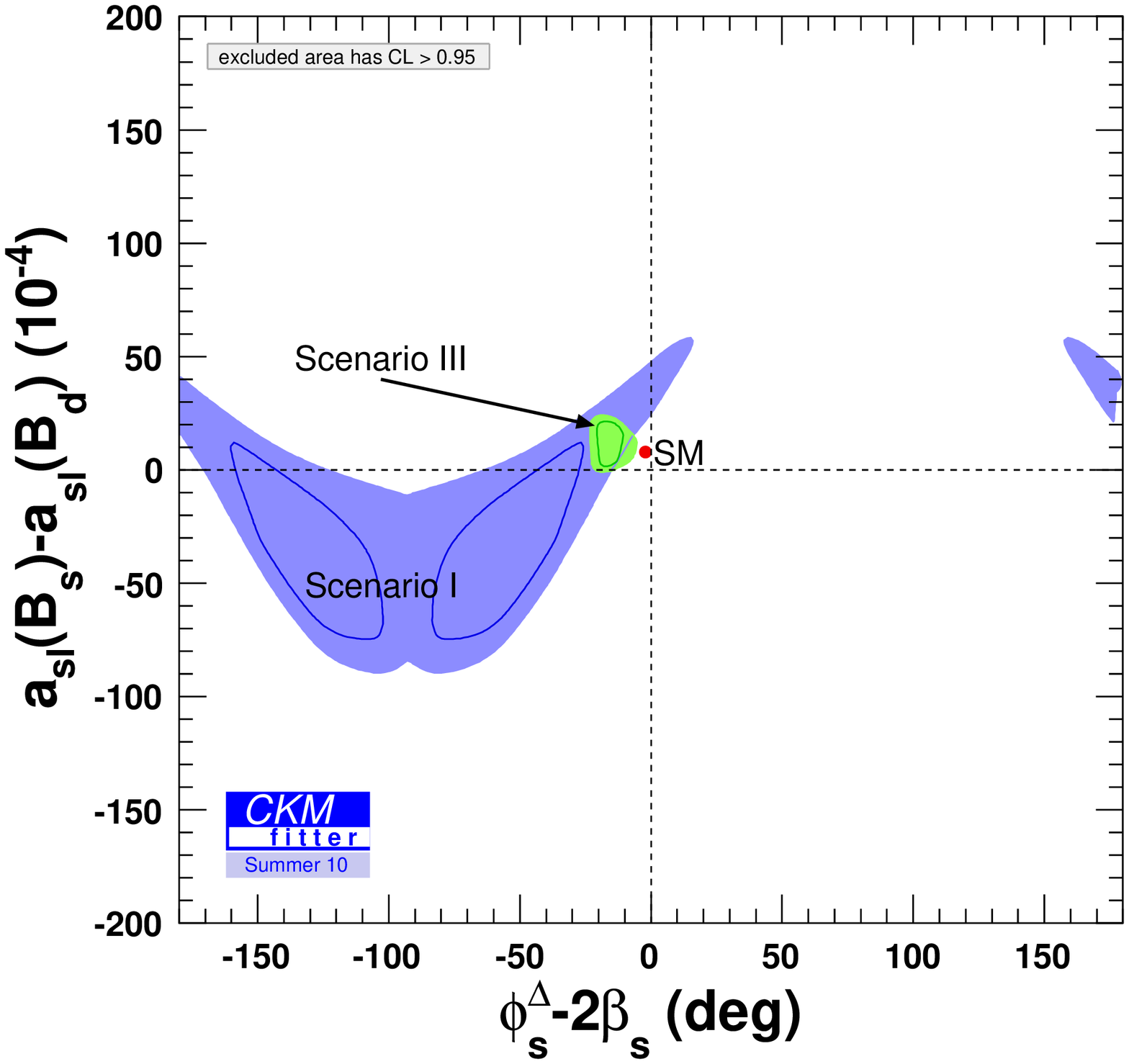}
  \caption{\small  Indirect fit prediction for the difference 
$a_{sl}^s-a_{sl}^d$ as a function of $\phi_s^\Delta-2\beta_s$, in the Standard Model and 
the New Physics Scenarios I and III. The allowed regions correspond to ${\rm CL} < 95.45~\%$. The 
prediction from Scenario II is not shown since it is very close to the SM. Note that for this plot, 
the direct measurement of $\phi_s^\Delta-2\beta_s$ in $B_s\to J/\psi\phi$ is removed from the inputs.
                  \label{fig-DeltaAsl-psiphi}}
\end{nfigure}

  On the other hand, significant improvements on
  the measurement of $\BRB{\tau\nu}$ can only be expected from a
  Super-B-like electron-positron machine since the \babar\ and Belle
  results do already rely on most of the available statistics
  collected at the B-factories PEP-II and KEKb. Since the theoretical
  translation of the ratio $\BRB{\tau\nu}/\dm_d$ into a constraint on
  CKM and New Physics parameters relies on the calculation of the decay
  constant $f_{B_{d}}$ and the bag parameter $\Bag_{B_{d}}$, future
    progress in Lattice QCD calculations is also very important. 
  The same remark applies to the hadronic matrix elements  
   entering $M_{12}^s$ and $\Gamma_{12}^s$. We hope that the current 
   exciting  experimental situation will stimulate novel activities
   in lattice gauge theory in this direction.

\section*{ACKNOWLEDGMENTS}
We thank D.~Becirevic, D.~Guadagnoli, L.~Lellouch, P.~Marquard, M.~Neubert and
N.~Uraltsev for stimulating discussions and comments. We are
  grateful to A.~Kagan for bringing a mistake in \eq{dectdett} in
  an earlier version of this paper to our attention.  We would also
like to thank the members of the CKMfitter group for their interest and
support for the various aspects of this article.

This work was partially supported by the ANR contract ANR-06-JCJC-0056 and
the EU Contract No. MRTN-CT-2006-035482, 
FLAVIAnet. The work of UN is supported by the DFG through project C6 of 
the collective research centre SFB-TR9 and by BMBF grant 
05H09VKF.

\appendix

\boldmath
\section{Relationship between $\epsilon_K$ and \kkm}\label{app:kappaepsilon}

\subsection{Corrections to the usual $\epsilon_K$ formula}
\unboldmath In Sect.~\ref{ssec:kkmbasics}, we have discussed the
connection between $\epsilon_K$ and \kkm. The resulting equation,
\eq{eq:epsilon0}, has been obtained thanks to several approximations
(concerning the phase $\phi_\epsilon$, neglect of $\xi$, computation
of $M_{12}$ from lowest-dimension operators of the effective
Hamiltonian).  The corrections to these approximations are encoded in
the deviation of the factor $\kappa_\epsilon$ from 1.  A series of
papers~\cite{run2,Andriyash:2003ym,bg,bgi} has assessed more precisely
the value of this factor in order to account for the terms neglected
by the previous approximations.

$\xi\neq0$ and $\phi_\epsilon\neq 45^\circ$ imply $\kappa_\epsilon \neq 1$, 
Ref.~\cite{bg} finds $\kappa_\epsilon = 0.92 \pm 0.02$.
In Ref.~\cite{LLVdW} a lattice QCD
calculation of $\imag A_2$ is combined with the experimental value of
$\epsilon'_K/\epsilon_K$ to compute $\xi$ and finds 
agreement with Ref.~\cite{bg} while quoting
an even smaller uncertainty: $\kappa_\epsilon = 0.92 \pm 0.01$. 
Finally, correcting for $\imag M_{12}^K\neq \imag
M_{12}^{(6)}$ by including higher-order terms of the operator product
expansion leads to $\kappa_\epsilon = 0.94\pm 0.02$.
Actually, in the correction factor $\kappa_\epsilon$, the three
approximations have to be treated in different ways, since they mix
uncertainties from experimental and theoretical origins. The correction
from $\phi_\epsilon$ is of experimental nature and can be treated
easily.

The correction involving $\xi$ could be computed by combining the
experimental value of ${\rm Re}\ A_0$ and the theoretical computation of
${\rm Im}\ A_0$ using the effective Hamiltonian $H^{|\Delta S|=1}$. 
However, the latter is dominated by the matrix element of the
QCD penguin in the $I=0$ channel $\langle (\pi\pi)_0 |Q_6|K\rangle$,
which is poorly known. Here we follow the method of Refs.~\cite{Andriyash:2003ym,bg,LLVdW} 
which uses $\epsilon'_K$ to correlate $\langle (\pi\pi)_0 |Q_6|K\rangle$ with
$A_2$. The latter amplitude involves the matrix element of the
electroweak penguin $Q_8$ in the $I=2$ channel, $\langle
(\pi\pi)_2|Q_8|K\rangle$, which has been computed using lattice
simulations and sum rules. Indeed, one finds \cite{bg}
\begin{equation}\label{eq:Omega}
\frac{\epsilon'_K}{\epsilon_K}= 
         -\omega\frac{\xi}{\sqrt{2}|\epsilon_K|}(1-\Omega),
\qquad\quad \mbox{with}~~~
\omega=\frac{{\rm Re}A_2}{{\rm Re A_0}},\qquad
\Omega=\frac{1}{\omega}\frac{{\rm Im}A_2}{{\rm Im} A_0}.
\end{equation}
$\omega =0.0450$ is obtained from experiment, whereas $\Omega$
describes the weight of the (imaginary part of the) $I=2$ contribution
with respect to the $I=0$ one. 

In practice, a numerical estimate of the contributions from other
(subleading) operators than $Q_6$ and $Q_8$ in $H^{|\Delta S|=1}$ has
been obtained~\cite{BurasLectures,BurasJamin,DG} under the assumptions
that ${\rm Im}\ A_0$ and ${\rm Im}\ A_2$ can be computed accurately
combining the effective Hamiltonian approach and experimental values for
${\rm Re}\ A_0$ and ${\rm Re}\ A_2$:
\begin{equation}\label{eq:epsprimeovereps}
\frac{\epsilon'_K}{\epsilon_K} = N_0 + N_1 \cdot R_6 + N_2 \cdot R_8, 
\qquad \qquad R_6=R_6[(\epsilon'/\epsilon)_{\rm exp},R_8]
\end{equation}
where $N_0, N_1, N_2$ are numbers, coming mainly from $\lambda_t$, the
experimental values for the real parts of $A_0$ and $A_2$, and $R_6$ and
$R_8$ are rescaled bag parameters.  Following the review of
ref.~\cite{Cirigliano}, the authors of ref.~\cite{BurasJamin} proposed 
the following conservative range
$R_8=1.0\pm 0.2$ which we will follow. Using
$\epsilon'_K/\epsilon_K=(1.65\pm 0.26)\cdot 10^{-3}$, one can use
eq.(\ref{eq:epsprimeovereps}) to determine $\Omega$, which
corresponds to the ratio between the $I=2$ and $I=0$ contributions to $\epsilon'_K/\epsilon_K$.
In principle, this would require to split $N_0$ into two pieces coming respectively from $I=0$ and $I=2$, and to assess the size of these contributions following Ref.~\cite{BurasLectures}. A quicker way to obtain a similar estimate consists in computing~\cite{DG}:
\begin{equation}
\qquad
\Omega_1 = \frac{N_2 \cdot R_8}{N_0 + N_1\cdot R_6}
\qquad 
\Omega_2 = \frac{N_0 + N_2 \cdot R_8}{N_1\cdot R_6}
\label{romega12}
\end{equation}
where $\Omega_1$ and $\Omega_2$ correspond to the (extreme) hypothesis that $N_0$ is saturated either by $I=0$ or $I=2$ contributions.
Eq.~(\ref{eq:Omega}) can be then used to compute $\xi$ in either hypotheses, and we will take the spread of the obtained values as a (conservative) systematic uncertainty in the determination of $\xi$. A more detailed analysis of the contributions to $N_0$ would allow us to decrease this systematic uncertainty.

The last correction comes from higher-dimension contributions to ${\rm
  Im} M_{12}$. As shown in ref.~\cite{bgi}, there are two different
corrections at $d=8$, corresponding to the $\Delta S=2$ $d=8$ operators
and the double insertion of $\Delta S=1$ operators connected by a $u,c$
loop. The first is expected to be very suppressed compared to the $d=6$
contributions, whereas the second one is essentially dominated by
long-distance pion exchanges estimated in Chiral Perturbation Theory for
weak processes, leading to:
\begin{equation}
\epsilon_K=\sin\phi_\epsilon e^{i\phi_\epsilon} \left[\frac{{\rm Im} M_{12}^{(6)}}{\Delta M} + \rho\xi \right], \qquad\qquad
 \rho=0.6\pm 0.3.
\end{equation}

For our purposes, it will be simpler to combine these corrections with
the experimental input for $\epsilon_K$:
\begin{equation}
  \epsilon_K^{(0)}=\epsilon_K\left(\frac{1}{\sqrt{2}\sin\phi_\epsilon}+\rho\frac{\epsilon'_K}{\epsilon_K}
    \frac{1}{\omega(1-\Omega)}\right)\equiv 
    \frac{\epsilon_{K,exp}}{\kappa_\epsilon}
\end{equation}
where $\epsilon_K^{(0)}$ denote the approximate value of $\epsilon_K$ in
eq.~(\ref{eq:epsilon0}) with $\kappa_\epsilon=1$. Depending on the
choice of $\Omega$ (i.e. the respective part of $I=0$ and $I=2$ contributions in
the formula for $\epsilon'/\epsilon$), we get
\begin{eqnarray}
\kappa_\epsilon^{(1)} &=& 0.943\pm 0.003 (\epsilon'/\epsilon)\pm 0.012 (\phi_\epsilon)\pm 0.004 (R_8)\pm 0.015 (\rho)\\
\kappa_\epsilon^{(2)}
  &=& 0.940 \pm 0.003 (\epsilon'/\epsilon)\pm 0.012 (\phi_\epsilon)\pm 0.004 (R_8)\pm 0.018 (\rho)
\end{eqnarray}
Combining the first two errors in quadrature for the Gaussian part and the last two errors linearly and taking the spread of the values into the Rfit part, we obtain the estimate
\begin{equation}
\kappa_\epsilon = 0.940 \pm 0.013 \pm 0.023
\end{equation}
in good agreement with 
% the fact that $\lambda_\epsilon $ is the inverse of $\kappa_\epsilon$,
% and with the estimate
$\kappa_\epsilon=0.94\pm 0.02$ in ref.~\cite{bgi}. This determination relies on the assumption that $\epsilon'_K$ is not affected by New Physics.

\subsection{Error budget for $\epsilon_K$}
Recently, it has been claimed that there is a discrepancy between the $\epsilon_K$
constraint and its prediction~\cite{bg,smtensions}. With the input values used in our fit 
and with the Rfit treatment of theoretical uncertainties we do not observe any 
sizeable discrepancy when comparing the prediction for $\epsilon_{K}$ 
(Table~\ref{tab:fitResults_SM1}) with its measured value 
(Table~\ref{tab:ExperimentalInputs}). 
This can also be seen in Fig.~\ref{fig-rhoeta_smfit} where the combined fit prefers 
a region in $\rhobar$ and $\etabar$ that is close to the edge of, though still inside
the $95~\%$ CL region of the $\epsilon_{K}$ constraint.

As illustrated in Fig.~\ref{fig-epsK} one can obtain a minor discrepancy 
at 1.2 $\sigma$ if one treats as Gaussian all the parameters involved in \eq{eq:epsilon0} 
(i.e., $|V_{cb}|$, $\Bag_K$, $\kappa_\epsilon$, the QCD correction terms $\eta_{cc,ct,tt}$ 
and the masses $\bar{m}_{c,t}$), but such a treatment of the systematics remains questionable.
Another way of seeing the absence of discrepancy is to compare the prediction for 
$\kappa_\epsilon$ and estimates of these quantity. One can see clearly from 
Fig.~\ref{fig-kappa} that the global fit can cope easily with a value of $\kappa_\epsilon$ 
down to 0.9, and that the  prediction from the global fit agrees well with the recent 
estimates of this quantity.

\begin{nfigure}{!Htb}
  \includegraphics[width=10cm]{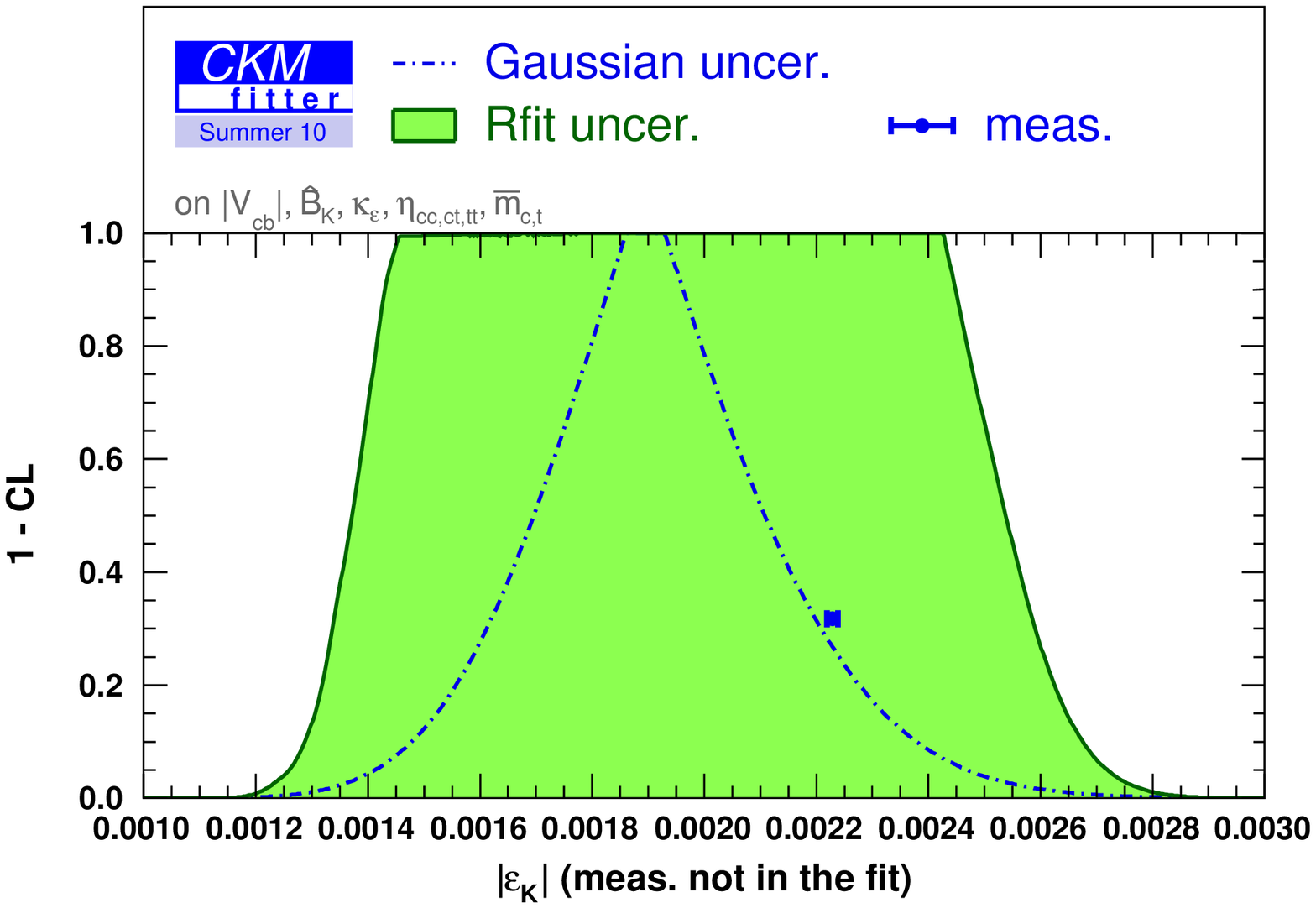}
  \caption{\small Constraint on $|\epsilon_K|$ predicted from the global fit,
 compared to the experimental value. The green curve is obtained with our Rfit 
 treatment of the uncertainties for the parameters entering \eq{eq:epsilon0}.  
 The dashed curve is obtained by treating all the errors as Gaussian.
\label{fig-epsK}}
\end{nfigure}
\begin{nfigure}{!Htb}
  \includegraphics[width=10cm]{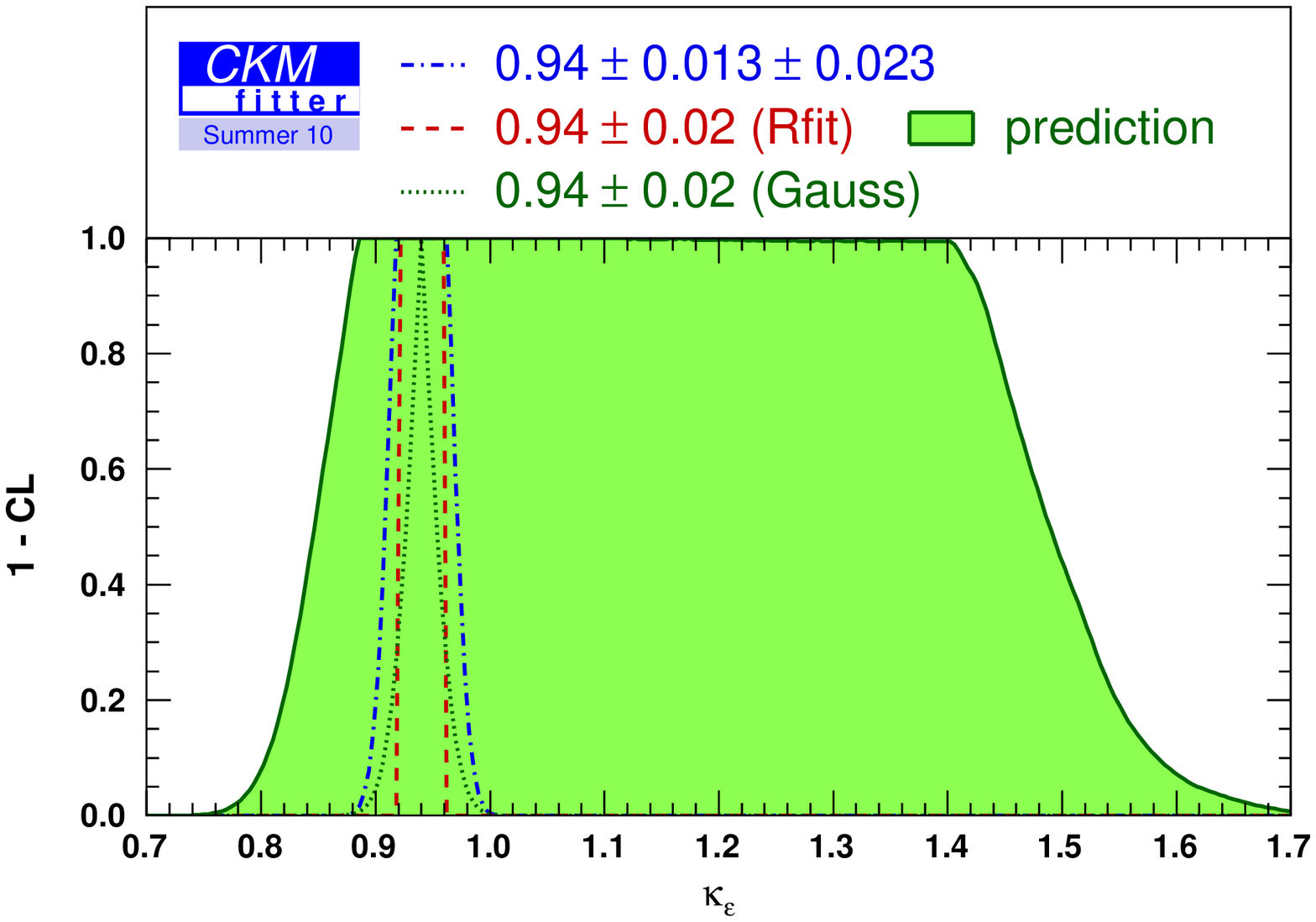}
  \caption{\small Constraint on $\kappa_\epsilon$ predicted from the global fit 
   using \eq{eq:epsilon0}, compared to three different theoretical inputs 
   $\kappa_\epsilon=0.94\pm 0.02$ with either a Rfit or a Gaussian treatment of 
   the error, or $\kappa_\epsilon=0.94\pm 0.013\pm 0.023$ corresponding to our 
   own treatment of the uncertainties involved in its evaluation. 
                 \label{fig-kappa}}
\end{nfigure}

In order to make the comparison of our $|\epsilon_K|$ analysis with
the one of Ref.~\cite{LLVdW} easier we treat all the errors as
Gaussian and calculate the error budget for the fit prediction of
$|\epsilon_K|$. With our inputs we find
\begin{eqnarray}
10^3|\epsilon_K|&=& 1.893 \pm 0.020 
  |_A \pm 0.020|_\lambda \pm 0.063 
    |_{(\bar\rho,\bar\eta)} \nonumber\\
                & &       \ \ \ 
  \pm 0.180|_{\Bag_K} \pm 0.019|_\mathrm{top} 
  \pm 0.084|_\mathrm{charm} \pm 0.054|_\mathrm{\kappa_\epsilon}
                \nonumber\\
                &=& 1.89^{+0.26}_{-0.23}
\end{eqnarray}
which is $1.2~\sigma$ away from the experimental measurement, while with the inputs of the last reference in~\cite{LLVdW} we find
\begin{eqnarray}
10^3|\epsilon_K|&=& 1.769 \pm 0.019|_A \pm 0.019|_\lambda \pm 0.061|_{(\bar\rho,\bar\eta)} \nonumber\\
                & &       \ \ \ \pm 0.062|_{\Bag_K} \pm 0.018|_\mathrm{top} \pm 0.067|_\mathrm{charm} \pm 0.033|_\mathrm{\kappa_\epsilon}
                \nonumber\\
                &=& 1.77 ^{+0.18}_{-0.16}
\end{eqnarray}
which is $2.4~\sigma$ away from the experimental measurement, in agreement with~\cite{LLVdW}. Hence we see that the difference between our analysis and
Ref.~\cite{LLVdW} mainly comes from our input for $\Bag_K$ that has a larger theoretical error, and from our central value for $|\epsilon_K|$ that is larger
because of the different analysis of the CKM parameters. We thus conclude 
that the potential anomaly in $|\epsilon_K|$ cannot yet be
precisely quantified independently of the theoretical inputs and 
therefore deserves further investigations.

\end{thebibliography}

\end{document}